\documentclass[12pt,a4paper]{article}
\pdfoutput=1
\usepackage{jheppub}

\usepackage{bm}
\usepackage{amsmath}
\usepackage{amssymb}
\usepackage{amsthm}
\usepackage{amsfonts}
\usepackage{ccaption}
\usepackage{booktabs}

\usepackage{mathrsfs}
\usepackage{subcaption}

\makeatletter
\@addtoreset{equation}{section}

\makeatletter
\renewcommand\section{\@startsection {section}{1}{\z@}%
                                   {-3.5ex \@plus -1ex \@minus -.2ex}
                                   {2.3ex \@plus.2ex}%
                                   {\normalfont\large\bfseries}}
\renewcommand\subsection{\@startsection{subsection}{2}{\z@}%
                                     {-3.25ex\@plus -1ex \@minus -.2ex}%
                                     {1.5ex \@plus .2ex}%
                                     {\normalfont\bfseries}}

  \captionnamefont{\bfseries}
  \captiontitlefont{\small\sffamily}
  \captiondelim{: }
  \hangcaption

\def\sect#1{\S\ref{#1}}
\def\fig#1{Fig.\,\ref{#1}}
\def\req#1{(\ref{#1})}

\definecolor{rust}{rgb}{0.8,0.2,0.2}

\def\rh{r_+}
\def\rs{r_s}

\def\SAdS{Schwarzschild-AdS}
\def\io{\alpha}
\def\ph{\varphi}
\def\length{\ell}

\def\CN{{\cal N}}

\def\ETEBA{ETEBA}
\def\eteba{ETEBA}

\title{Holographic probes of collapsing black holes}

\author{Veronika E. Hubeny}
\author{ \& Henry Maxfield}

\affiliation{ Centre for Particle Theory \& Department of Mathematical Sciences,\\
Science Laboratories, South Road, Durham DH1 3LE, UK.}\emailAdd{veronika.hubeny@durham.ac.uk}
\emailAdd{h.d.maxfield@durham.ac.uk}

\abstract{
We continue the programme of exploring the means of holographically decoding the geometry of spacetime inside a black hole using the gauge/gravity correspondence.  To this end, we study the behaviour of certain extremal surfaces (focusing on those relevant for equal-time correlators and entanglement entropy in the dual CFT) in a dynamically evolving asymptotically AdS spacetime, specifically examining how deep such probes reach.  To highlight the novel effects of putting the system far out of equilibrium and at finite volume, we consider spherically symmetric Vaidya-AdS, describing black hole formation by gravitational collapse of a null shell, which provides a convenient toy model of a quantum quench in the field theory.
 
Extremal surfaces anchored on the boundary exhibit rather rich behaviour, whose features depend on dimension of both the spacetime and the surface, as well as on the anchoring region.  The main common feature is that they reach inside the horizon even in the post-collapse part of the geometry.  In 3-dimensional spacetime, we find that for sub-AdS-sized black holes, the entire spacetime is accessible by the restricted class of geodesics whereas in larger black holes a small region near the imploding shell cannot be reached by any boundary-anchored geodesic.  In higher dimensions, the deepest reach is attained by geodesics which (despite being asymmetric) connect equal time and antipodal boundary points soon after the collapse; these can attain spacetime regions of arbitrarily high curvature and simultaneously have smallest length.  Higher-dimensional surfaces can penetrate the horizon while anchored on the boundary at arbitrarily late times, but are bounded away from the singularity.

We also study the details of length or area growth during thermalization. While the area of extremal surfaces increases monotonically, geodesic length is neither monotonic nor continuous.
} 

\arxivnumber{1312.6887}

\begin{document}


\maketitle

\flushbottom
\renewcommand{\thefootnote}{\arabic{footnote}}

\section{Introduction and summary}
\label{s:intro}

The gauge/gravity correspondence\footnote{
For definiteness we'll mostly focus on the prototypical case of the AdS/CFT correspondence \cite{Maldacena:1997re} 
which relates the four-dimensional $\CN=4$ Super Yang-Mills  (SYM) gauge theory to a IIB string theory (or supergravity) on asymptotically AdS$_5 \times S^5$ spacetime.
}  has proved invaluable in providing useful insights into the behaviour of strongly coupled field theories, yet the converse quest of using the field theory to understand quantum gravity in the bulk is still far from reaching its fruition.
Already from the outset, one of the key obstacles is our incomplete understanding of bulk locality.  Questions of how the field theory encodes bulk geometry and causal structure, or how it describes a local bulk observer, remain opaque despite intense efforts of the last 15 years.\footnote{
Early investigations of bulk locality from various perspectives include
\cite{Banks:1998dd,Polchinski:1999ry,Susskind:1998vk,Giddings:1999jq,Horowitz:2000fm,Hamilton:2005ju}
whereas more recent developments and reviews are given in e.g.\
\cite{Gary:2009mi,Heemskerk:2009pn,Penedones:2010ue,Kabat:2011rz,Fitzpatrick:2011jn,Heemskerk:2012mn,Raju:2012zr,Raju:2012zs}.
}
While the scale/radius duality provides us with valuable intuition in the asymptotic bulk region, the mapping becomes far more obscure deeper in the bulk, and inapplicable for bulk regions which are causally separated from the boundary.  The question {\it `how does the gauge theory see inside a bulk black hole?'} has been foremost from the start.

Causal considerations aside, the context of black hole geometry holds a particularly sharp testing ground for quantum gravity, as the curvature singularity inside a black hole is a region near which classical general relativity breaks down.  Nevertheless, the gauge theory contains the full physics
-- it understands what resolves or replaces this classical singularity in the bulk.  But for learning the answer of the gauge theory, we first need to understand what question to ask in that language:
in what field theoretic quantity can we isolate the near-singularity behaviour?  

One of the approaches aimed at elucidating the encoding of bulk geometry in the dual field theory observables was recently undertaken in \cite{Hubeny:2012ry}, which explored how much of the bulk spacetime is accessible to certain field theory quantities related to specific geometrical probes in the bulk.  In particular, \cite{Hubeny:2012ry} focused on field theory probes characterised by bulk geodesics and more general extremal surfaces,  anchored on the AdS boundary.  Being geometrical by nature, such probes are well-suited for decoding the bulk geometry,\footnote{  
For example, following  \cite{Hubeny:2006yu}, the works \cite{Hammersley:2006cp,Bilson:2008ab} demonstrated that one can reconstruct the bulk metric for an arbitrary static and spherically symmetric bulk geometry, simply from knowing the proper length/area/volume of such surfaces along with where they end on the boundary.
} at least at the classical level.
At the same time, they are related to well-defined CFT quantities: certain types of correlators for spacelike geodesics or bulk-cone singularities for null geodesics \cite{Hubeny:2006yu}; Wilson-Maldecena loops  for 2-dimensional surfaces \cite{Maldacena:1998im,Rey:1998ik}; and entanglement entropy for codimension-two surfaces \cite{Ryu:2006bv,Ryu:2006ef,Hubeny:2007xt}.
Hence, postponing for the moment the discussion of the subtleties of the actual relation between these CFT `observables' and the corresponding bulk geometrical constructs, we will follow the approach of \cite{Hubeny:2012ry} in asking how much these bulk geometrical quantities, i.e.\ extremal surfaces anchored on the AdS boundary, know about the geometry, now specifically focusing on the black hole interior.

Perhaps the most intriguing result of  \cite{Hubeny:2012ry}  was  that extremal surfaces anchored on the boundary of AdS cannot penetrate through the horizon of a static black hole.  Of the several arguments provided, the most general of these, which analysed the equation of motion near its turning point, 
applies to an extremal surface of any dimension, anchored on any shape of simply-connected boundary region, in any static asymptotically-AdS spacetime with planar symmetry and event horizon.  Nevertheless, as emphasised in \cite{Hubeny:2012ry}, extremal surfaces are able to penetrate the event horizon of a dynamically-evolving black hole.  This is simply because the event horizon is a global construct whose location depends on the full future evolution, whereas the location of extremal surfaces is determined by the local geometry.

Indeed, this  observation formed the basis of \cite{Hubeny:2002dg} which used it to argue that the event horizon by itself is not an obstruction to precursor-type CFT probes.  In particular, \cite{Hubeny:2002dg} presented a simple gedanken experiment wherein a thin null shell implodes from the AdS boundary and forms a large black hole.
The bulk geometry is pure AdS to the past of the shell and Schwarzschild-AdS to its future; however the horizon generators (outgoing radial null geodesics which define the late-time static horizon) originate at the center of AdS prior to the shell.  
In particular, a bulk constant-time\footnote{
In a static part of the spacetime, there is a geometrically unique bulk foliation by `constant time' surfaces that are anchored at a fixed boundary time.
}
slice, anchored on AdS boundary shortly before the creation of the shell, passes through the AdS region enclosed by event horizon. 
Spacelike geodesics as well as higher-dimensional extremal surfaces in AdS which are anchored at this time will lie along the same time slice, and will therefore penetrate the black hole as long as their anchoring region is sufficiently large.

However, it not clear in this example what bulk regions remain inaccessible to such probes. In particular it is unclear whether it is possible to probe past the event horizon in the more genuine black hole geometry to the future of the shell, and to penetrate near the curvature singularity, and therefore to be useful in addressing the most interesting question of what happens there.\footnote{
In fact, it would even be interesting to sample the late-time horizon itself, in the recently-explored context of firewalls \cite{Almheiri:2012rt,Almheiri:2013hfa}, where semiclassical physics breaks down already at the horizon.}
Nevertheless, this argument makes it clear that we should be able to use extremal surfaces to penetrate the horizon even after the shell has formed the black hole, as long as the black hole is still evolving.  This is the question we set out to explore: {\it how deep into the collapsed black hole, and especially how close to the curvature singularity, can extremal surfaces penetrate?}

Although obtaining strongly time dependent black hole solutions in general relativity is typically a daunting process due to the non-linearity of Einstein's equations, there are certain solutions with sufficient symmetry which are known analytically.  Perhaps the simplest and best-known of these is the Vaidya (in our gauge/gravity context,  Vaidya-AdS) class of solutions.  These describe a spherically symmetric collapsing null dust, where we are free to specify the radial (or equivalently temporal) profile of the shell.  Early-time geometry (inside or before the shell) is pure AdS, while at late times (outside or after the shell), it is \SAdS.  The `dust particles' making up the shell follow ingoing radial null trajectories, so the black hole forms maximally rapidly.  This is particularly useful in the present context: since we seek a feature which is absent in static geometries, we are more likely to see a large effect for geometries which are as far away from being static as possible.

There is another motivation for probing duals of such collapse geometries coming from field theory. Since a large black hole in the bulk corresponds to a thermal state in the dual field theory, collapse to a black hole describes the process of thermalization.  Moreover, if the collapse is rapid, the dynamics describes a far-from-equilibrium process.  While we typically have a good handle on equilibrium situations, out of equilibrium processes are more interesting but far less understood.
Sudden changes in the field theory Hamiltonian, known as  ``quantum quenches", and subsequent equilibration have been much studied in field theory, 
and have recently received mounting attention from holographic studies.
The analysis of thermalization using  (global, i.e.\ spherically symmetric) Vaidya-AdS as toy-model for quantum quench initiated in \cite{Hubeny:2007xt} was extended in the planar case by 
\cite{AbajoArrastia:2010yt,Aparicio:2011zy} in 3-dimensional bulk, by
\cite{Albash:2010mv} in 4 dimensions, by
\cite{Balasubramanian:2010ce,Balasubramanian:2011ur} in 3,4 and 5 dimensions 
(the latter having used entanglement entropy as well as equal time correlators and Wilson loop expectation values), and more recently by \cite{Liu:2013iza,Liu:2013qca} with more general considerations.\footnote{
See also \cite{Lowe:2008ra,Albash:2011nq,Albash:2012pd,Baron:2012fv,Caceres:2012em,Galante:2012pv,Fischler:2013fba} and references therein for other  explorations of holographic entanglement entropy as a probe in different contexts.
For a more extensive review of the earlier work, see e.g.\ \cite{Hubeny:2010ry} and references therein.}
While most of these works focused on the thermalization aspect, less attention was paid to the question of how much of the collapsing black hole can such probes access, along the lines motivated by \cite{Hubeny:2012ry}.  Hence, apart from the interest in further exploration of holographic thermalization, the question of probing inside the black hole motivates us to continue the study of (global) Vaidya-AdS, using spacelike geodesics and codimension-two extremal surfaces as probes.
The use of these probes was more fully justified in \cite{Hubeny:2012ry} and many of the references mentioned above; here we simply employ the same rationale in exploring them further.

Perhaps the greatest novelty in our findings stems from the fact that, motivated by creating a black hole with compact horizon, we are working with asymptotically globally AdS spacetimes, rather than the (geometrically simpler and more often studied) asymptotically Poincar\'e AdS spacetimes.  While the global case includes the planar Poincar\'e case as a special limit, the converse is not true:  
the possibility of geodesics and surfaces which can `go around' a spherical black hole allows for a vastly richer structure.  This was evident already in the recent study \cite{Hubeny:2013gta} involving extremal surfaces in the static spherical Schwarzschild-AdS black hole, where it was demonstrated that in a wide region of parameter space, there are infinitely many of extremal surfaces anchored on the same boundary region, unlike the planar case where there is just one.
Correspondingly, working with field theory on the spatially compact Einstein Static Universe describing the boundary of global AdS allows us to explore interesting finite-volume effects, which would have been absent in the non-compact case.

Having motivated the spacetime of interest, specifically the global Vaidya-AdS class of spacetimes (with shell thickness, final black hole size, and dimension of the spacetime left as free parameters that we can dial), let us now specify our probes.  From previous studies such as \cite{Hubeny:2012ry}, it is clear that extremal surfaces of different dimensionality can behave qualitatively differently from each other; nonetheless there is a certain `monotonicity' of the behaviour in terms of dimension.  The greatest qualitative difference occurs between geodesics and higher-dimensional surfaces, and the greatest difference from the geodesic behaviour occurs for surfaces of highest possible dimension.  This motivates us to focus on spacelike geodesics and codimension-two extremal surfaces, whose lengths and areas characterize certain types of correlators and entanglement entropy respectively in the dual CFT.
Note that in 3-dimensional bulk, spacelike geodesics coincide with codimension-two extremal surfaces. 
Although in this special case many features trivialize (and as has already been well-appreciated, the BTZ black hole singularity behaves fundamentally differently from higher dimensional black hole singularities \cite{Fidkowski:2003nf}), it will nevertheless be instructive to include Vaidya-AdS$_3$ in our explorations, in order to draw contrast with the higher-dimensional case.


The plan of the paper is as follows.  
In \sect{s:Vaidya} we describe the class of bulk geometries which we will use, namely the Vaidya-AdS  spacetimes, and explain the coordinates for presenting our results graphically. 
We then turn to examining the CFT probes of this geometry, starting with spacelike geodesics in  \sect{s:geods}, first focusing on the higher dimensional case in \sect{VaidyaSAdSgeods} and then contrasting this with the Vaidya-BTZ case in \sect{VaidyaBTZgeods}, where we can supplement our numerical results by closed-form expressions for the key quantities.
 In \sect{s:surfsVaidya} we turn to bulk codimension-two extremal surfaces in higher-dimensional Vaidya-AdS, and we conclude with a discussion in \sect{s:disc}.  The more involved technical details are relegated to the appendices so as to avoid breaking the flow of the presentation. In the remainder of this section we give a preview of the main results.
 
In \sect{s:geods} we consider geodesics with both endpoints anchored on the boundary, which we dub `boundary-anchored' geodesics. We observe that every point in the bulk spacetime (in the higher dimensional cases) lies along some boundary-anchored spacelike geodesic.  However, the closer this point lies to the singularity, the more nearly-null will such a geodesic be, which in turn means that its endpoints will in general be temporally separated.  This motivates us to restrict attention  to spacelike geodesics, whose endpoints lie at equal time on the boundary (dubbed `equal-time-endpoint boundary-anchored' or \ETEBA\ for short), which are the ones relevant for encoding equal-time correlators in the field theory. One way to achieve this is for the geodesic to be symmetric under swapping the endpoints, though we find that this is not the only option.

When both endpoints are located prior to the shell, the entire geodesic remains in the AdS part of the spacetime.  This means that such \ETEBA\  geodesics are constant-time geodesics in the pure AdS part of the spacetime, which cannot penetrate near the singularity. For a short time soon after the quench, we find that there exists a class of geodesics which are not symmetric under reversing their affine parameter, and yet still have endpoints at equal times.  This class is not only more novel than the symmetric geodesics (since it does not appear in static spacetimes), but also important, as in a certain regime of the parameter space such asymmetric geodesics can reach closer to the singularity, and furthermore are shorter than the symmetric ones.  Indeed, if the endpoints are taken to be antipodal and occur arbitrarily soon after the shell, the corresponding shortest geodesic is nearly null, crossing the shell near its implosion at the origin, and has arbitrarily small length. While the asymmetric geodesics exist only up until some finite endpoint time, there are symmetric \ETEBA\ geodesics reaching the boundary at arbitrarily late times, yet sampling the interior of the horizon. These necessarily cross the shell to circumvent the arguments of \cite{Hubeny:2012ry}. 

We then restrict the search further, to consider only the shortest geodesics for given endpoints. The intricate nature of the results is illustrated in \fig{f:antipodal}, which plots the regularised proper length $\length$ along all families of geodesics which join antipodal points at time $t$, as a function of $t$. Curiously, the minimum $\length$ for this set of curves jumps discontinuously, not once, but in fact four times (twice down and twice up), before the thermal value is achieved.

Having mapped the space of initial conditions for the geodesics to the space of corresponding boundary parameters (namely the length and position of the endpoints), we turn to identifying what part of the spacetime is actually probed by shortest \ETEBA\ geodesics.
 We find that even this most restrictive class allows access to a spacetime region inside the horizon and simultaneously to the future of the shell, as indicated in \fig{f:geodesicRegion}.  However, this region is limited to relatively short time after the shell, and late-time near-singularity regions remain inaccessible.  

These results are in contrast to the analogous results for the 3-dimensional (Vaidya-BTZ) spacetime.   In this case, geodesics exhibit qualitatively different behaviour for small black hole ($\rh<1$) as opposed to large black hole ($\rh>1$) spacetimes. 
We first specialise to radial \ETEBA\ (which in this case implies symmetric) geodesics.  These only probe a part of the spacetime (except for the special case of $\rh=1$ when the entire spacetime is accessible), though the character of the unprobed region changes depending on whether the black hole is small or large.
This behaviour is illustrated on spacetime plots in \fig{f:VBTZ_rad_geods} and  on the corresponding Penrose diagrams in \fig{f:VBTZ_rad_geods_PD}.
Adding angular momentum however has a dramatic effect: in the case of small black holes, the entire spacetime becomes accessible, even by \ETEBA\ geodesics.  On the other hand, for large black holes, the deepest boundary-anchored symmetric radial geodesic in fact bounds the region accessible to {\it any} boundary-anchored geodesic.   In other words, for large black holes, a certain region of spacetime still remains inaccessible.
However this region has very different -- and almost complementary -- character from its higher-dimensional counterpart.  Here it is confined to the vicinity of the shell inside the horizon, while the late-time near-singularity regions are fully accessible.   (However, probing late-time near-singularity regions is not as useful as it would be in the higher-dimensional case, since the spacetime is locally AdS, so we cannot use such geodesics to directly probe the interesting strong-curvature effects; cf.\ \cite{Kraus:2002iv}.)

The behaviour of the length along shortest  \ETEBA\ geodesics, as function of time, is likewise very different for the BTZ case, as illustrated in \fig{f:AntipodalLengths}.
For any-sized black hole, the length increases monotonically from the AdS value to the BTZ value, without exhibiting any remarkable features.  This is consistent with the  expectations for the behaviour of CFT correlator during thermalization, which we expect to be directly extractible from the shortest lengths geodesics for this 3-dimensional case (as argued in a similar context in \cite{Shenker:2013pqa}); we revisit this point in \sect{s:disc}.  

Indeed, the consideration of boundary-anchored geodesics in 3-dimensional spacetime can be thought of as a special case of codimension-two extremal surfaces anchored on the boundary.  Hence the shortest boundary-anchored geodesics have a bearing both on certain CFT correlators, as well as on entanglement entropy corresponding to a certain region (bounded by the geodesic endpoints). The results are compared quantitatively with the results of \cite{Liu:2013iza,Liu:2013qca} and \cite{Hubeny:2013hz}, and agree with these in the regimes of early quadratic growth and intermediate linear growth of entanglement entropy.

In \sect{s:surfsVaidya} we turn to considering codimension-two extremal surfaces in the higher-dimensional case of Vaidya-AdS$_{d+1}$.  While these share certain features in common with the 3-dimensional case, there are also important differences, as already exemplified by the fact that even in the static black hole geometry, higher dimensional surfaces have richer structure \cite{Hubeny:2013gta}.
We demonstrate, both analytically and numerically, that for arbitrarily  late boundary time, one can construct extremal surfaces anchored at that time which penetrate the black hole, an example of which is presented in \fig{f:esLate}.  These surfaces lie along a specific maximal area surface inside the horizon, as observed recently in a related context by \cite{Hartman:2013qma,Liu:2013iza}.  By studying the linearized  perturbations of the surfaces away from this point, we obtain a good handle on what region of the bulk can be probed, indicated in \fig{f:ESRegion}.
We find that while we can probe to a finite depth inside the event horizon for arbitrarily late times, the near-singularity region of the geometry remains inaccessible.  In this respect, the codimension-two extremal surfaces  appear to be less suitable probes of the singularity than geodesics. This of course persists when we restrict to the surfaces of smallest area, which reach only a very limited region inside the horizon.

We also consider the evolution of the area of these surfaces for a fixed boundary region, which is directly related to the thermalization of the entanglement entropy,
analogously to the recent examination by \cite{Liu:2013qca,Liu:2013iza} in the planar context. Since we expect the global geometry to offer richer structure than its planar limit, we focus on nearly-hemispherical boundary regions.  This is presented in \fig{f:EEthermalization} and (perhaps disappointingly) offers no new surprises:  the entanglement entropy increases smoothly, and monotonically interpolates between the vacuum and thermal value. The growth is linear at intermediate times, controlled by the surface hugging the maximal area constant-$r$ surface, in agreement with \cite{Liu:2013iza,Liu:2013qca}, and for a sufficiently thin shell also exhibits the early-time quadratic growth derived therein.

%
\section{The Vaidya-AdS spacetime}
\label{s:Vaidya}

To model a simple holographic thermalization process, we consider a bulk geometry given by a global Vaidya-AdS$_{d+1}$ spacetime, mostly for $d=2,4$.
This is a solution to Einstein's equations with negative cosmological constant and a stress tensor for a spherically symmetric null gas, obtained by expressing the \SAdS\ metric in ingoing coordinates, and then allowing the mass to depend on the ingoing time $v$.
The metric can be written as
\begin{equation}
ds^2 = - f(r,v) \, dv^2 + 2\,dv\,dr + r^2\,d\Omega^2_{d-1} \ ,
\label{VaidyaMet}
\end{equation}
where $d\Omega^2_{d-1}$ is the round metric on the unit $S^{d-1}$,
\begin{equation}
 f(r,v) = r^2 +1 - \vartheta(v) \, \left(\frac{\rh}{r}\right)^{d-2}\,(\rh^2 +1) \ ,
\label{Vaidyafrv}
\end{equation}
and $\vartheta(v)$ is monotonic function, increasing from 0 in the past to 1 in the future, characterising the profile of a spherical null shell collapsing from the boundary. We take the shell to be concentrated around $v=0$, with a thickness of order $\delta$, taking $\vartheta'(v)$ as a function with compact support $[0,\delta]$. We also consider the limit of a thin shell, for which $\delta\to0$.

The metric interpolates between pure AdS inside (or to the past of) the shell and \SAdS\ outside (or to the future of) the shell. Concretely, away from $v=0$, the metric inside and outside can be expressed separately in static coordinates,
\begin{equation}
ds_\io^2 = - f_\io(r) \, dt_\io^2 + \frac{dr^2}{f_\io(r)} + r^2 d\Omega^2_{d-1} \ , 
\label{metio}
\end{equation}
where the subscript $\io$ stands for $i$ inside the shell and $o$ outside. The event horizon is at $r=\rh$ at late times, and it originates from $r=0$ at some $v=v_h<0$. The origin of spherical coordinates $r=0$ is smooth for $v<0$, but forms a curvature singularity for $v>0$.

\paragraph{Coordinates for spacetime plots:}
It is convenient to compactify the radial coordinate such that AdS boundary is drawn at finite distance, using $\rho \in ( 0, \pi/2 )$ defined by
\begin{equation}
\rho = \tan^{-1} r.
\end{equation}	
A natural temporal coordinate is one which makes ingoing radial null curves always at 45 degrees.  In terms of $v$ and the compact radial coordinate $\rho$, the new temporal coordinate is 
\begin{equation}
t = v - \rho + \frac{\pi}{2},
\label{}
\end{equation}	
the last term ensuring that $t$ coincides with $v$ on the AdS boundary. 
In pure AdS, $t = t_i$ is in fact the usual static coordinate, and both ingoing and outgoing radial null curves are at 45 degrees.  However, in the black hole geometry, $t$ is different from the static coordinate, $t \ne t_o$: indeed, $t$ is a good coordinate on the whole spacetime, whereas the static $t_o$ blows up at the horizon.

In plotting spacetime diagrams, with all relevant directions visible, we will use coordinates 
$(\rho\, \cos \psi \, , \, \rho \,  \sin \psi \, , \, t )$, where we use $\psi$ as a shorthand for an angular variable on the $S^{d-1}$.  In particular, $\psi$ will be either a longitude $\ph$, or the colatitude $\theta$, as appropriate to the symmetries of the problem in question. It will often be more convenient to consider 2-d projections by suppressing $t$ or $\psi$.  In particular, we will use  ingoing Eddington-Finkelstein plots $(\rho \, , \, t )$ and  `Poincar\'e disk' plots $(\rho\,\sin\psi \, , \, \rho\,\cos\psi)$.

The Eddington diagrams are useful because the static nature of the pre- and post-collapse geometries is made manifest. It should be appreciated that they can be quite misleading in that they distort the geometry, particularly near to the shell, and do not represent the causal structure.

For these reasons, we complement them by showing plots on Carter-Penrose diagrams, in which radial null geodesics, both ingoing and outgoing, appear as 45 degree lines.
\begin{figure}
\begin{center}
\includegraphics[width=.35\textwidth]{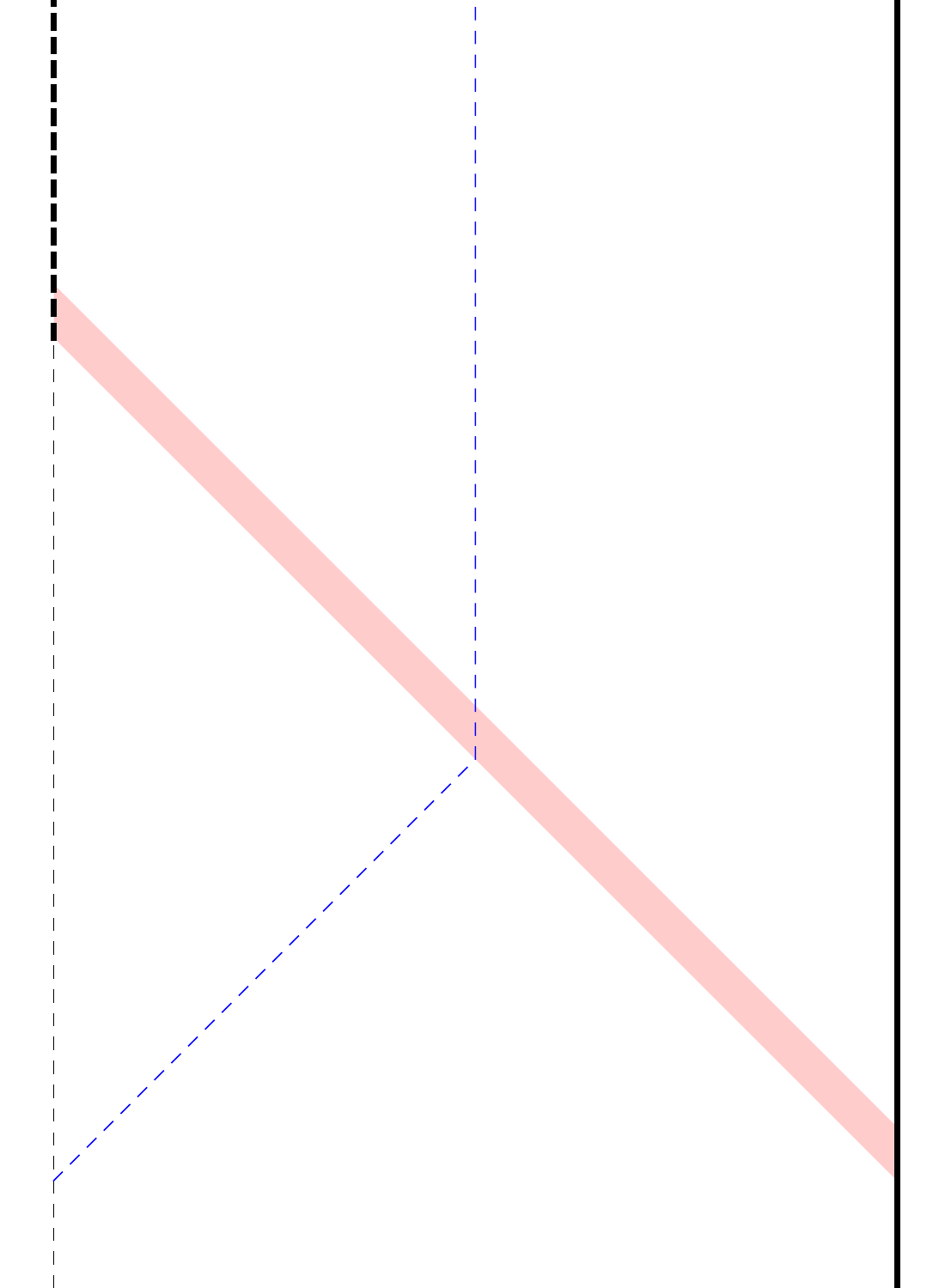}
\hspace{1cm}
\includegraphics[width=.4\textwidth]{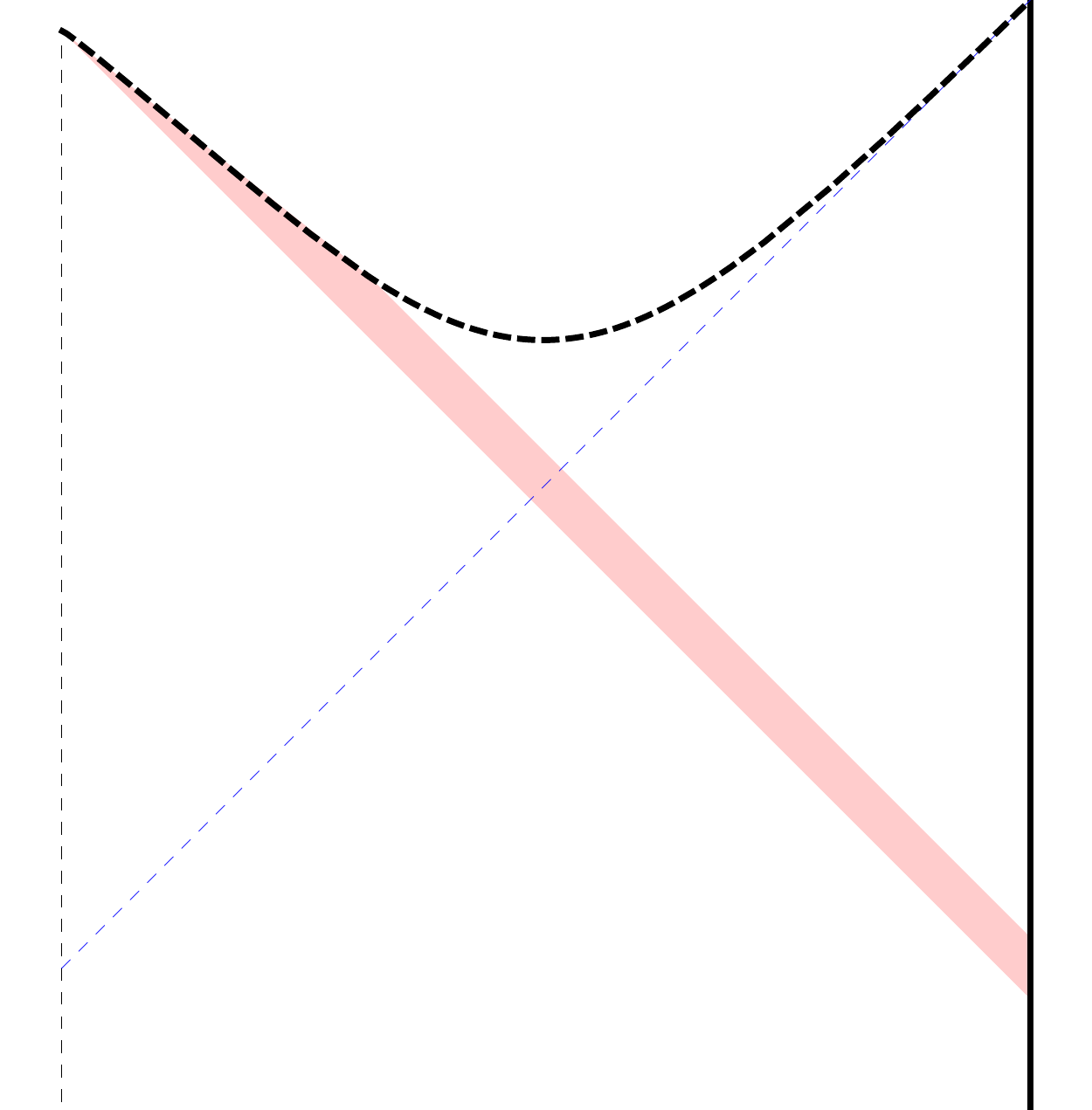}
\caption{
Eddington-Finkelstein (left) and Carter-Penrose (right) diagrams for the AdS-Vaidya spacetime, with $d=4$ and $\rh=1$. The vertical black dashed line on the left side in each panel is the origin of spherical coordinates before the shell begins, and the thick dashed curve the singularity. The AdS boundary is the solid thick line on the right. The infalling shell of matter is indicated by the red shading (its width indicating the shell thickness $\delta$ used in our numerical calculations), and the blue dashed line denotes the event horizon.
}
\label{f:penrose}
\end{center}
\end{figure}
To construct the necessary lightcone coordinates $(U,V)$ for this purpose, we observe that every radial null geodesic has a past endpoint on the boundary. The lightcone coordinates are obtained by assigning a number to each outgoing and each ingoing radial null geodesic, which can be done in a very general procedure.

Given a spacetime point $p$, we define $U$ by constructing the radial outgoing null geodesic through $p$ into the past, and noting the value of $v$ when it meets the origin $r=0$. The coordinate $V$ can then be chosen to be constant along ingoing null geodesics, and such that the boundary lies at $V=U+\pi$. Operationally, this can be done by finding where an outgoing null radial geodesic ending on the boundary passes through the origin, which gives $V-\pi$ for that boundary point.

This construction has the advantage of ensuring that both the AdS boundary, and the smooth origin at $r=0$ before the shell collapse begins, are straight lines on which $U-V$ is constant. The resulting coordinates in the pre-collapse part of the spacetime are identical to the standard (radially compact) coordinates on pure AdS, and hence noncompact in the past. The diagram is however compact in the future, terminating at the singularity.

For the lowest-dimensional case of Vaidya-BTZ, with $d=2$, in the limit of a thin shell, this coordinate change can be computed explicitly, and the result is reproduced in Appendix \ref{VaidyaBTZapp}. The metric in this case takes a particularly simple form:

\begin{equation}
ds^2=\frac{-dT^2+dR^2}{\cos^2R}+r(T,R)^2 d\phi^2,
\end{equation}
where $T=\frac{V+U}{2}$ and $R=\frac{V-U}{2}$, and the radius of the $\phi$ circle is given by
\begin{equation}\label{BTZPenroseMetric}
r(T,R)= 
  \begin{cases}
   \frac{(1-\rh^2)\sin R-(1+\rh^2)\sin T}{2\cos R} & \text{if } R+T>0\\ 
   \tan R       & \text{if } R+T<0
  \end{cases}
\end{equation}
with the coordinate ranges bounded by origin at $R=0$, the boundary at $R=\frac{\pi}{2}$, and the singularity at $(1-\rh^2)\sin R=(1+\rh^2)\sin T$.

While the causal structure is made manifest in these diagrams, they hide the fact that the post-collapse geometry is static, and can be distorting because late times are compressed into a corner of the diagram.

\section{Geodesics}
\label{s:geods}

We begin by studying geodesics, as the simplest example of extremal surfaces. We will restrict our considerations to spacelike geodesics, since these can end at the boundary but need not remain outside the event horizon.
In contrast, null curves are causally prevented from entering the horizon and reemerging out to the boundary, while timelike geodesics cannot even reach the boundary.

By spherical symmetry, we can restrict to geodesics lying in the equatorial plane, reducing the spherical directions to a single relevant longitude $\ph$. 
This simplifies the problem to a 3-dimensional one, described by Lagrangian
\begin{equation}
\mathcal{L}=-f \, \dot{v}^2+2\, \dot{r}\, \dot{v}+r^2 \, \dot{\ph}^2,
\end{equation}
where the dots denote differentiation with respect to an affine parameter $s$, chosen so that $\mathcal{L}$ is constant at $+1$ on the curve. This parameter $s$ is then an arclength.

The spherical symmetry also supplies us with a first integral for the angle, given by the conservation of angular momentum
\begin{equation}\label{angmom}
L=r^2 \, \dot{\ph}.
\end{equation}
Away from the shell, the spacetime is locally static, so there is in addition a conserved energy
\begin{equation}
E=f \, \dot{v}-\dot{r},
\end{equation}
though it is important that this is constant only locally in regions where $f$ is independent of $v$, and changes whenever the shell is encountered.

The $v$ equation of motion can be written to express this change, in the form
\begin{equation}\label{EJump}
\dot{E}=\frac{1}{2} \, f_{,v} \, \dot{v}^2 \leq 0
\end{equation}
where the inequality uses the fact that the profile function $\vartheta$ is nondecreasing. 
As well as telling us the sign of the jump in the energy, it also confirms the natural expectation that it should be greater when the shell is more dense, at smaller $r$, when $d>2$.
In the limit of a thin shell, the discontinuity can be calculated exactly, as
\begin{equation}
E|_{v=0^+}-E|_{v=0^-} = \frac{1}{2} \,  ( f|_{v=0^+}-f|_{v=0^-} )  \, \dot{v}|_{v=0}
\end{equation}
and is equivalent to the condition that $\dot{v}$ is continuous.

For numerics, we use second order equations of motion for $v$ and $r$, given by
\begin{align}
\ddot{v} &= -\frac{1}{2} \, f_{,r} \, \dot{v}^2+\frac{L^2}{r^3} \nonumber\\
\ddot{r} &= \frac{1}{2}(f_{,v}-f\,  f_{,r}) \, \dot{v}^2+f_{,r} \, \dot{r} \, \dot{v}+f \, \frac{L^2}{r^3},
\end{align}
integrating the definition of the angular momentum \req{angmom} to solve for $\ph$.

One useful fact that can be seen immediately from the equations of motion is that whenever $\dot{v}$ vanishes, $\ddot{v}$ must be positive. This, along with the fact that $v$ must be increasing as the boundary is approached, implies that $v$ has exactly one local (and hence also global) minimum along the geodesic. 
The uniqueness makes this a convenient point from which to start numerical integration.

\paragraph{Effective potential:}

Much of the qualitative behaviour of the geodesics in the static parts of the geometry can be understood from expressing the radial motion in the form of an effective potential, by eliminating $\dot{v}$ and $\dot{\ph}$ in favour of the conserved quantities $L$ and $E$:
\begin{equation}
\dot{r}^2=E^2-V_\mathrm{eff}(r),\quad\text{where}\quad V_\mathrm{eff}(r)=\left(\frac{L^2}{r^2}-1\right)f_\io(r).
\end{equation}

Again the subscript $\io$ refers to $i$ or $o$ to distinguish the pre- and post-collapse static parts of the geometry. Most important is the form of this potential in the static \SAdS\ spacetime. Excepting the special case of radial geodesics ($L=0$), it is unbounded from below as $r$ tends to both zero and infinity, has exactly two zeroes at $r=\rh,L$, and has a single maximum between them.

\begin{figure}
\begin{center}
\includegraphics[width=.9\textwidth]{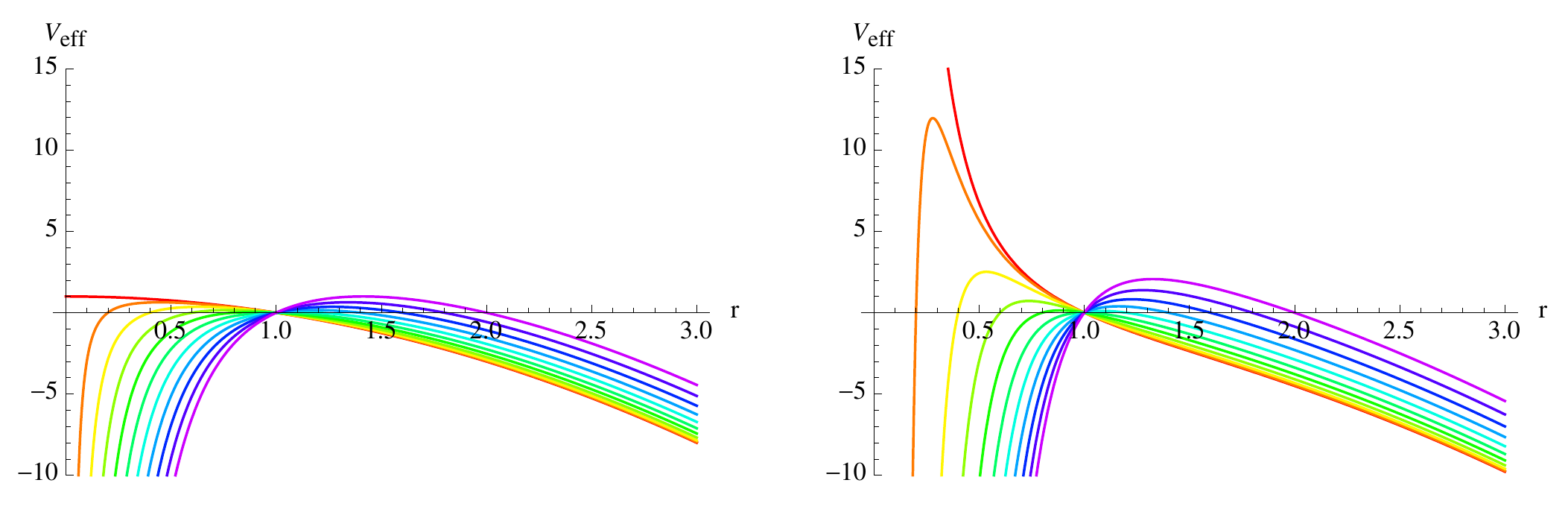}
\caption{
Effective potentials for spacelike geodesics in the BTZ (left) and  \SAdS$_5$ (right) geometry with horizon radius $\rh=1$, for various values of the angular momenta: $L=0$ (red) to $L=2$ (purple), in increments of 0.2.  The two cases are qualitatively different for low-$L$ values. 
}
\label{f:Veffs}
\end{center}
\end{figure}

As is well-known (see e.g.\ the discussion in \cite{Fidkowski:2003nf}), the 2+1 dimensional case is qualitatively different from the higher dimensional cases.  This is because in 3 dimensions, the BTZ black hole has locally the same geometry as pure AdS, so geodesics are not cognisant of the curvature singularity at $r=0$.  We can see this explicitly from the form of the effective potential, plotted in \fig{f:Veffs}.
For $d>2$, the height of the maximum grows without bound as $L$ is taken to zero, with the potential for radial geodesics unbounded from above for small $r$. 
(For example, in $d=4$, the maximum of $V_\mathrm{eff}$ scales as $\frac{\rh^2 (\rh^2+1)}{2\, L^2}$ at small $L$.)
The consequence is that the singularity repels even nearly-null geodesics, if they are sufficiently close to being radial. This simple observation turns out to be crucial to our considerations.
The $d=2$ (BTZ) case is qualitatively different from this, with the potential reaching a maximum of $(\rh-L)^2$, which is bounded for small $L$. This means that geodesics must have small energies, or end up in the singularity, so the nearly null geodesics of relevance in higher dimension will not be relevant for the $d=2$ case.

\paragraph{Lengths of geodesics:}

One natural observable associated with spacelike geodesics is their proper length. Since we are using arclength as a parameter, in principle we merely need to read off the difference $\Delta s$ between the initial and final value on the curve. This is complicated by the fact that the length is infinite: $r$ asymptotes to $e^{\pm s}$ as $s\to \pm \infty$. We will only need to compare lengths of geodesics with matching endpoints, so we need not worry about the details of choosing a renormalization scheme. We regulate in a simple way, cutting off at a large radius $r_c$, and subtract off the divergent piece $2\log(2r_c)$. Whenever length $\length$ is referred to, it may be taken to mean this regularized version.

\paragraph{Terminology:}
As mentioned above, for purposes of relating the geodesic (length) to a natural CFT observable (i.e.\ a two-point function of high-dimension operators with insertion points at the geodesic endpoints on the boundary), we wish to restrict attention to spacelike geodesic with both endpoints on the (same) boundary.  We will refer to these as {\it boundary-anchored} geodesics\footnote{
In \cite{Hubeny:2012ry} these were referred to as `probe geodesics'.}
and we will be primarily interested in the question of what part of the bulk spacetime is reached by the set of boundary-anchored geodesics.  In particular, how deep into the black hole, and how close to the curvature singularity, can such boundary-anchored geodesics penetrate.  Since one is often most interested in equal-time correlators on the CFT side, we will find it convenient to further refine our class of boundary-anchored geodesics to ones with both endpoints lying at the same boundary time; we will call these \eteba\ (for `equal-time-endpoint boundary-anchored') geodesics.

\paragraph{Initial condition space:}
A preliminary task is to find a convenient parameterization of the set of all geodesics. For this, we use the fact that on any given geodesic, $v$ has exactly one stationary point (in contrast to $r$, for example, which may have more). 
With this in mind, we parameterize the set of geodesics by three parameters $(v_0,r_0,E_0)$, respectively corresponding to the initial values of $v$ and $r$ when $\dot{v}=0$, and the energy $E=f\dot{v}-\dot{r}$ at that point.
We take $E_0$ to be nonnegative, since choice of sign corresponds only to choosing the direction of parameterization. The angular momentum follows from these; in fact $L=r_0$ (another choice of sign here corresponds to choosing the direction of increase of $\ph$). These parameters are sufficient to give an initial unit tangent vector, from which the geodesic may be found, and no two different sets of parameters will give rise to the same geodesic. Of course, some of these will end up in the singularity, so can be disregarded. We have thus put the set of all geodesics (modulo symmetries) into one-to-one correspondence with the set of $(v_0,r_0,E_0)\in \mathbb{R}\times[0,\infty)\times[0,\infty)$, which we will henceforth refer to as `initial condition space'.

\subsection{Geodesics in higher dimensions}
\label{VaidyaSAdSgeods}

As already noted, the lowest dimensional case of BTZ is qualitatively different from higher dimensions, and the questions we are considering have correspondingly different answers. This section will focus on the case of Vaidya-\SAdS$_{d+1}$ with $d\ge3$, postponing the discussion of Vaidya-BTZ to \sect{VaidyaBTZgeods}.

\begin{figure}
\begin{center}
\includegraphics[width=.35\textwidth]{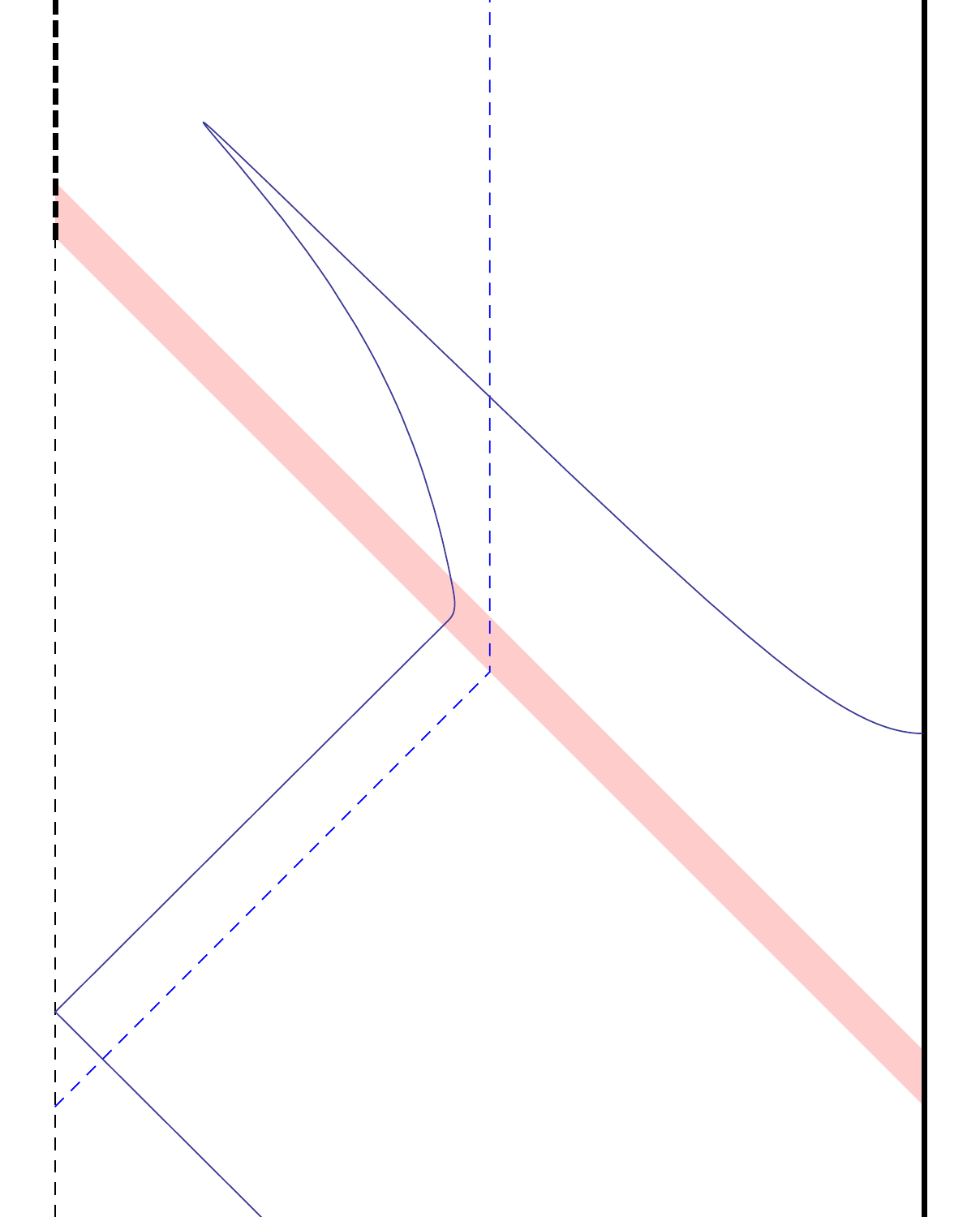}
\hspace{1cm}
\includegraphics[width=.4\textwidth]{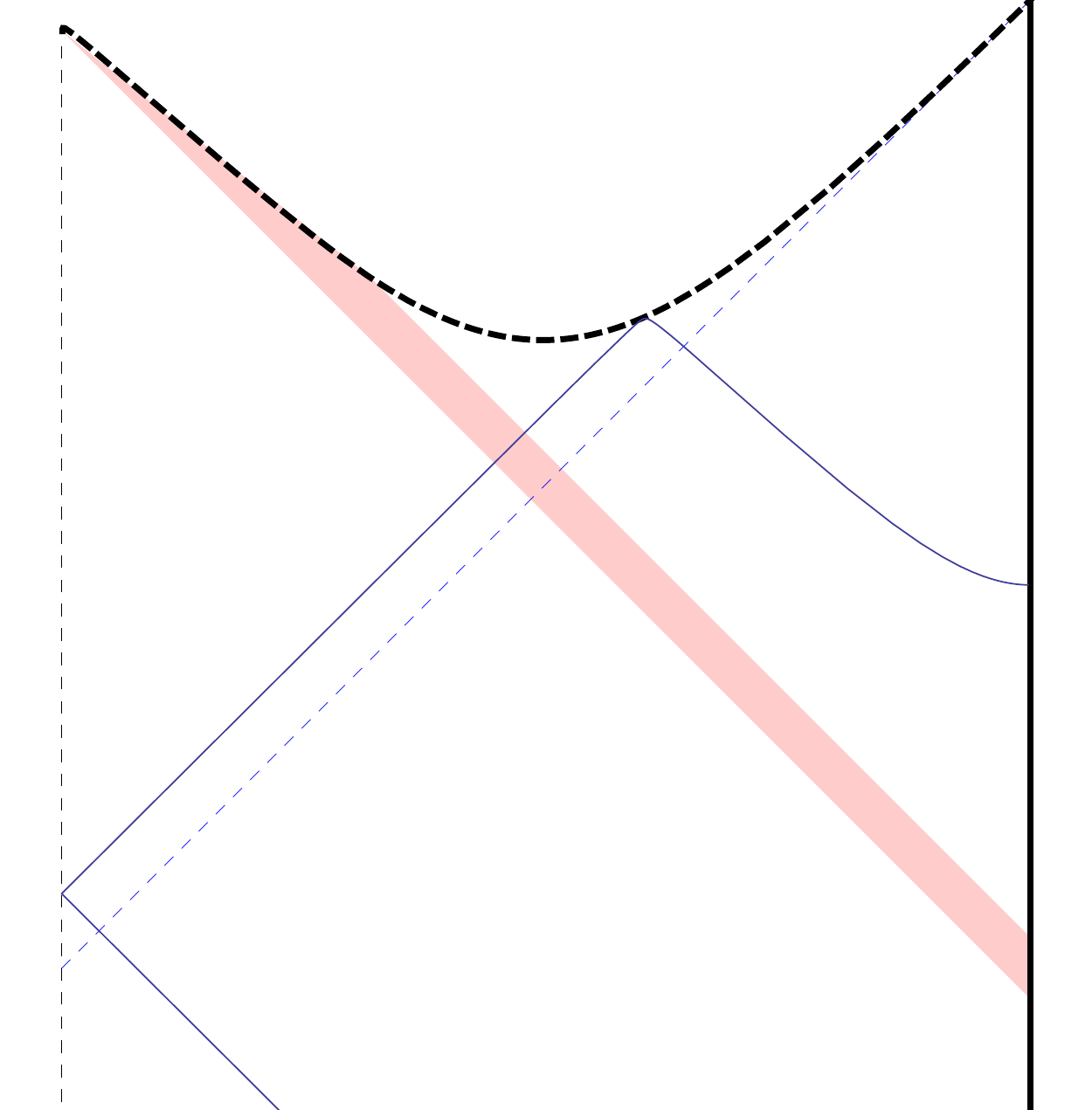}
\caption{
A radial geodesic (solid blue curve) with $v_0=-1.4$ and $E_0=12$, in \SAdS$_5$ with $\rh =1$, plotted on Eddington (left) and Penrose (right) diagrams, as described in \fig{f:penrose}.
We have cut off the uninteresting bottom part of the geodesic;
 its continuation approaches the boundary in a similar manner to the top part.
Note that on the Penrose diagram in the right panel, the geodesic looks like it reaches the singularity, but this is misleading effect of the coordinates, as evident in the Eddington diagram on the left panel.
}
\label{f:unequalGeodesic}
\end{center}
\end{figure}

One natural question to ask is what spacetime region is accessible to spacelike geodesics with both endpoints anchored on the AdS boundary.
Our first observation is that the answer to this question is in fact very simple:
Every point in the spacetime has a boundary-anchored spacelike geodesic passing through it.  For example, given any point $(r_0,v_0)$ inside the horizon and after the shell, one may take a radial geodesic, picking the energy such that $E^2=V_\mathrm{eff}(r_0)$, so that $r$ is at a local minimum. Constructing the geodesic in the maximally extended static black hole spacetime, it would join opposite asymptotic regions, geometrically encoding correlations between the two halves of the thermofield double state.\footnote{
In that context, such a geodesic would not qualify as boundary-anchored geodesic since it connects different boundaries; indeed, as argued in \cite{Hubeny:2012ry} for any static spacetime, boundary-anchored geodesics can only probe the spacetime region outside the black hole.
}
In the Vaidya-AdS spacetime, one end of this is altered since the second asymptotic region is replaced inside the shell by part of pure global AdS. Both ends must then lie on the (single) boundary, since there is no other boundary and no way it can reach the singularity since the effective potential is unbounded there. An example of such a geodesic is shown in \fig{f:unequalGeodesic}.

There are two noteworthy points.
Firstly, geodesics reaching very close to the singularity must be nearly null, and will hence have arbitrarily short lengths. Indeed, this was the key observation used in the eternal \SAdS\ context in \cite{Fidkowski:2003nf} to probe the black hole singularity (upon suitable analytic continuation).
Secondly, geodesics reaching inside the horizon at late times will have one end close to $v=v_h$, the value of $v$ at which the horizon is first formed. 
For strictly null geodesics, this observation was used  in the context of bulk-cone singularities \cite{Hubeny:2006yu} to detect the horizon formation event:  radial null geodesics whose earlier endpoint approaches $v_h$ have the other endpoint $v \to \infty$.  However, bulk-cone singularities arise from individual null geodesics which cannot penetrate the black hole by the usual causality constraints, so they are more limited probes of the bulk geometry \cite{Hubeny:2012ry}.

With these observations made, we now restrict attention to the case of \ETEBA\ geodesics, whose endpoints lie at matching times, corresponding to equal-time CFT correlators. 

\paragraph{\ETEBA\ geodesics:}

One obvious way to ensure a geodesic will have endpoints lying at equal times is to impose a $\mathbb{Z}_2$ symmetry under reflection, i.e.\ under swapping the endpoints. This is equivalent to setting $E_0=0$, so the initial conditions at the earliest part of the geodesic have this enhanced symmetry. In a globally static geometry, energy conservation implies that this is the only option, but it no longer needs to be the case in spacetimes with nontrivial time evolution. Indeed, the Vaidya geometry admits geodesics with equal-time endpoints which do {\it not} respect this symmetry.

A further refinement, relevant in cases with multiple geodesics joining the same endpoints, is to restrict to geodesics of shortest length for given time and angular separation of the endpoints, which are expected\footnote{
This expectation is subject to the assumption that this dominant saddle point lies on the path of steepest descent.  For nearly-null geodesics bouncing off the singularity in the eternal \SAdS\ spacetime this does not happen as discussed in  \cite{Fidkowski:2003nf}, so accessing the signature of this geodesic directly from the field theory is more subtle.  We revisit this point in \sect{s:disc}.
} to dominate the CFT correlators.

The classification of these classes of geodesics amounts to the following procedure:
\begin{enumerate}
\item Characterize the set of geodesics with both endpoints on the boundary.
\item Identify those geodesics with endpoints at equal times.
\item Compare lengths of such geodesics with matching endpoints.
\end{enumerate}

Having identified the initial condition space $(v_0,r_0,E_0)$, we must first find the region of this space for which both ends of the associated geodesic reach the boundary, and then find the two-dimensional surface in initial condition space for which the endpoints are at equal times. This is the level set $\Delta t=0$, where $\Delta t$ is the difference of times at final and initial endpoints (being the limits of $v$ as $s\to\pm\infty$). This is a 2-parameter set of geodesics. One part of this surface will be the portion of the plane $E_0=0$ for which the geodesic reaches the boundary. Then, we find  the time $t_\infty$ and angular separation $\Delta\ph$ of the endpoints for each such geodesic, along with the length $\length$. This amounts to finding the map from initial condition space to `boundary parameter space' $(t_\infty,\Delta\ph,\length)$, which collects all the field theory data associated with a given geodesic. The image of the equal-time geodesics under this map is a two-dimensional surface in boundary parameter space, and comparing lengths for given endpoints will amount to understanding different branches of this surface.

\paragraph{Initial condition surface of \eteba\ geodesics:}

We numerically undertook a systematic study of the geodesics in the Vaidya-\SAdS$_5$ spacetime, to find a representative sample of \eteba\ geodesics. This was done by taking a fine grid of initial points $(v_0,r_0)$, and for each of these points finding every initial energy which gives an appropriate geodesic, in the following process:
\begin{enumerate}
\item Identify the range of energies for which geodesics reach the boundary at both ends. This turns out to be an interval (possibly empty), which can be understood from the effective potential: the geodesic hits the singularity when the energy exceeds the maximum of $V_\mathrm{eff}$. This maximum energy is found by progressively bisecting between energies reaching the boundary or hitting the singularity.
\item Take a sample of geodesics reaching the boundary, and identify when the endpoints swap temporal order between adjacent energies. Each such occasion identifies an interval of energies containing a root of $\Delta t$.
\item Use a root-finding algorithm to find the appropriate initial energy within each such interval.
\end{enumerate}
Each energy $E_0$ found in this manner gives a point $(v_0,r_0,E_0)$ in the equal-time surface $\Delta t=0$ of the initial condition space. Sufficiently many such points build up a complete picture of this surface.

\begin{figure}
\begin{center}
\includegraphics[width=.6\textwidth]{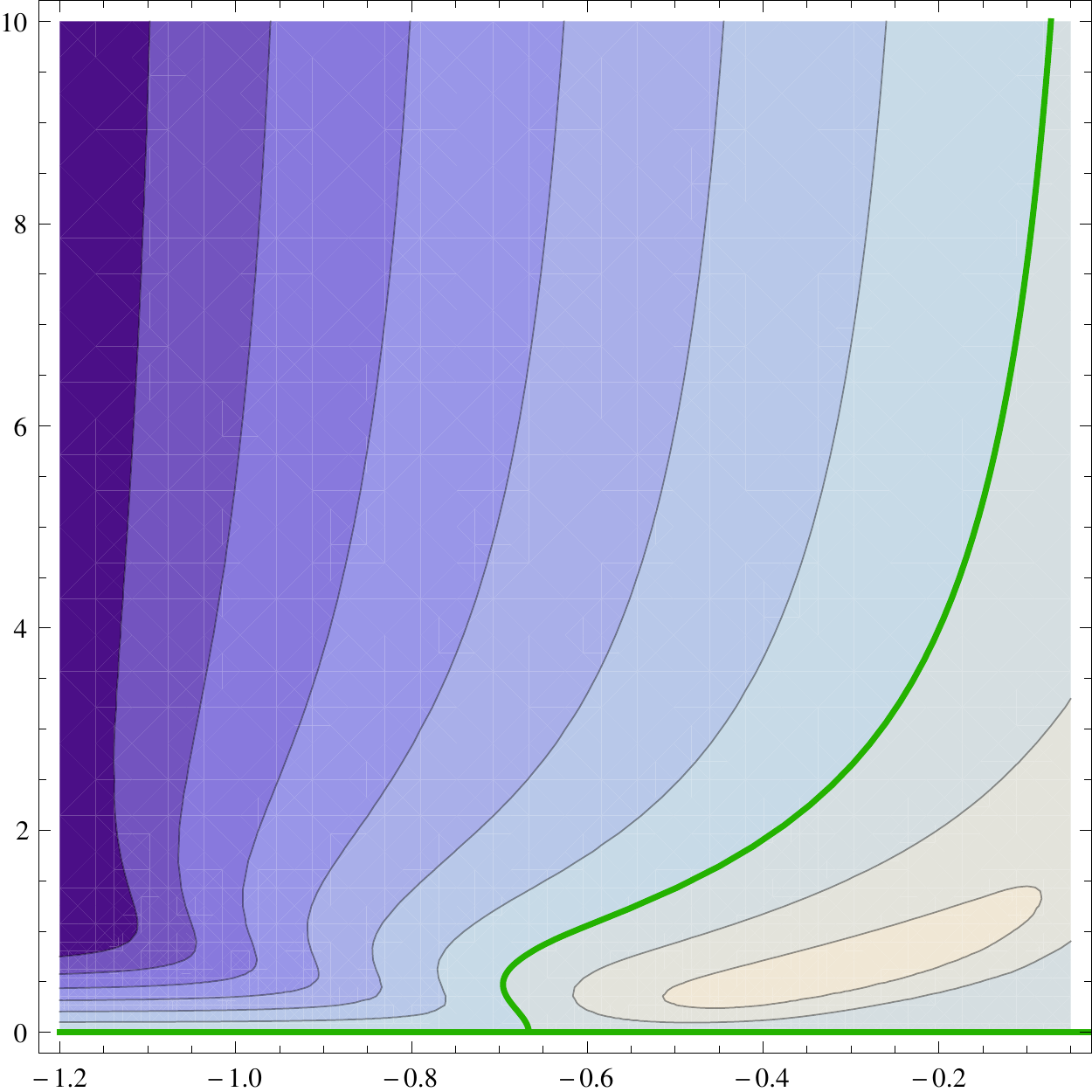}
\begin{picture}(0,0)
\setlength{\unitlength}{1cm}
\put (0.2,0) {$v_0$}
\put (-11.2,10.2) {$E_0$}
\end{picture}
\caption{
Contours of $\Delta t$ for radial geodesics. They are parameterised by the value of $v=v_0$ and the energy $E_0$ when they pass through the origin. The green lines give the $\Delta t=0$ contours, corresponding to \eteba\ geodesics.
}
\label{f:radialContours}
\end{center}
\end{figure}
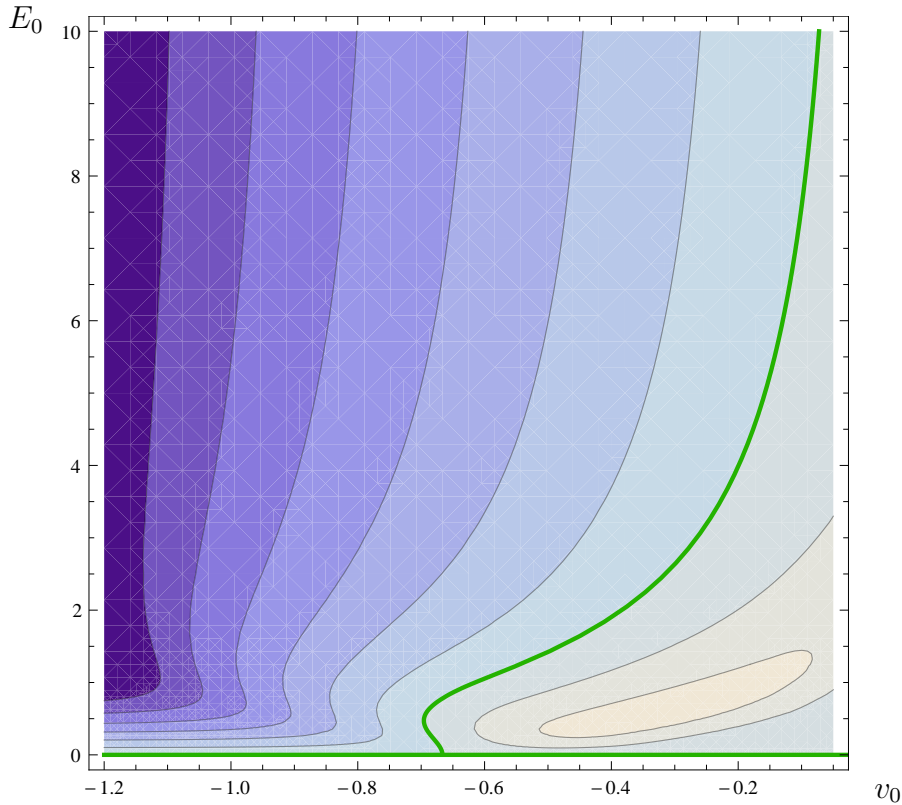

The first piece of the picture can be obtained from looking at radial geodesics, for which $L=r_0=0$. Provided the initial point is regular (meaning $v_0<0$ here), these always end at the boundary, since the \SAdS\ effective potential is unbounded as $r\to0$ in this case (cf.\ the red curve in right panel in \fig{f:Veffs}). The restriction to radial geodesics leaves us with two parameters to specify, namely $(v_0,E_0)$, and the equal-time radial geodesics give a curve in this space. This turns out to have two branches, as shown in \fig{f:radialContours}, one the symmetric $E_0=0$ case, and another at nonzero initial energy.

The reason for the latter is a trade-off between two competing effects. At nonzero energy, as it goes away from the origin the geodesic moves into the future or past depending on which direction is taken, and this separates the two branches in time. In a globally static geometry, the conservation of energy means that this separation persists to the boundary. This argument fails in the evolving geometry, but as long as the time-dependence is not too strong, this effect should still dominate.

The second effect is that the future branch encounters the shell of matter later and closer to the origin, when it has collapsed more, is more dense, and causes stronger curvature. This strongly influences the geodesic, and there is a large jump in energy as the shell is crossed, as implied from equation \req{EJump}. The future branch of the geodesic becomes nearly null, and hugs very closely to the shell. If this effect is different enough for past and future branches, it can cause the endpoints to exchange order in time.

\begin{figure}
\begin{center}
\includegraphics[width=.4\textwidth]{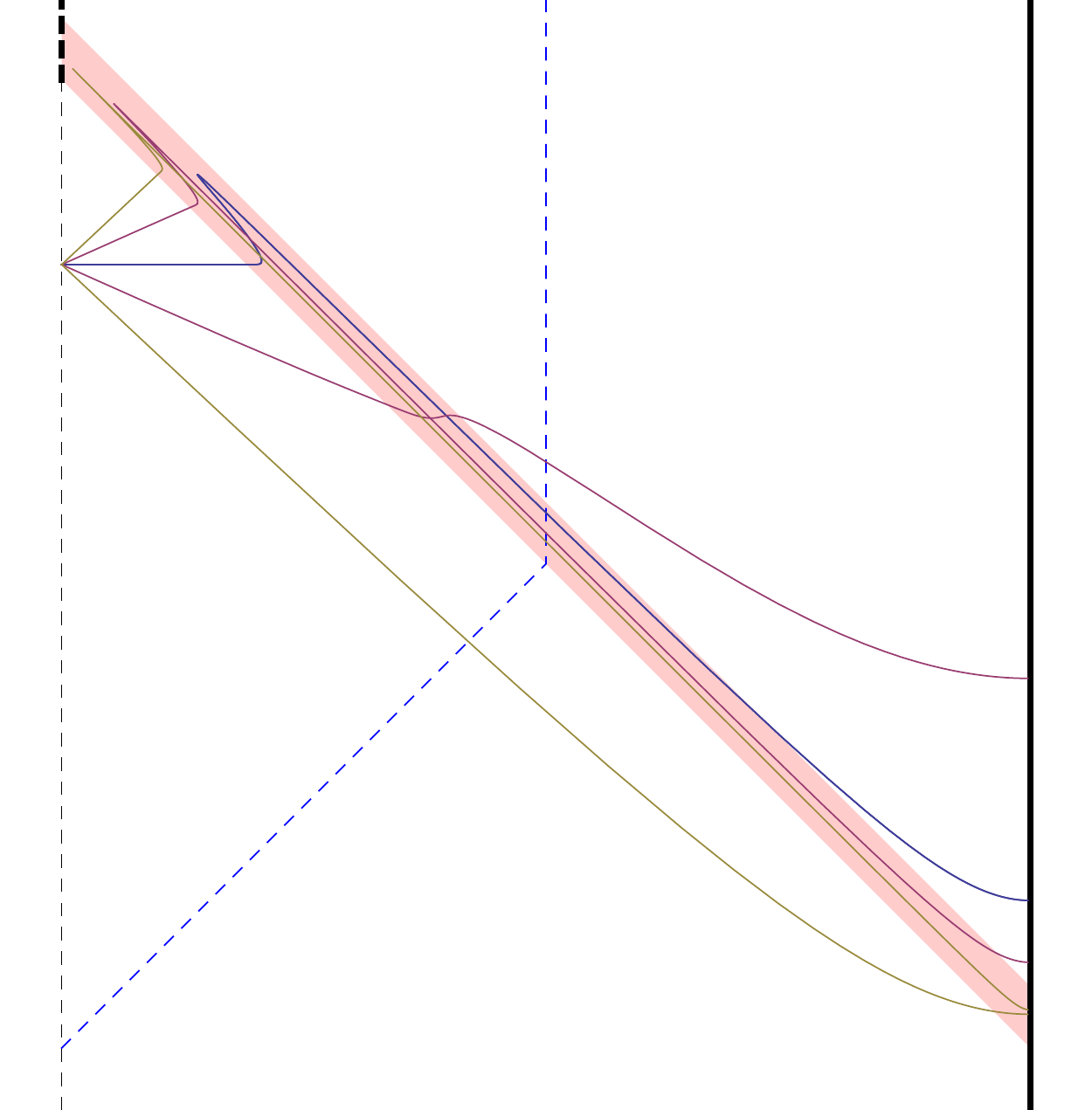}
\hspace{1cm}
\includegraphics[width=.4\textwidth]{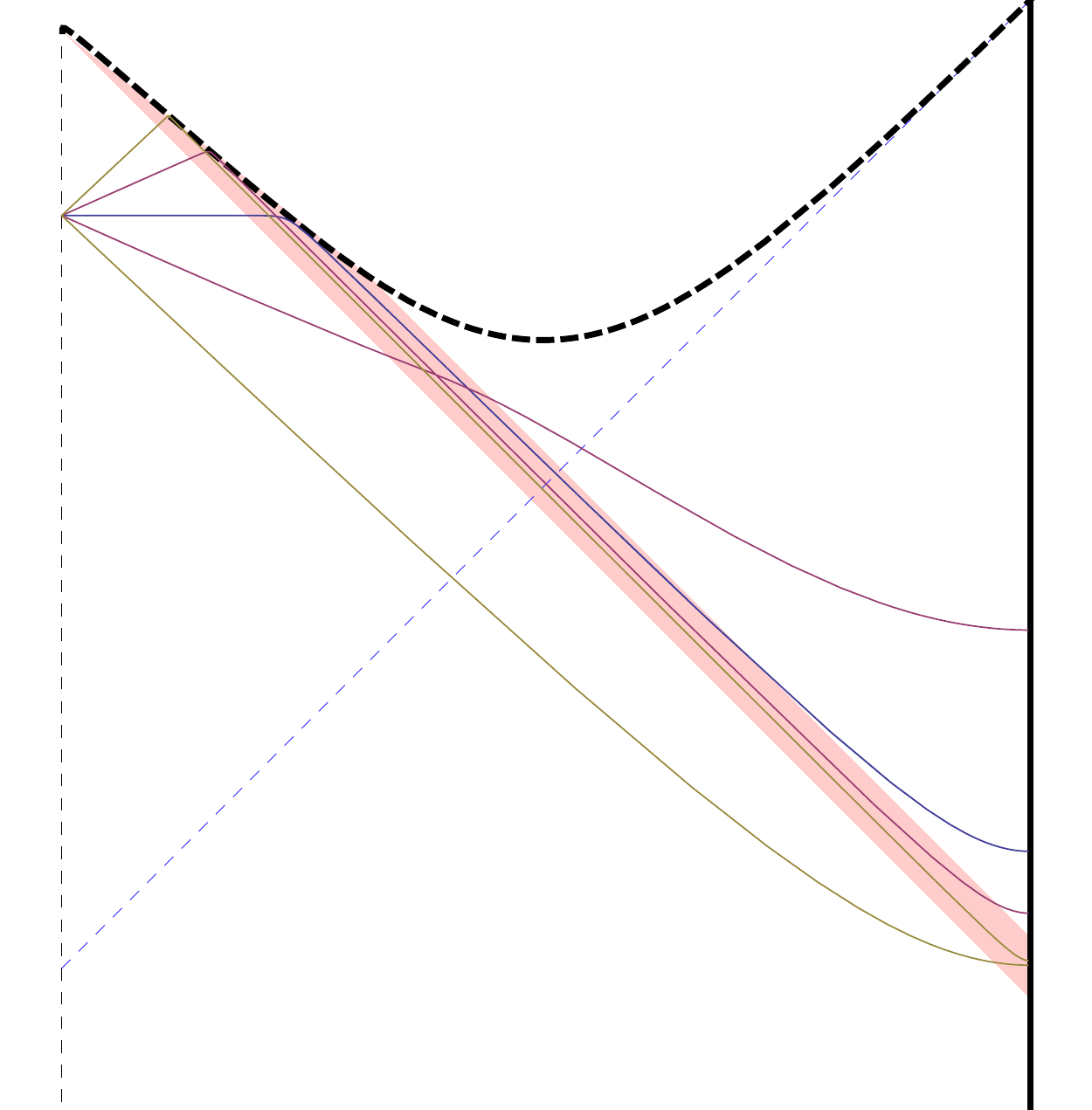}
\caption{
Radial geodesics passing through the origin at $v_0=-0.3$, with increasing energy, plotted on Eddington diagram (left) and Penrose diagram (right), with $d=4$ and $\rh=1$. The blue curve has zero initial energy, so is symmetric, the purple has initial energy $E_0=0.5$, and the yellow has $E_0=2.7$, close to the energy required to give equal-time endpoints.
}
\label{f:increasingEnergy}
\end{center}
\end{figure}

It turns out that for sufficiently late initial conditions, it is the latter effect which dominates at low energies, the former taking over when the geodesic is nearly null, as illustrated in \fig{f:increasingEnergy}.

This intuition for the existence of asymmetric equal-time geodesics also gives an indication of when they are unlikely to exist. Firstly, as we will argue in \sect{VaidyaBTZgeods}, they do not exist in a 3-dimensional bulk. The effect of the shell on the energy is independent of the time at which the geodesic crosses it, because of the slow fall-off of gravity, so the competition is absent. Related to this, even radial geodesics of sufficient energy will not be prevented from ending in the singularity. Secondly, moving back to higher dimensions, the competition relies on the high energy, nearly null geodesics, which will fail for appreciable angular momentum. The maximum of the effective potential must be high enough to reflect the geodesics away from the singularity, but this maximum is reduced as $L$ is increased. The result is that asymmetric geodesics only exist joining points of the boundary sphere that are close to antipodal.

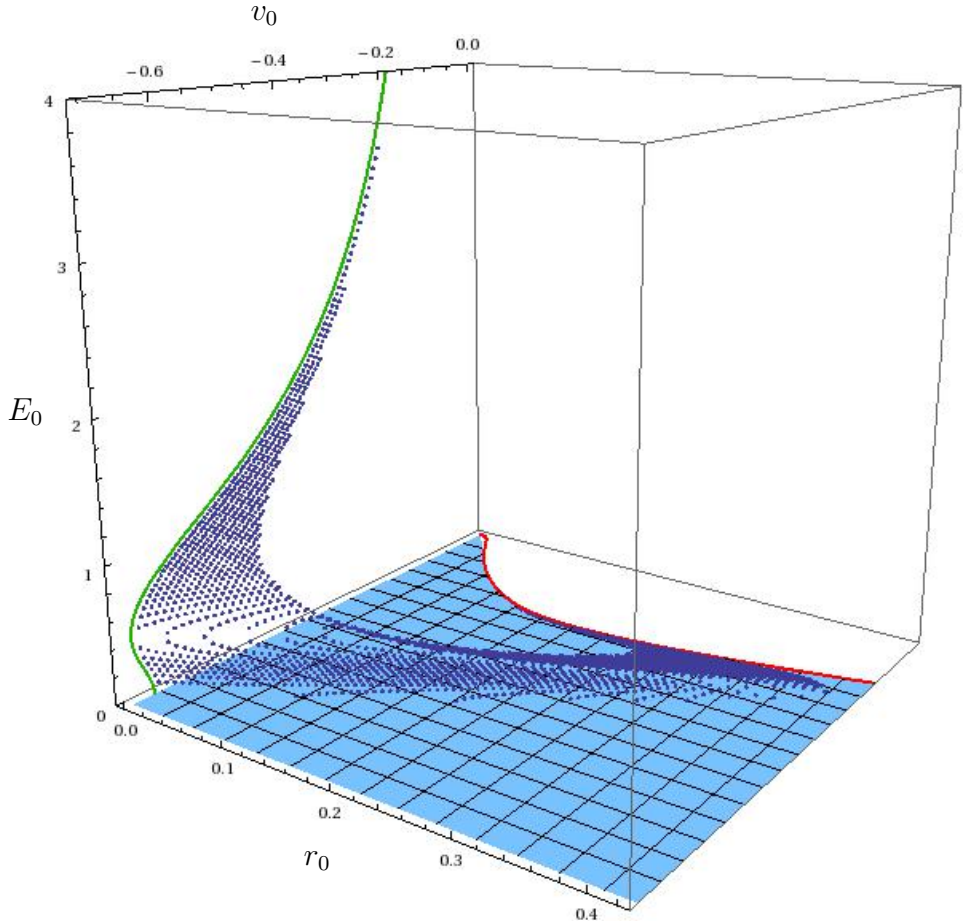
\begin{figure}
\begin{center}
\includegraphics[width=.7\textwidth]{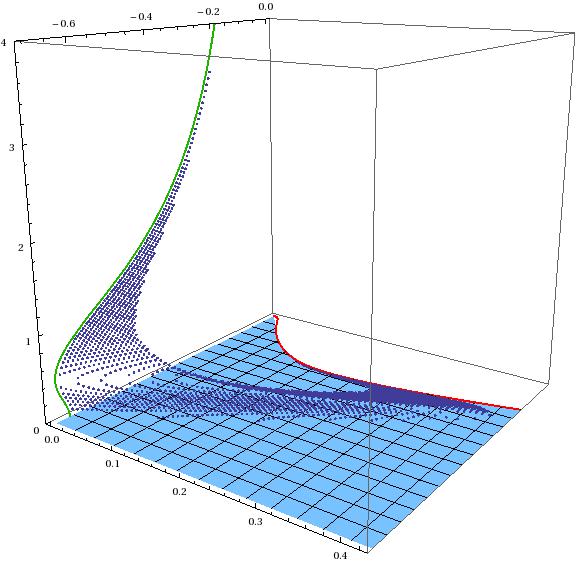}
\begin{picture}(0,0)
\setlength{\unitlength}{1cm}
\put (-9.5,12) {$v_0$}
\put (-8.8,0.8) {$r_0$}
\put (-12.7,6.7) {$E_0$}
\end{picture}
\caption{
The surface in initial condition $(v_0,r_0,E_0)$ space corresponding to \ETEBA\  geodesics. The part of the plane $E_0=0$ for which geodesics are boundary-anchored, bounded by the red curve, gives symmetric geodesics. The blue points give asymmetric geodesics, and the curve for those which are radial is shown in green (c.f.\ \fig{f:radialContours}).
}
\label{f:equalTimeICs}
\end{center}
\end{figure}

The full surface of initial conditions corresponding to \eteba\ geodesics is shown in \fig{f:equalTimeICs}.

\paragraph{Length of geodesics:}

\begin{figure}
\begin{center}
\includegraphics[width=.8\textwidth]{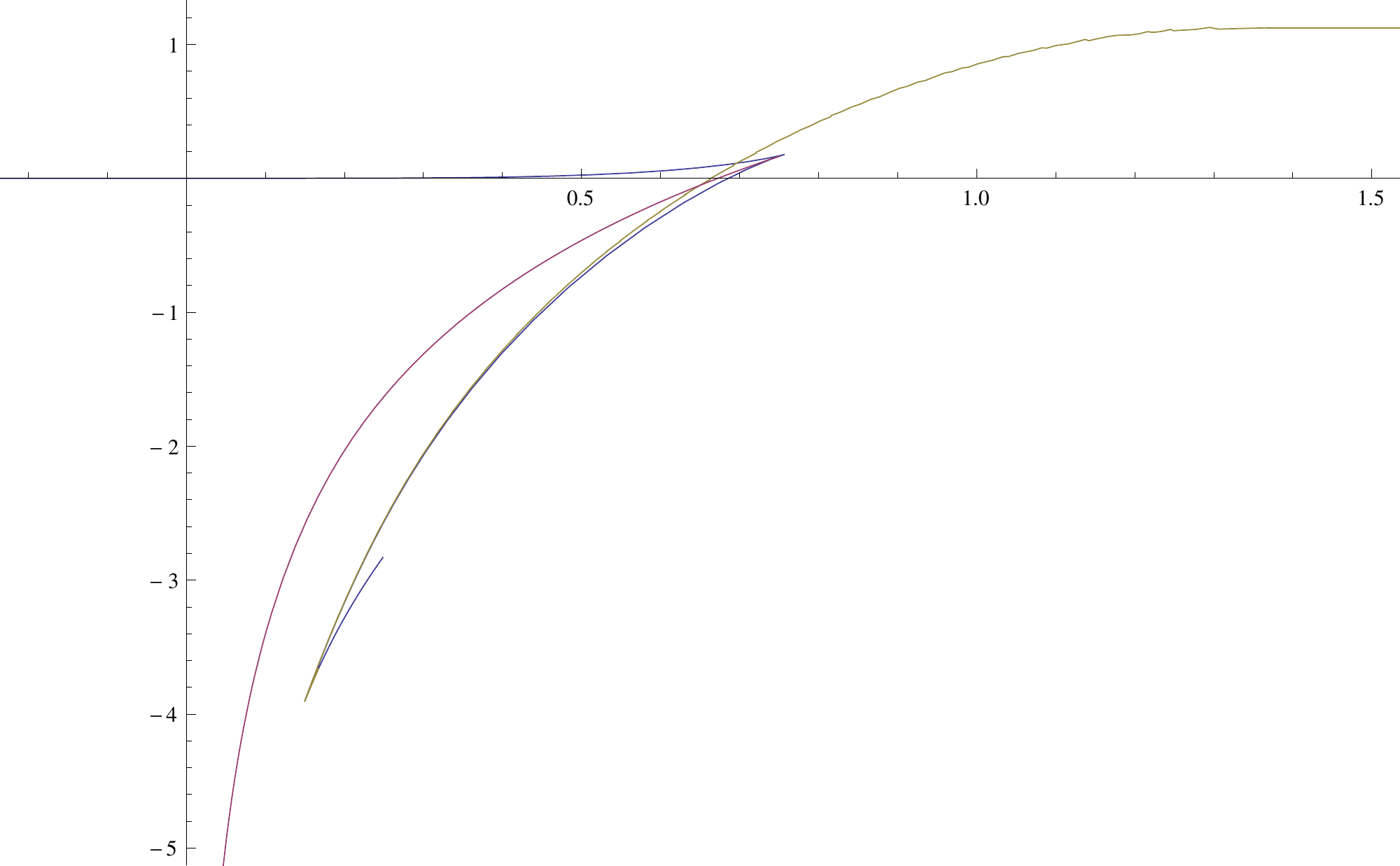}
\begin{picture}(0,0)
\setlength{\unitlength}{1cm}
\put (-12.5,8.6) {$\length$}
\put(0.2,6.6){$t$}
\end{picture}
\caption{
Regularised length of \eteba\ geodesics joining antipodal points, plotted against the time at which the boundary is reached. The blue curve corresponds to radial, symmetric geodesics; the yellow curve to symmetric but not radial, and the purple curve to radial but not symmetric ones.
}
\label{f:antipodal}
\end{center}
\end{figure}

The next stage is to map this surface (of initial conditions corresponding to \ETEBA\ geodesics) into the boundary parameter space $(t_\infty,\Delta\ph,\length)$. This gives a complicated, multi-branched surface, but many of the salient features are revealed from taking a cross-section at $\Delta\ph=\pi$, which corresponds to the set of geodesics joining antipodal points of the boundary sphere at equal times, shown in \fig{f:antipodal}. At early times, before the collapse begins, the only possibility is a simple straight line through the middle of AdS; this geodesic is both symmetric and radial. At late times, the only possibilities are again symmetric geodesics, lying at constant \SAdS\ time $t_o$, but these are not radial as they cannot penetrate the event horizon. This regime is then dominated by a geodesic simply deformed to one side of the horizon.\footnote{
There are infinitely more possibilities, since the geodesic may wrap around the horizon arbitrarily many times, but such geodesics are of course longer.} In the intermediate region, these families can be continued, and indeed meet, but there is also the additional possibility of the asymmetric geodesics presented in \fig{f:radialContours} and the accompanying discussion. This additional family dominates for a short time immediately after the collapse; indeed for sufficiently early times the lengths may be arbitrarily short as the geodesics become very nearly null. On the other hand, there are no antipodal \ETEBA\ geodesics which are neither symmetric nor radial.

The structure that \fig{f:antipodal} reveals is surprisingly intricate.  In the course of thermalization (i.e.\ between $t=0$ when the shell starts imploding and $t\approx1.3$ when the antipodal \ETEBA\ geodesic remains entirely to the future of the shell), there are 4 `jumps' in the shortest length as different branches start or terminate.  There are also several points where families of geodesics exchange dominance, but these kinks are hidden by the shorter $\length$ families.
The field theory interpretation of \fig{f:antipodal} is, on the face of it, quite strange. It would seem to suggest that, during thermalization, the equal-time correlators of high-dimension operators of antipodal points correspondingly undergo no less than four discontinuous jumps. Furthermore, the first of these, at the start of thermalization, is an unbounded increase.  However, as discussed in \sect{s:disc}, the shortest $\length$ real-time geodesics may not actually be the ones to dominate the CFT correlator; such contingency arises in the simpler context of the eternal \SAdS\ geometry \cite{Fidkowski:2003nf}.  Nevertheless, even if the correlator is not dominated by these geodesics,  their rich structure should still be subtly encoded in the correlation functions, possibly extractible by suitable analytic continuation.

We expect the geometry leading to this unexpected behaviour to be robust to changing many details of the collapse, depending rather only on the main features: spherical symmetry, and the formation of a spacelike singularity.\footnote{
The singularity however has to be `black-hole-like' in the sense that it repels at least some class of spacelike geodesics; if the singularity were of the big crunch type (wherein all the transverse directions contract as the singularity is approached), then our spacelike geodesics would simply terminate in it.  This observation indicates that probing cosmological singularities (and correspondingly the resolution of a cosmological singularity in quantum gravity) would be expected to be drastically different from that of a black hole singularity.
} This is because any such geometry allows for nearly null radial geodesics, essentially following light rays except close to the singularity where they are repelled, with equal-time endpoints, by sending them into the corner of the Penrose diagram where the singularity is formed. 

For geodesics joining points which are far from antipodal, the picture is simpler, with one family having the shortest length for all time, smoothly and monotonically interpolating between vacuum and thermal values. The asymmetric geodesics are absent entirely, this family disappearing very quickly on moving away from $\Delta\ph=\pi$. The other parts of the curves visible in \fig{f:antipodal} split into two families, one dominant, and the other corresponding to geodesics passing round the far side of the black hole.
In the limit of large black hole and small angular separations, which recovers the planar black hole case, the picture becomes even simpler, since then even the possibility of passing on the other side of the black hole is not present.  Hence the intricate structure observed in \fig{f:antipodal} relies on both the black hole having compact horizon and on the geodesic connecting sufficiently far-separated points within the spherical boundary.

\paragraph{Region of geometry probed:}

The final task is to identify the region of spacetime covered by the \ETEBA\  geodesics, both in totality, and also restricting to the shortest length for given endpoints. The latter region gives the part of the bulk on which the associated field theory observable is most sensitive. We find that the deepest probing geodesics are those connecting antipodal points $\Delta \ph=\pi$, so restricting to these alone will not reduce the accessible region.

\begin{figure}
\begin{center}
\includegraphics[width=.4\textwidth]{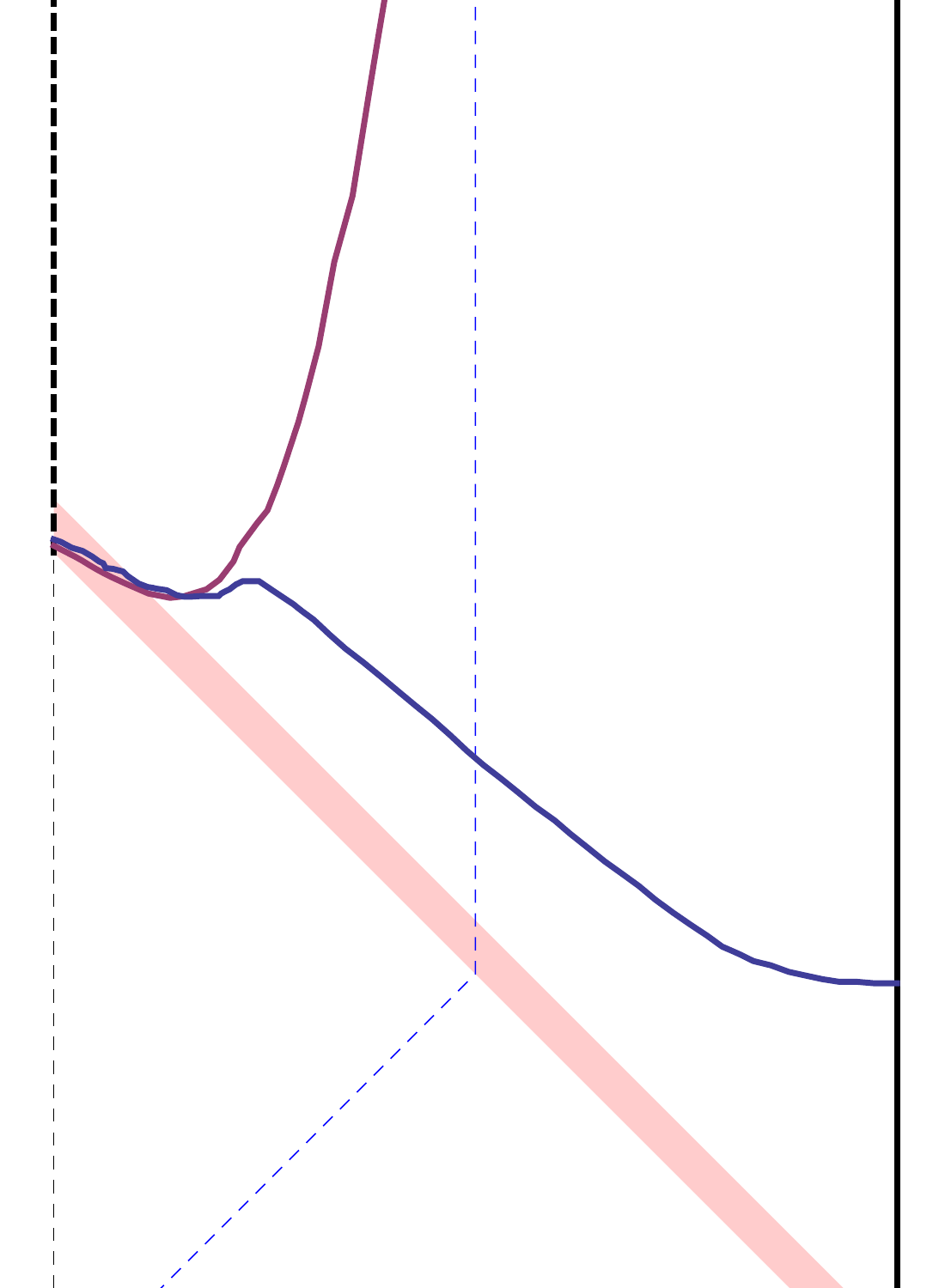}
\hspace{1cm}
\includegraphics[width=.4\textwidth]{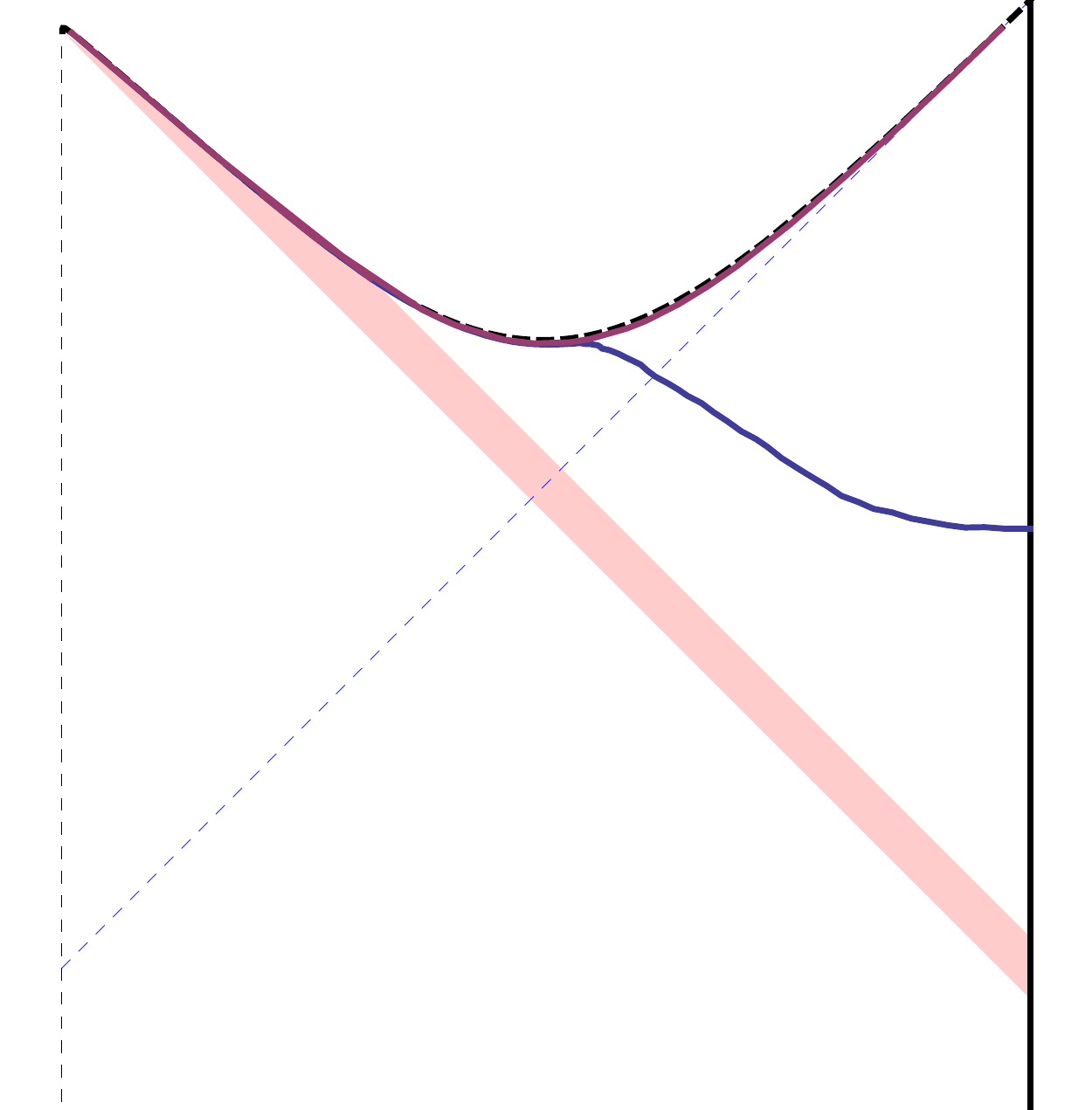}
\caption{
The region covered by all \eteba\ geodesics, on Eddington and Penrose diagrams. The purple curves indicate the boundary of the region covered by only the symmetric geodesics, and the blue curves the region covered by the asymmetric geodesics. In particular, the asymmetric geodesics reach deeper, but only in a very small region.
}
\label{f:geodesicRegion}
\end{center}
\end{figure}

The region covered by the geodesics as a whole is illustrated in \fig{f:geodesicRegion}, which shows the deepest points reached by asymmetric and symmetric geodesics. The symmetric geodesics are adequate to cover almost all of the accessible region. In particular they reach inside the horizon at arbitrarily late times, though only by a small distance, shrinking to zero as $v\to\infty$. They also cover the entirety of the spacetime inside the shell ($v<0$), which includes points arbitrarily close to the singularity. From our numerics, it appeared that these geodesics did this in such a way as to remain at bounded curvature (considering, for example, the Kretschmann scalar $R_{abcd}R^{abcd}$, which goes like $r^{-8}\vartheta(v)^2$). This computation is rather sensitive to the fine details of the profile of the shell, so it is not clear how robust the conclusion is. Indeed, taking the limiting case of a shell of zero thickness, it is clear from considering symmetric radial geodesics passing through the origin immediately before collapse that unbounded curvature can be obtained.

This region close to the singularity is the only place where one may do better by including the asymmetric geodesics. These reach the region of small $r$ to only slightly later times, but crucially appear to be able to get arbitrarily close to the singularity at some strictly positive $v$, where the curvature may become arbitrarily strong.

Including the restriction of considering only the shortest geodesics, we do not lose access to much of the region soon after formation of the black hole. In particular, the same asymmetric geodesics that reach to regions of arbitrary curvature are also those of arbitrarily short length, and thus dominate.

\begin{figure}
\begin{center}
\includegraphics[width=.4\textwidth]{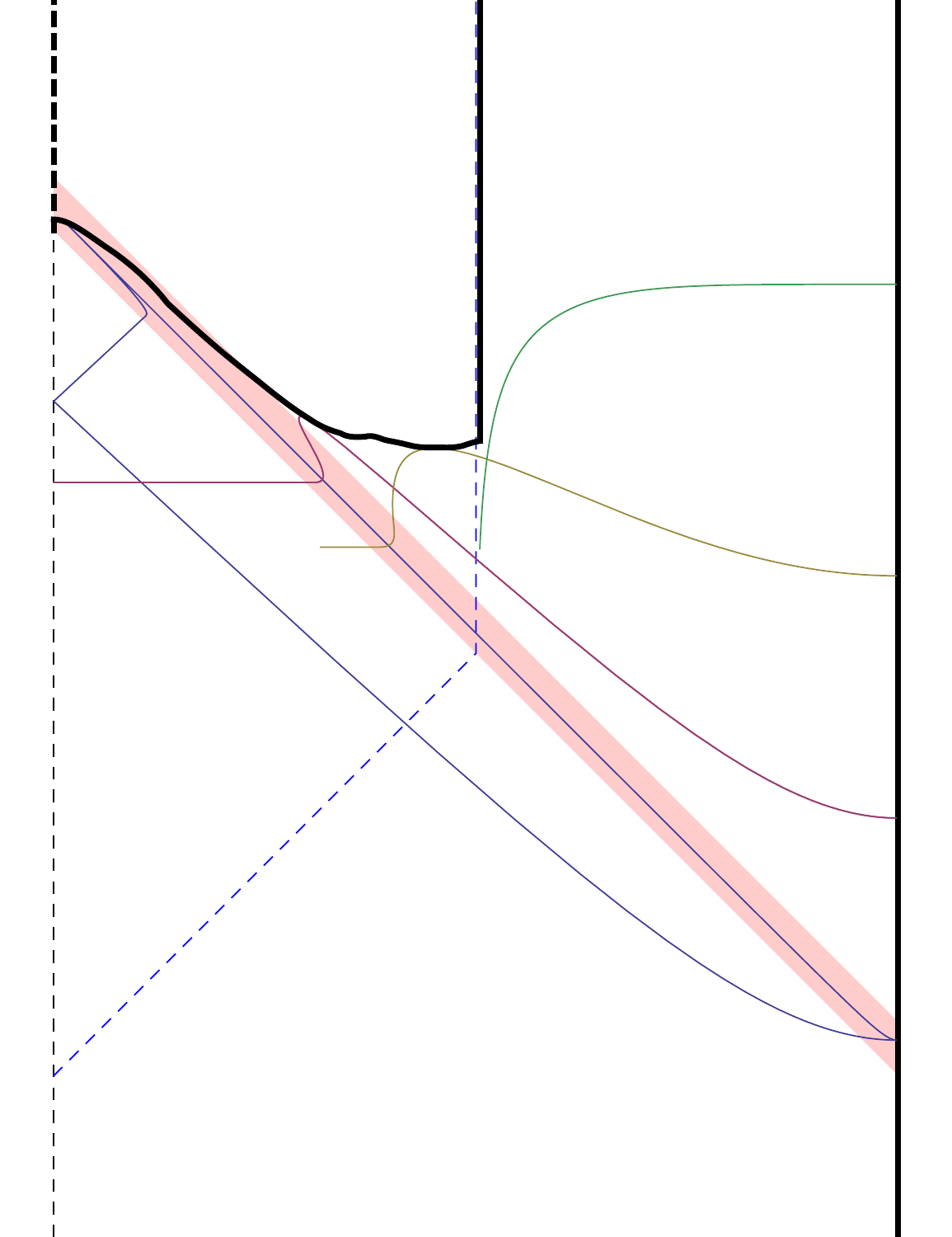}
\hspace{1cm}
\includegraphics[width=.4\textwidth]{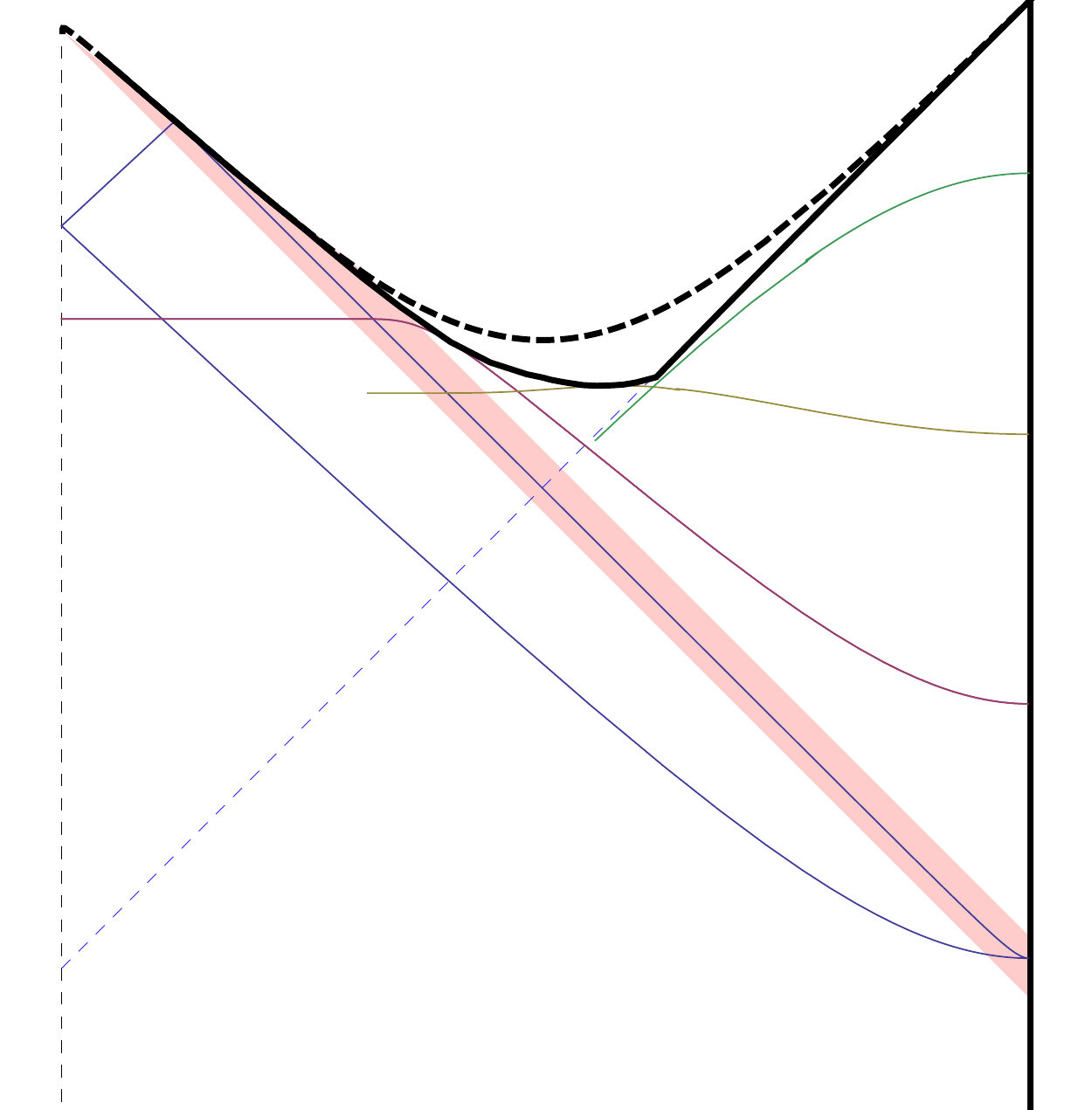}
\caption{
The region covered by shortest \eteba\ geodesics, on Eddington and Penrose diagrams, bounded by the black curve, and examples of each family of such geodesics. Moving from early to late time, the blue curve is asymmetric and radial, the purple curve symmetric and radial, and the yellow and green symmetric but not radial. The green curve lies entirely in the \SAdS\ part, reaching not to the horizon but only to $r\approx1.014$ in this case ($d=4$, $\rh=1$)
}
\label{f:geodesicShortestRegion}
\end{center}
\end{figure}

Thereafter, we must consider what happens as dominance is exchanged between various families, as illustrated in \fig{f:antipodal}. The result is that we must exclude the geodesics reaching inside the horizon at late times, so the region after the shell and inside the horizon covered by shortest \eteba\ geodesics is very limited, as shown in \fig{f:geodesicShortestRegion}. For example, in the case of $\rh=1$, $d=4$, the latest time a shortest geodesic reaches the interior of the horizon is at $v\approx0.4$. Thereafter, it should be emphasised that they reach not the whole exterior of the horizon, but only to the deepest radius of the shortest antipodal geodesic in \SAdS, which is at a finite though small distance above the horizon.

  Apart from this region inside and close to the black hole, there is a distinct region which is not reached by shortest-length \eteba\ geodesics. The shortest geodesics jump after $t=0$ with the transition to the nearly-null geodesics, and because of this, a part of the pure AdS section of the geometry is also missed. This is the one place where including the geodesics which are not antipodal will allow access to a larger region. Despite this, there is still a small region remaining inaccessible, close to $r=0$ and for some intermediate range of times, well after formation of the horizon but well before formation of the singularity.

\subsection{Geodesics in Vaidya-BTZ}
\label{VaidyaBTZgeods}

We have seen in \sect{VaidyaSAdSgeods} that not every point in the Vaidya-AdS$_5$ spacetime is reached by the  equal-time-endpoint boundary-anchored geodesics. 
In particular, events inside the black hole at late time (large $v$) do not lie on any \ETEBA\ geodesic.
 However, at the same time, our geodesics probe arbitrarily close to the singularity just after its formation (though only traversing regions of bounded curvature).  Here we wish to contrast this with the analogous set-up in $2+1$ bulk dimensions, i.e.\ the Vaidya-BTZ geometry.  While as pointed out previously, this case is qualitatively different since the geometry is locally AdS$_3$ everywhere outside the shell and singularity, this case is most tractable by analytical means and most amenable to direct comparison to field theory.  To take full advantage of the former, we also take the limit of a thin shell ($\delta\to0$) in order to write simple closed-form expressions.

An additional curiosity in the case of BTZ is that for a time, the singularity is timelike. This can be seen from looking at outgoing radial geodesics: they may move away from $r=0$ as long as $f(r=0,v)>0$, which happens for some window during the collapse. By making the collapse very slow, the singularity may even be made naked.  Indeed, if the shell does not carry enough energy, having a BTZ black hole final state is not an option.\footnote{
See however \cite{Bizon:2013xha} for a numerical study of scalar collapse in AdS$_3$ inducing turbulent instability which nevertheless remains regular.
} In the Vaidya case, while it starts out timelike, the singularity is of a particularly mild type, being only a spatial conical defect.

\paragraph{Symmetric radial geodesics:}
Let us first consider the simplest case of symmetric radial geodesics in Vaidya-BTZ, starting at the origin $r_0=0$ before the implosion of the shell with $v_0 = -\tau$ where $\tau \in (0,\frac{\pi}{2})$, and with initial energy $E_0=0$. To simplify the computations, it turns out to be convenient to parameterize the final black hole size by a parameter $\mu$ defined by
 $r_+ =\sec \mu + \tan\mu$,
 where  $\mu \in (-\frac{\pi}{2},\frac{\pi}{2})$.  Note that $r_+ = 1$ corresponds to $\mu=0$, which is a critical size separating qualitatively distinct types of behaviour.
 
The radial equation of motion outside the shell can be written as
\begin{equation}
\dot{r}^2 = r^2 +\frac{\sin^2\mu-\sin^2\tau}{(1-\sin\mu)^2} \ ,
\label{}
\end{equation}	
from which it is clear that when $\tau\geq|\mu|$, the geodesic can never reach the singularity at $r=0$ since $\dot{r}^2$ would be negative for small $r$. When this fails ($\tau<|\mu|$), $r$ has no turning points, so the fate of the geodesic depends on the sign of ${\dot r}$ just after crossing the shell: it will end in the singularity or on the boundary if it is negative or positive respectively. The calculations give 
\begin{equation}
\dot{r} |_{v=0^+} =  \frac{\sin^2\tau-\sin \mu}{\cos \tau \, (1-\sin\mu)} \ ,
\label{}
\end{equation}	
which for small black holes ($\mu<0$) is automatically positive, so the geodesic continues to the boundary.  On the other hand, large black holes $\mu>0$ allow a regime for sufficiently small $\tau$ (i.e.\ later starting point, closer to the implosion of the shell) where $\dot{r}<0$ outside the shell, so that the geodesic initially recedes to smaller $r$.  If $\tau<\mu$, $\dot{r}$ remains negative for all $r$ so the geodesic crashes into the singularity.  On the other hand, if $\mu<\tau<\arcsin\sqrt{\sin\mu}$, it turns around at $r_{tp}$, where
\begin{equation}
r_{tp} \equiv \frac{\sqrt{\sin^2\tau-\sin^2\mu}}{1-\sin\mu} \ .
\label{}
\end{equation}	
This can be made arbitrarily small by letting $\tau \to \mu^+$, so such boundary-anchored geodesic gets arbitrarily close to the singularity.  Moreover, since $\dot{r}^2$ gets correspondingly small, the geodesic can remain in this vicinity for arbitrarily long span in $v$, and consequently make it out to the boundary arbitrarily late.  In particular, the time at which it attains the boundary is given by 
\begin{equation}
t=\frac{1-\sin\mu}{\cos\mu}\log\left[\frac{\cos\left(\frac{\tau+\mu}{2}\right)}{\sin\left(\frac{\tau-\mu}{2}\right)}\right] \ ,
\label{tbdy}
\end{equation}	
which is logarithmically divergent as $\tau \to \mu^+$.

\begin{figure}
\begin{center}
\includegraphics[width=.25\textwidth]{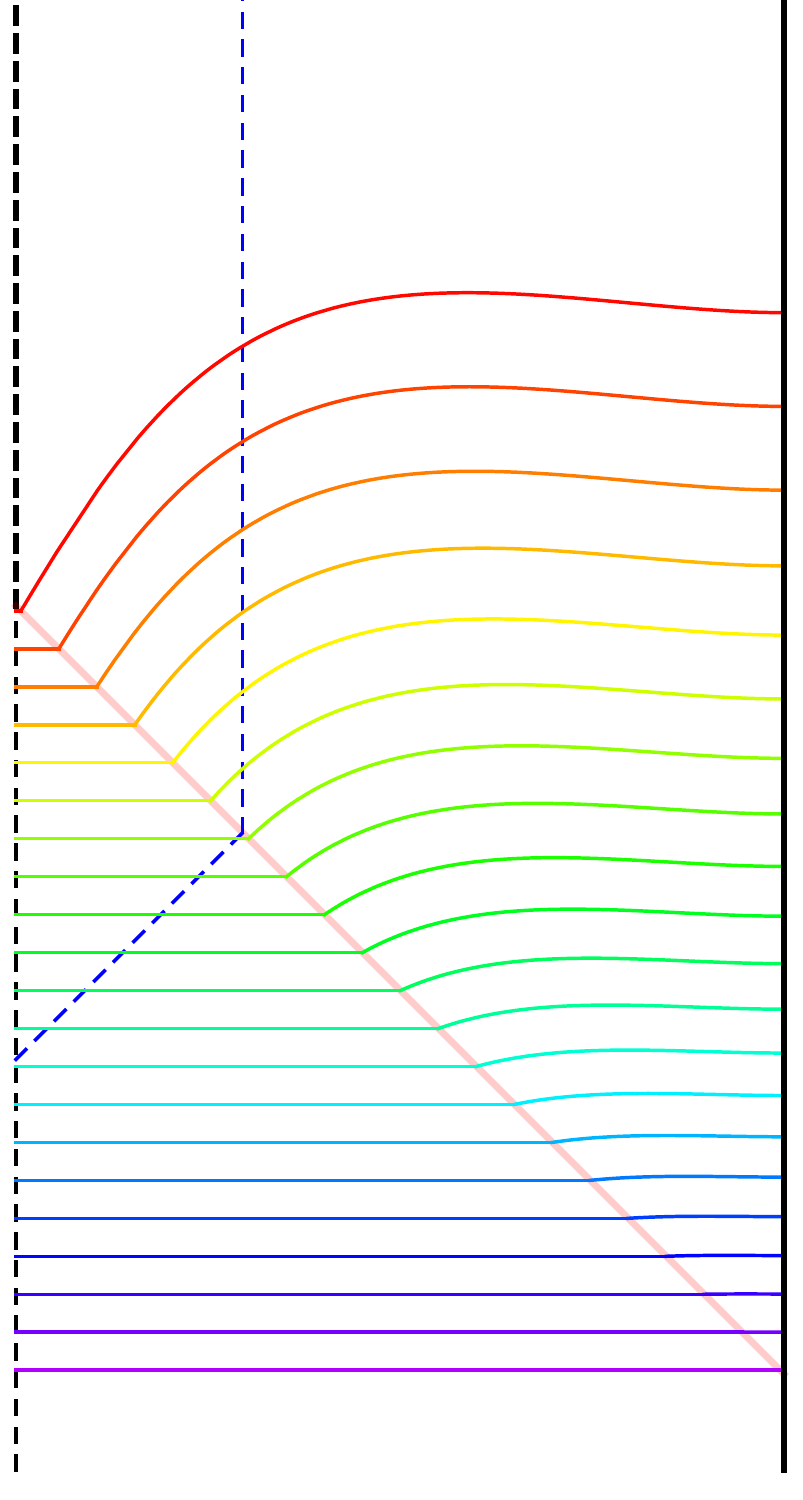}
\hspace{1cm}
\includegraphics[width=.25\textwidth]{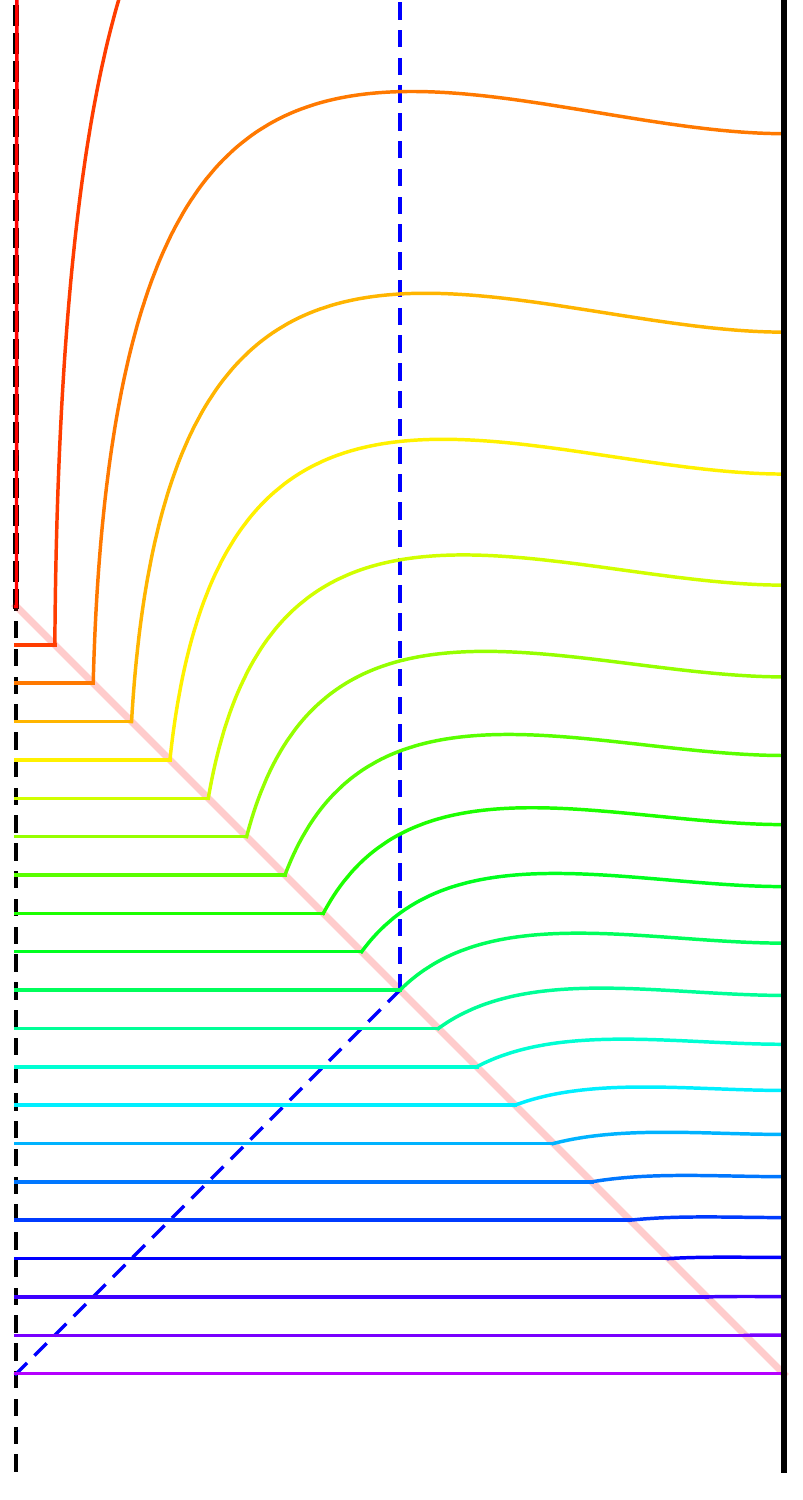}
\hspace{1cm}
\includegraphics[width=.25\textwidth]{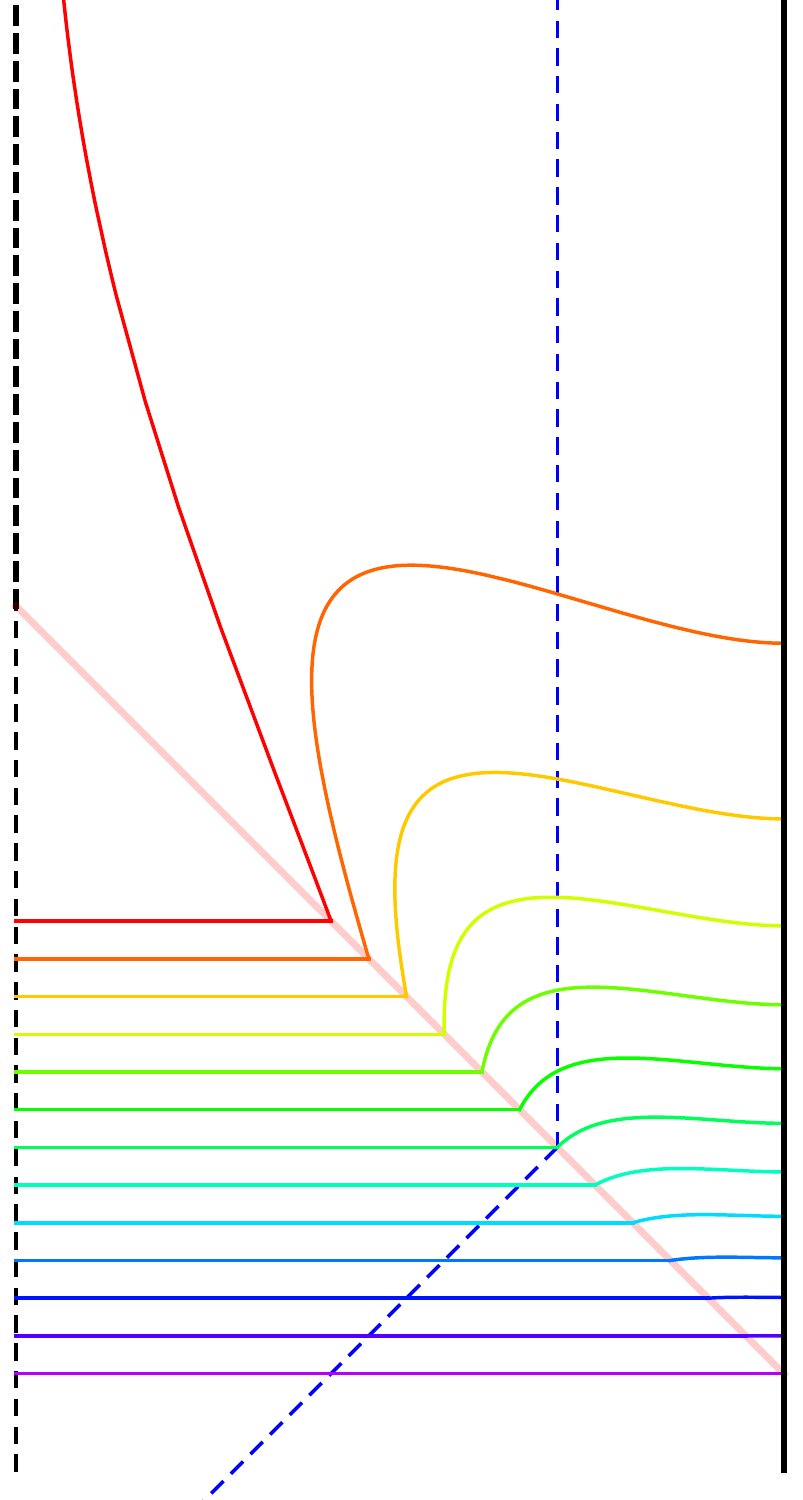}
\caption{
Radial symmetric \ETEBA\ geodesics in Vaidya-BTZ, with horizon size 
$\rh = 1/2$ 
 (left), 
$\rh = 1$ 
 (middle), and 
$\rh = 2$ 
  (right) black holes.  The red geodesic bounds the spacetime region which is attainable to this class of geodesics.  We see that the unattainable region is above and to the left of this curve; for $\rh=1$ (i.e.\ $\mu=0$) the entire spacetime is accessible.}
\label{f:VBTZ_rad_geods}
\end{center}
\end{figure}

\begin{figure}
\begin{center}
\includegraphics[width=.25\textwidth]{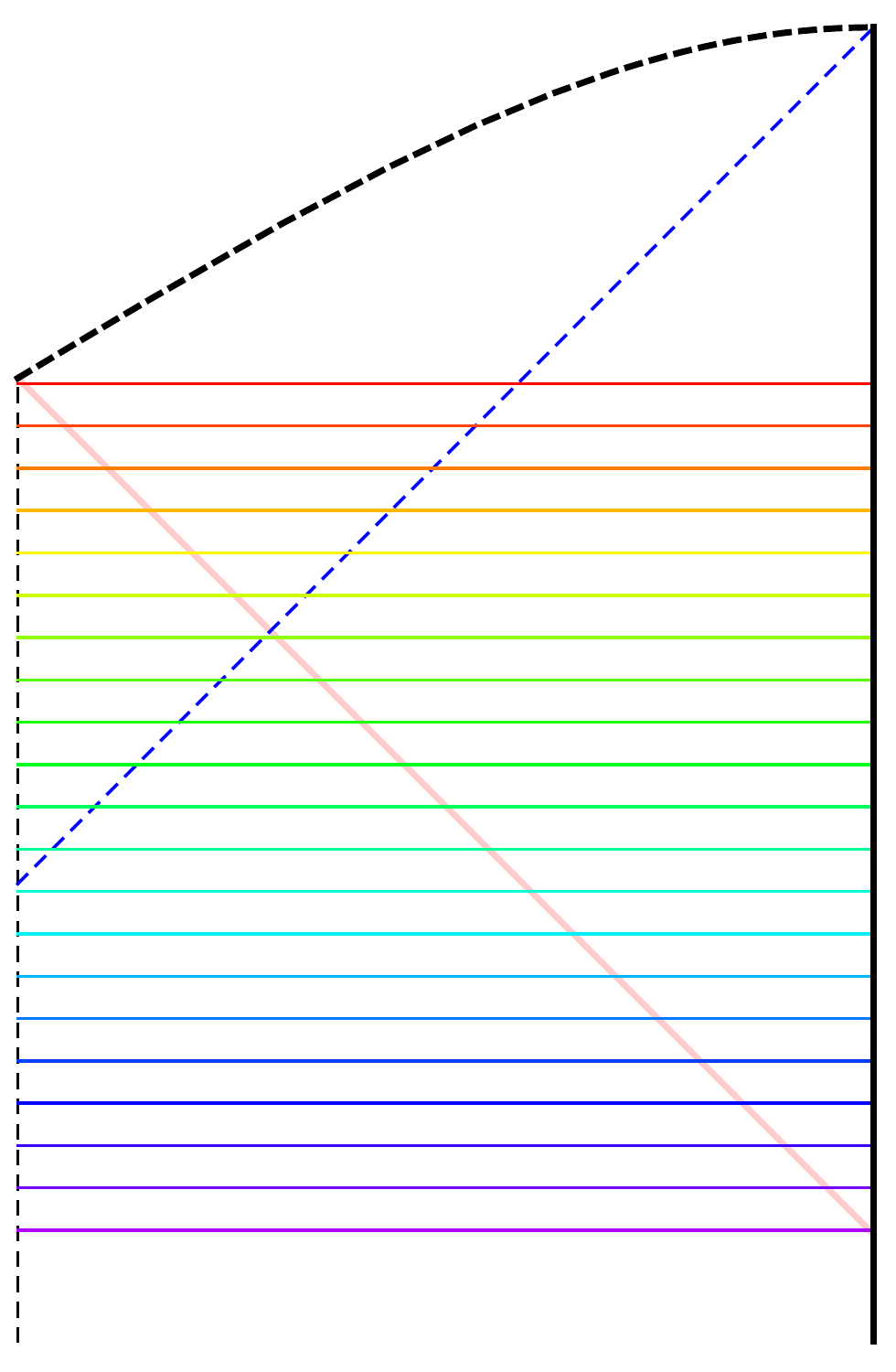}
\hspace{1cm}
\includegraphics[width=.25\textwidth]{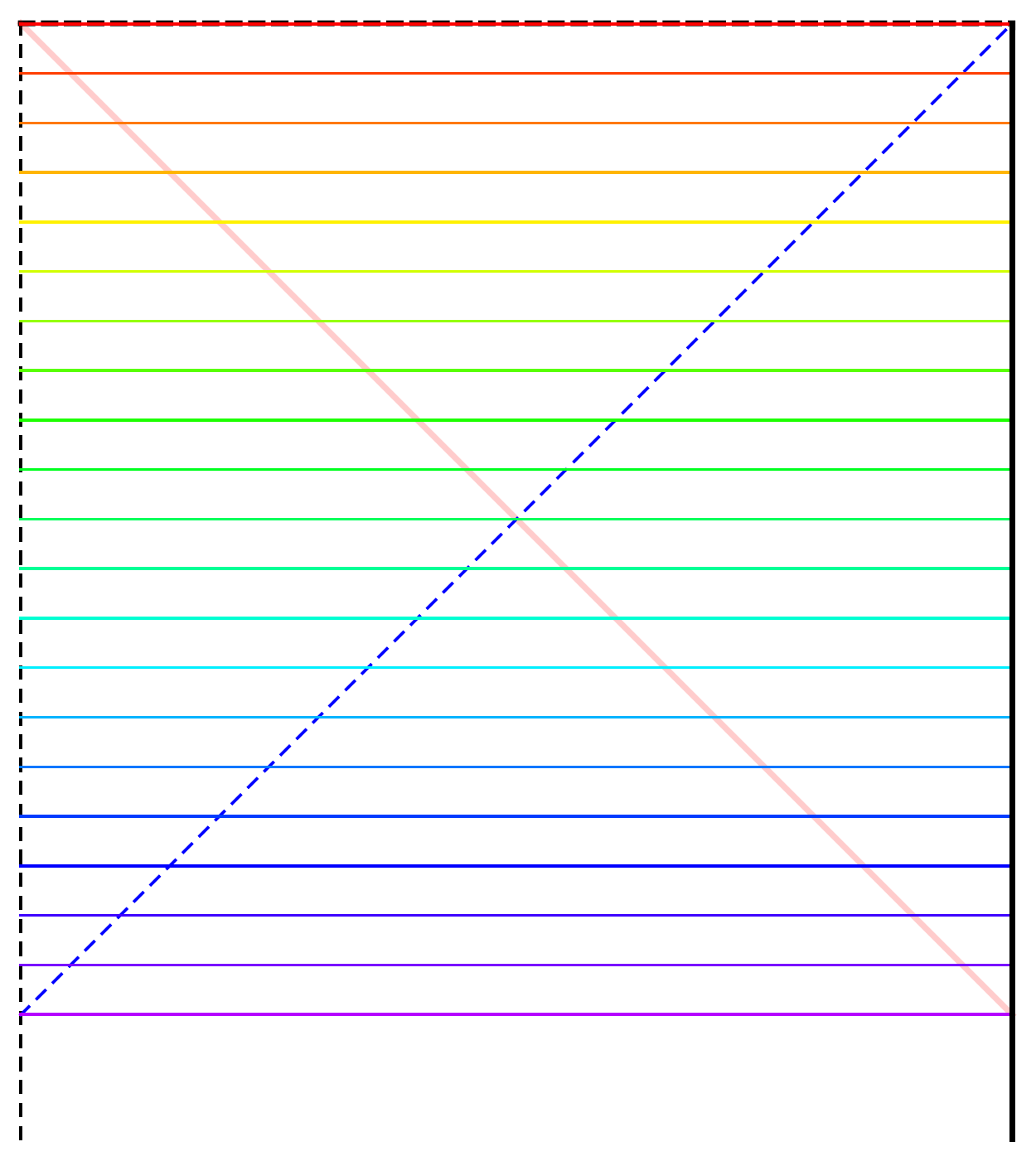}
\hspace{1cm}
\includegraphics[width=.25\textwidth]{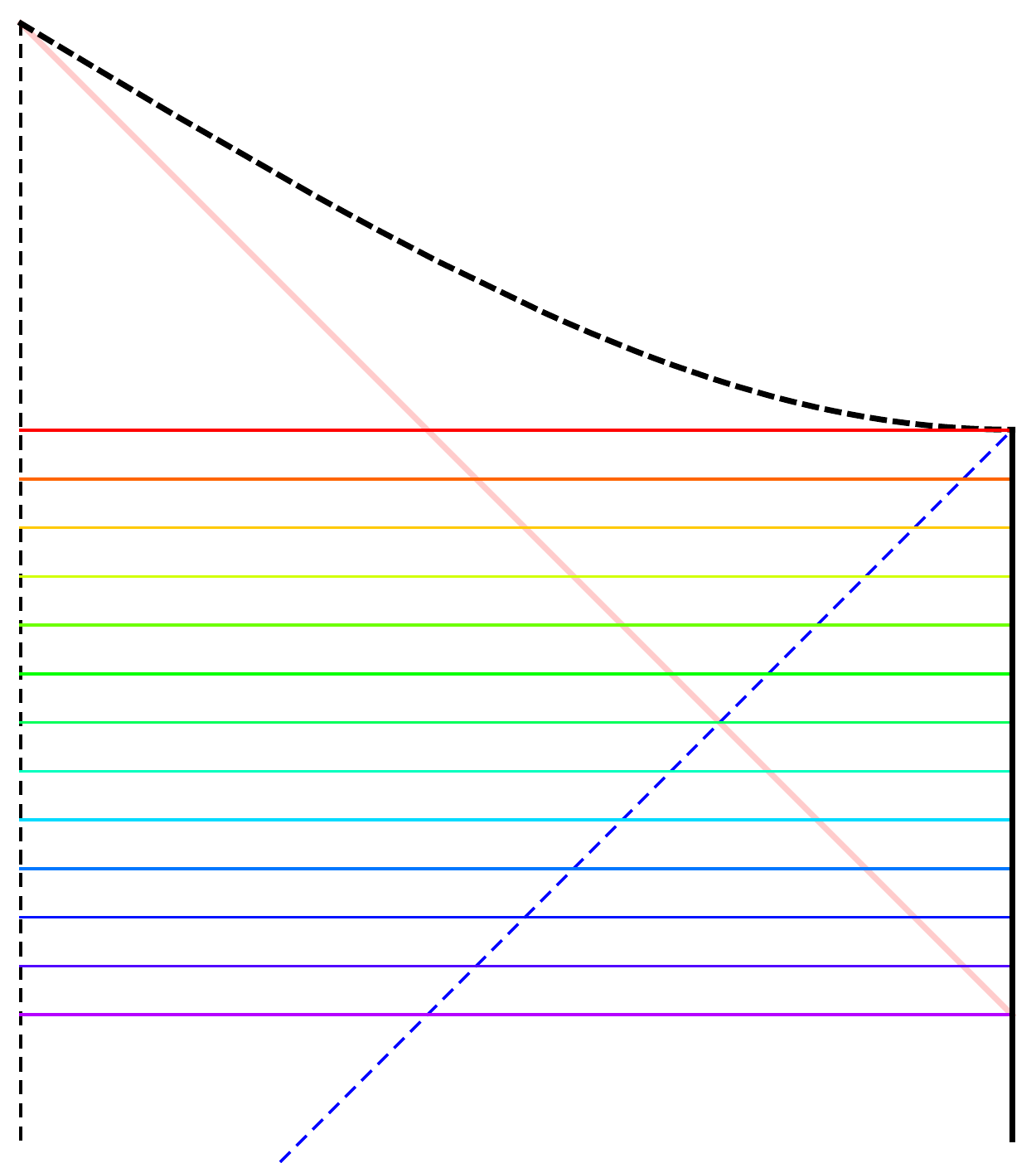}
\caption{
Radial symmetric \ETEBA\ geodesics in Vaidya-BTZ as in \fig{f:VBTZ_rad_geods}, now plotted on the Penrose diagram.}
\label{f:VBTZ_rad_geods_PD}
\end{center}
\end{figure}

From these considerations we can now determine what part of the spacetime is probed by these symmetric radial geodesics.  The attainable region is bounded by the latest such geodesic, which originates inside the shell at $\tau \to 0^+$ for small black holes (i.e.\ when $\mu<0$) and at $\tau \to \mu^+$ for large black holes (i.e.\ when $\mu>0$).  The limit $\mu \to 0$ agrees from both directions, and in this special case the entire spacetime is attainable.  However, when $\mu \ne 0$, some spacetime regions are missing, the character of which depends on whether $\mu$ is positive or negative.  
  This behaviour is illustrated in \fig{f:VBTZ_rad_geods} for small (left), intermediate (middle), and large (right) black holes on Eddington diagram, and in \fig{f:VBTZ_rad_geods_PD} on the corresponding Penrose diagrams.
%

\begin{description}
\item[Small black holes ($\mu<0$):]
The inaccessible region occurs to the future of the geodesic from $\tau\to 0^+$ (which is outgoing everywhere).  This includes the entire interior of the black hole to the future of this geodesic.  
As $\mu \to -\pi/2$, this region is described by the line\footnote{
This relation is simple in the tiny black hole limit since the spacetime region inside the horizon is so small that we can treat is as flat (recall that in BTZ the curvatures do not grow as the singularity is approached).   On AdS scales the curvature is felt, though, and this bounding geodesic attains the boundary at $v=2$.
Note that, in contrast to the Eddington spacetime diagram, all geodesics are in fact straight horizontal lines in the Penrose diagram.
} $v>2 \,  \tan^{-1} r - \frac{\pi}{2}$ on the Eddington plot.
On the other hand, as $\mu \to 0^-$, the initial slope $\frac{dv}{dr}$ increases, and the time at which the boundary is attained \req{tbdy} diverges logarithmically.  In this limit the unattainable region at large $v$ gets pushed off to infinity.
\item[Large black holes ($\mu>0$):]
Now the inaccessible region occurs to the future of the geodesic from $\tau\to \mu^+$, which is initially ingoing, and turns around arbitrarily near the singularity, with arbitrarily small velocity.  This means that the only unattainable region is the one between the shell and this geodesic.   In the limit of very large black hole, $\mu \to \pi/2$, this region is described by the triangle bounded by $r=0$, $v=0$, and $v= \tan^{-1} r - \frac{\pi}{2}$, while as $\mu \to 0^+$ the region receded towards and gets elongated along the singularity $r=0$.
\end{description}

These conclusions are made very clear by using the Penrose coordinates, which give the metric of equation \ref{BTZPenroseMetric}. In particular, it is manifest that the radial geodesics will follow identical curves to the case of pure AdS, and for the symmetric geodesics these are horizontal lines of constant $T$. The only remaining requirement is to know the shape of the singularity, given by $(1-\rh^2)\sin R=(1+\rh^2)\sin T$, which depends on the size of the black hole. For small black holes, this is at increasing $T$ as $R$ increases toward the boundary; for $\rh=1$ it is the horizontal line $T=0$; and for large black holes it lies at decreasing $T$ moving towards the boundary. Concretely, the singularity is between $R=T=0$, and $R=\pi/2, T=2\tan^{-1}\rh-\pi/2$. This alone is enough to reproduce the plots of figure \fig{f:VBTZ_rad_geods_PD} and the associated conclusions.

The very restricted set of symmetric radial geodesics is a good starting point, but is too constraining. In particular, one might naturally expect that the region of spacetime covered will be increased by including more general classes of geodesic.  As we demonstrate below, this expectation is only realised for small black holes.

For the small black holes, the result is analogous to the higher dimensions, in that boundary-anchored geodesics will cover the whole spacetime,
though for a rather different reason -- the mechanism can no longer rely on  geodesics bouncing off the singularity.
 Indeed, we can use the same construction used in the previous section, of picking a radial geodesic passing through an arbitrary point at a local minimum of $r$, though it requires more work to argue that it will avoid the singularity. In fact, we can do better still in this case, since we can reach the same conclusion even with the restricted class of symmetric geodesics, once angular momentum is allowed. In particular, this means that \eteba\ geodesics cover the whole spacetime.

This conclusion can be reached by considering a family of geodesics with initial conditions close to the singularity formation, with a small angular momentum. To fix notation, we will generalize the definition of the initial time $\tau$ to correspond to minus the initial AdS time, so that the shell is always reached at $\rs = \tan \tau$.  This means that we must restrict $L < \tan \tau$ so that the geodesic actually starts inside the shell. We then consider the family of geodesics with angular momentum $L=(-\sin\mu)\tau$. Taking $\tau$ to be small, there is a parametric separation between the radius where the geodesic crosses the shell $\rs$, the circular orbit radius $r_0$ at which the effective potential reaches its maximum, and the horizon $\rh$. Asymptotically as $\tau\to0$:
\begin{equation}
\rs \sim \tau 
\ll
r_0 \sim \sqrt{\frac{-\sin\mu \, \cos\mu}{1-\sin\mu}} \, \sqrt{\tau}
\ll
\rh = \frac{\cos \mu}{1-\sin\mu}.
\label{}
\end{equation}	
Moreover, the difference between $E^2$ and the maximum of the effective potential, which is the minimum of $\dot{r}^2$, is asymptotically
\begin{equation}
E^2-V(r_0) \sim \frac{-2\, \sin\mu \, \cos\mu}{1-\sin\mu} \, \tau > 0.
\label{}
\end{equation}	
This is positive but small, which means that the geodesic stays in the vicinity of $r_0 \sim \sqrt{\tau}$ for an arbitrarily long time $\Delta v$ as $\tau \to 0$ before reaching the boundary.  Since we can make $r_0$ arbitrarily small (and parametrically inside the horizon even for arbitrarily small black holes), and the radial velocity there likewise arbitrarily small, such geodesics penetrate arbitrarily close to the singularity at arbitrarily late time $v$. Details of the computation are included in \sect{VaidyaBTZapp}.

For the large black holes, the situation is entirely different, since the coverage of the radial symmetric geodesics is not improved by including even the most general boundary-anchored geodesics. The region covered by all geodesics is thus bounded by the innermost symmetric radial geodesic. This region includes points arbitrarily close to the singularity at late times, but is bounded away from its formation. This conclusion is easy to reach by using the Penrose coordinates once more. The equation of motion for geodesics associated with $T$ in the BTZ part of the spacetime is
\begin{equation}
\ddot{T}+2\dot{R}\dot{T}\tan R =\frac{1+\rh^2}{2}\frac{L^2}{r(T,R)^3}\cos R \cos T,
\end{equation}
and the right hand side is positive for the corresponding range of coordinates. If $\dot{T}=0$, $\ddot{T}\geq0$, with equality only for the radial ($L=0$) geodesics, so $T$ can never have a local maximum on the geodesic. This means that if a geodesic lies above the critical curve $T=2\tan^{-1}\rh-\pi/2$ for any of its length, it must end in the singularity in at least one direction. The conclusion is that boundary-anchored geodesics see no more of the spacetime than the symmetric radial ones, namely the region $T\leq2\tan^{-1}\rh-\pi/2$.

\paragraph{Asymmetric \eteba\ geodesics:}

In higher dimensions, we saw the novel feature of geodesics with endpoints at equal times, but nonetheless having no reflection symmetry. Our intuition for their existence relied on competition between two effects, one of which required nearly-null radial geodesics to be repelled from the singularity. In the case of BTZ, this effect is absent, since the effective potential is bounded, so it is a natural expectation that this class of geodesics does not exist.

If asymmetric \eteba\ geodesics were to exist, it is expected that they would appear amongst radial geodesics, to give the largest potential barrier away from the singularity. With this simplification of assuming zero angular momentum, it is immediate from the metric in terms of Penrose coordinates that they may not exist. As already noted, in these coordinates the radial geodesics are identical to those in pure AdS (with the restriction that they must avoid the singularity), which move monotonically in $T$.

Allowing for angular momentum, this straightforward argument fails since $T$ may have a minimum in the interior of the spacetime. The possibility that there may be asymmetric \eteba\ geodesics is not in principle ruled out, but it seems highly unlikely that they would only appear for some intermediate $L$. This conclusion is supported by numerical calculations such as performed in higher dimensions, from which we find that for $d=2$ there are indeed no asymmetric \eteba\ geodesics.

\paragraph{Regions probed by \eteba\ geodesics, and lengths:}

Our previous remarks have already answered the question of the region probed by \eteba\ geodesics, being in the case of small black holes the entire spacetime, and in the case of large black holes the region outside the latest boundary-anchored radial symmetric geodesic. The final part of the picture is the refined question of the region covered by the shortest \eteba\ geodesics.

The question of which geodesics dominate by virtue of having shortest length for given endpoints was investigated numerically, and turns out to have a simple answer, in contrast to the higher-dimensional cases. Because the only \eteba\ geodesics are the symmetric ones, we need only look at a two-parameter initial condition space, characterized by the location of the minimum of $v$.

We begin with the geodesics connecting antipodal points. There are two obvious candidates for such geodesics. Firstly, radial geodesics, with initial condition at $r=0$, will automatically fall into this class. Secondly, in the static BTZ geometry there are antipodal geodesics passing outside the event horizon, with closest approach at $r=r_\mathrm{min}$, so in Vaidya-BTZ they must exist at late times, along with a continuation of the family to earlier times. This family in fact joins up continuously with the radial geodesics. Before this time, the only choice is the radial family, but after the nonradial family appears, there is a choice of two, of which the nonradial is always shorter. This means that the shortest antipodal geodesics follow a continuous curve in initial condition space as boundary time increases, starting at $r_0=0$, moving to nonzero $r_0$ when the new family appears, and following this to join the static BTZ geodesics at $r_0=r_\mathrm{min}$. This outermost contour in initial condition space of $\Delta\phi=\pi$ turns out to be a boundary between initial conditions of shortest geodesics, lying outside it, and longer ones, lying inside it. In particular, the geodesics approaching close to the singularity are never shortest.

The region probed by these shortest geodesics is again covered by those with antipodal endpoints, with others reaching no deeper. In the case of small black holes $r_+\leq 1$, it is simple to characterize, being bounded by two curves. The first is the latest radial geodesic of shortest length, with initial conditions at the critical point at which nonradial antipodal geodesics appear. The second curve is the deepest reach of the surfaces contained entirely in the static BTZ, at $r=r_\mathrm{min}$. In particular, from the time when nonradial geodesics become dominant, they never see deeper than the last radial geodesic, excepting for later points outside the minimal radius $r_\mathrm{min}$.

For large black holes, the situation is similar, with the difference that at intermediate times the geodesics `cut the corner' inside these two curves, passing through a small additional portion of the spacetime.

\begin{figure}
\begin{center}
\includegraphics[width=.25\textwidth]
{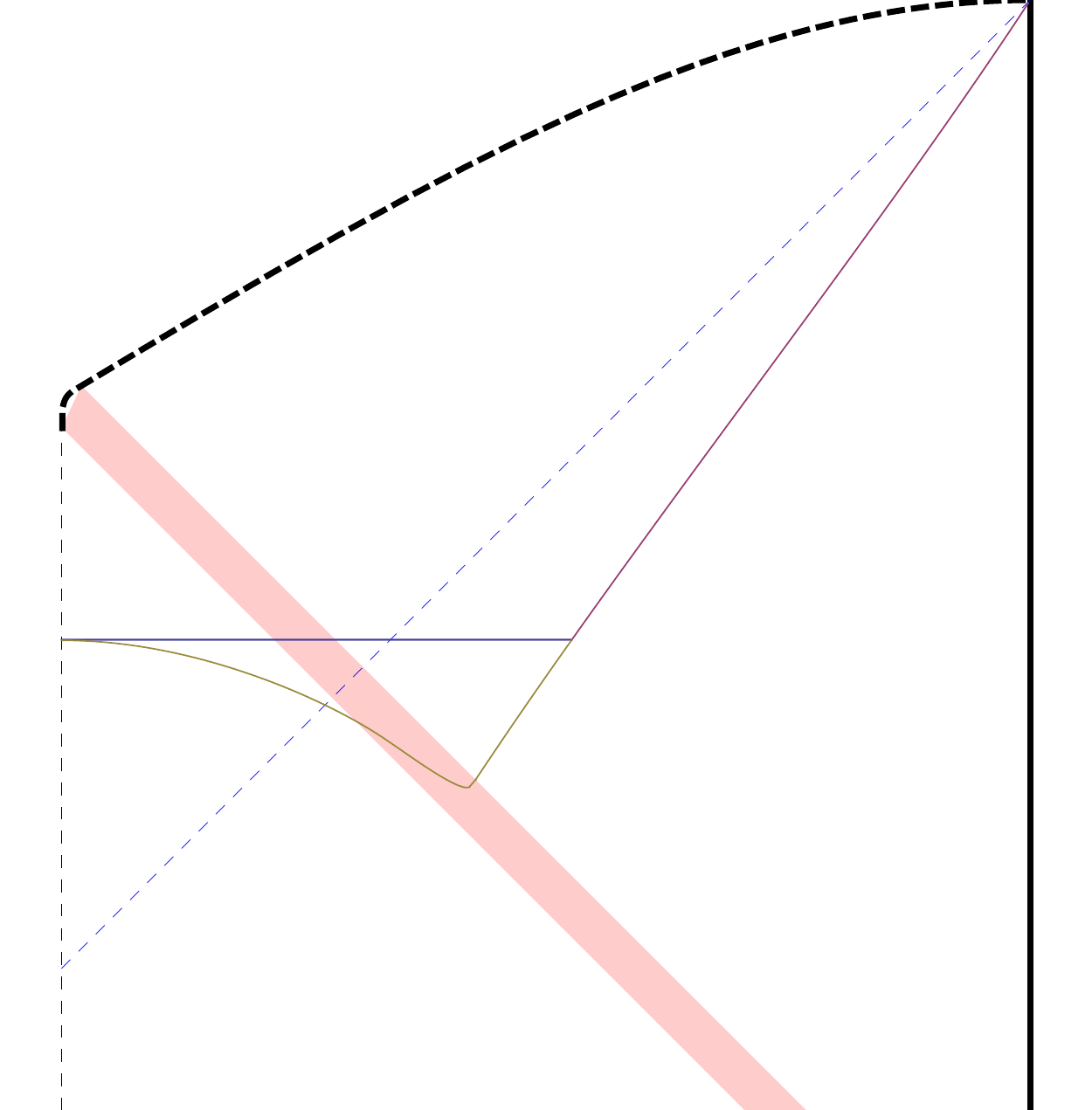}
\hspace{1cm}
\includegraphics[width=.25\textwidth]
{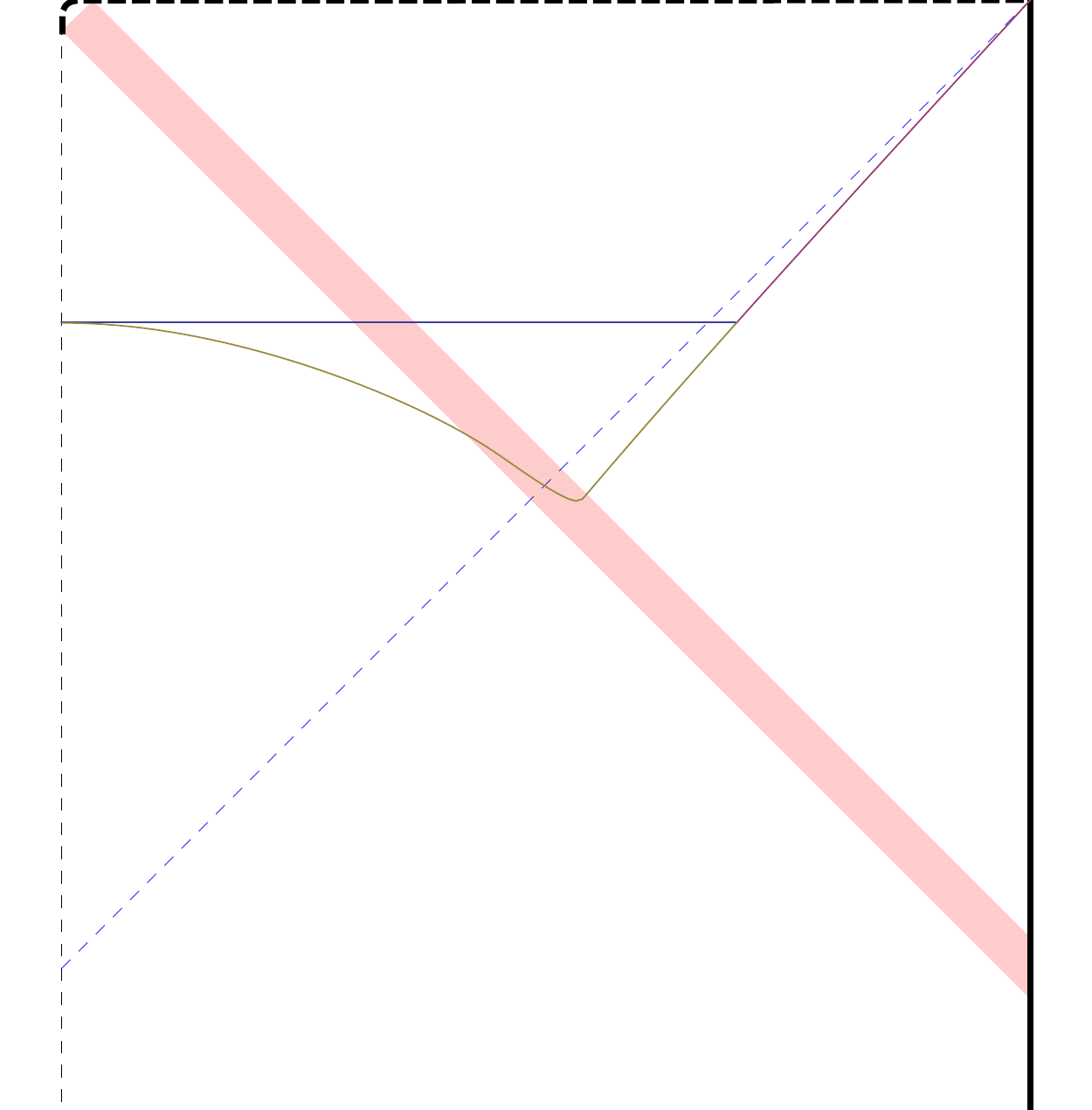}
\hspace{1cm}
\includegraphics[width=.25\textwidth]
{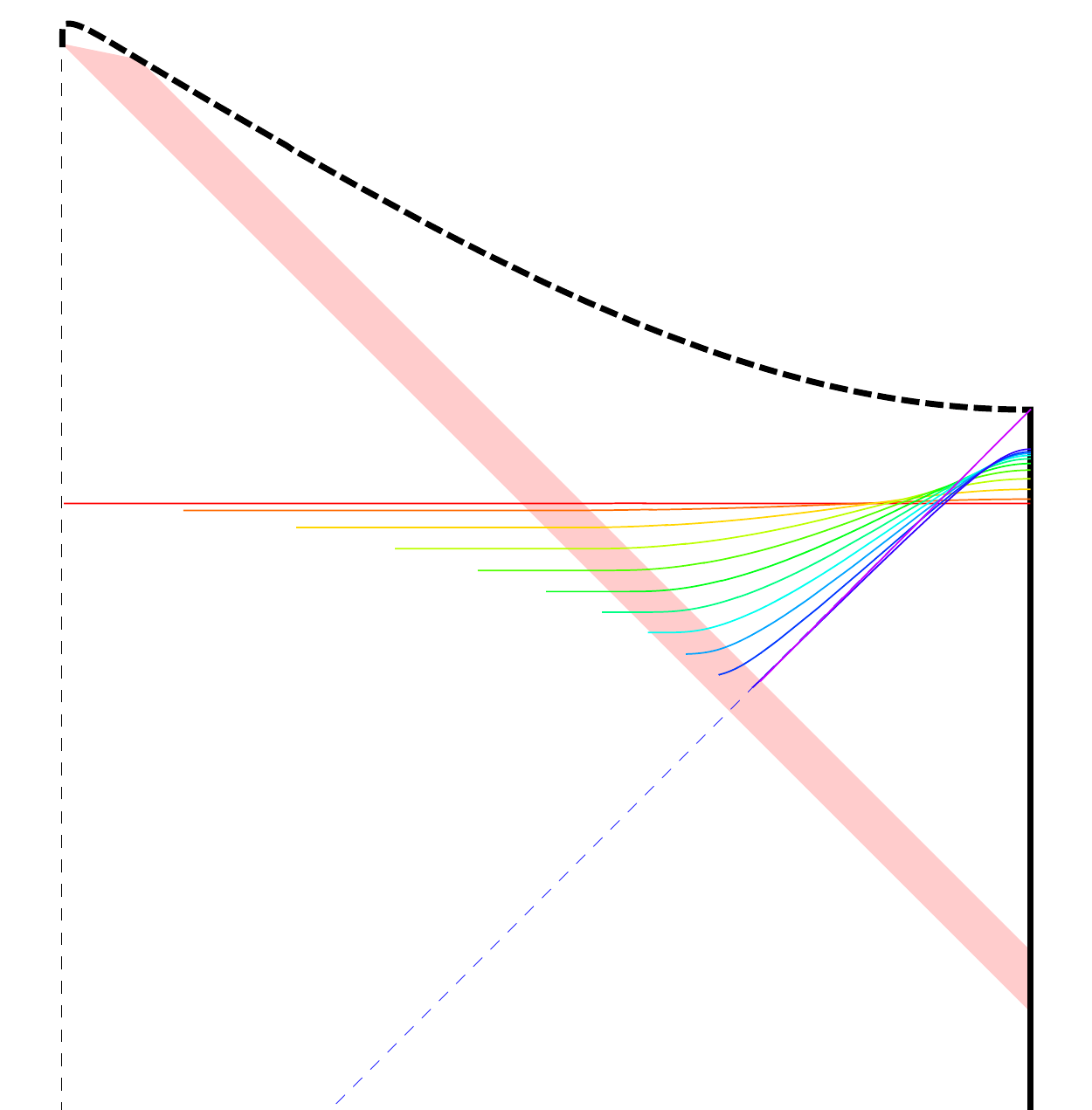}
\caption{
Region accessible by shortest \eteba\ geodesics in Vaidya-BTZ as in \fig{f:VBTZ_rad_geods},  plotted on the Penrose diagram. For large black hole, individual geodesics are plotted to illustrate the rounding of accessible region.}
\label{f:VBTZ_shortest_geods_PD}
\end{center}
\end{figure}

These regions covered are shown in \fig{f:VBTZ_shortest_geods_PD} for small, critical and large black holes, along with the curve of initial conditions giving antipodal geodesics.

\begin{figure}
\begin{center}
\includegraphics[width=.8\textwidth]{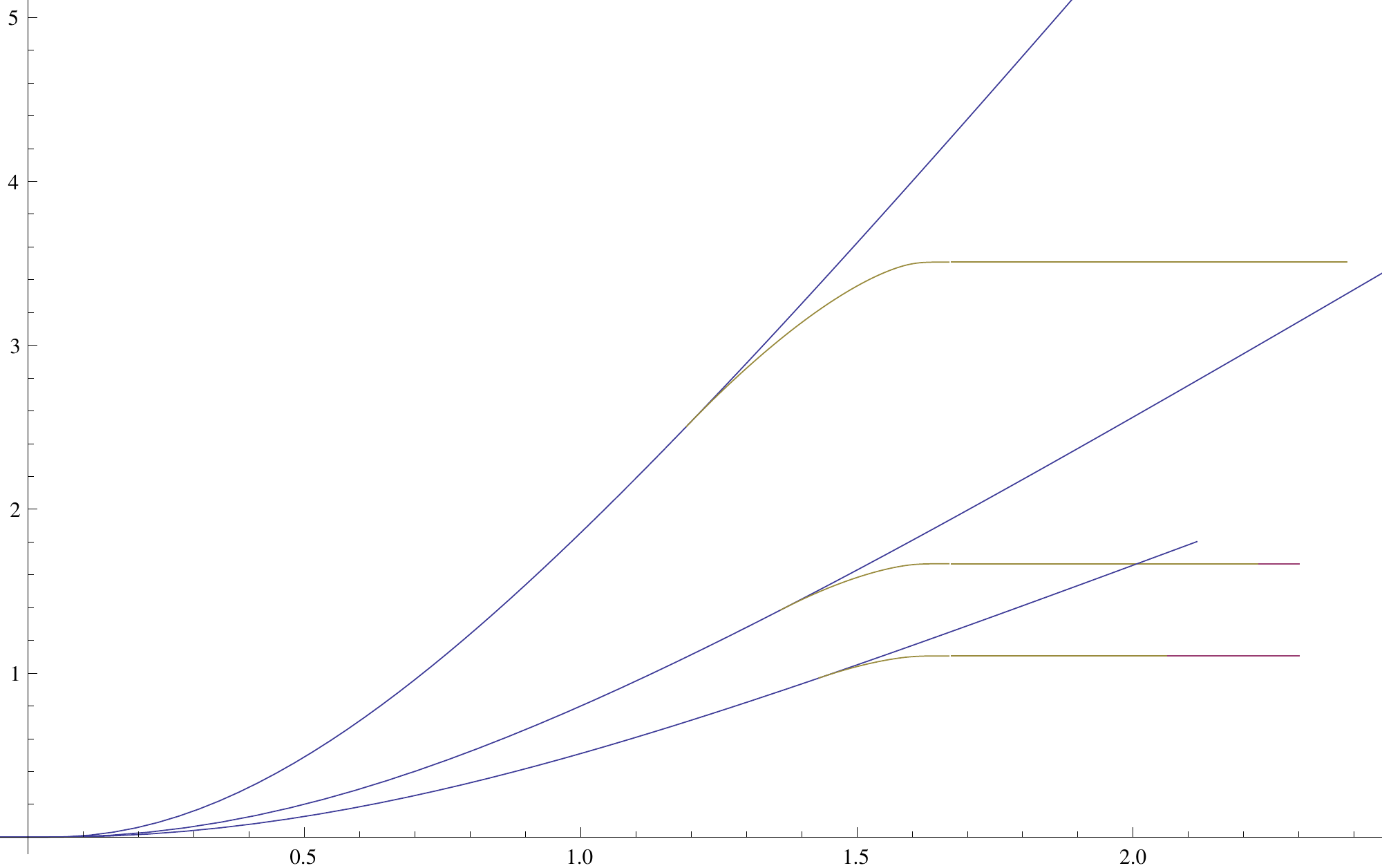}
\begin{picture}(0,0)
\setlength{\unitlength}{1cm}
\put (0.2,0) {$t$}
\put (-14.3,8) {$\length$}
\end{picture}
\hspace{1cm}
\caption{
Regularised proper lengths along \eteba\ geodesics in Vaidya-BTZ, plotted as a function of boundary time.  Blue curves correspond to  the radial geodesic branch, the others to the non-radial branch.  The three sets of curves (top to bottom) correspond to $\rh = 2, 1$, and $0.5$, respectively.
}
\label{f:AntipodalLengths}
\end{center}
\end{figure}
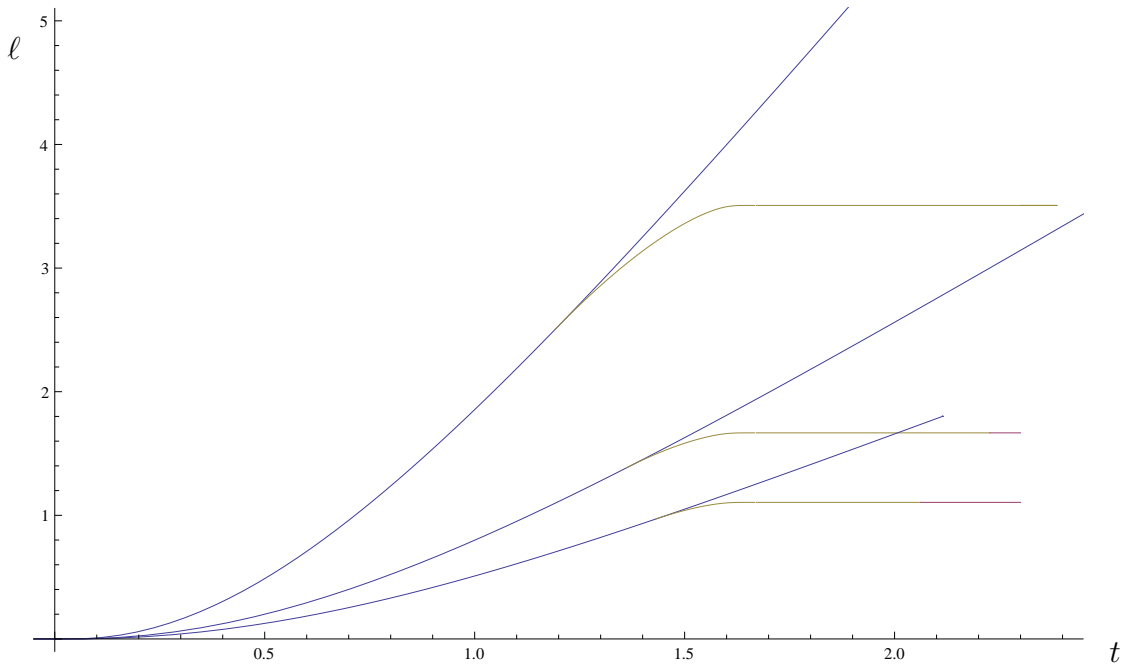

Finally, we take the opportunity to note how the lengths of the geodesics evolve with boundary time $t_\infty$. In stark contrast to the higher-dimensional case, the length increases monotonically and smoothly with time, as shown in \fig{f:AntipodalLengths}. This is fortunate, as we have a more direct field theory interpretation for the observable associated with these lengths, postulated to be the entanglement entropy of the region between the endpoints. 
Furthermore, the early time growth, which in the case of antipodal points can be extracted from the expression in equation \ref{BTZradiallengths}, agrees precisely with the results of \cite{Liu:2013iza,Liu:2013qca}.

\newpage

\section{Codimension-two extremal surfaces}
\label{s:surfsVaidya}

Having considered the properties of \ETEBA\ geodesics (which are simply one-dimensional extremal surfaces) in the previous section, we now turn to codimension-two extremal surfaces.  As remarked previously, the 3-dimensional Vaidya-BTZ set up studied in \sect{VaidyaBTZgeods} is a special case of these.  Here we generalise this case to higher dimensions, keeping the codimension fixed.
We restrict exclusively to surfaces anchored to $(d-2)$-spheres at constant latitude, to retain an $O(d-1)$-subgroup of the $O(d)$ spherical symmetry. Further, we consider only surfaces that respect this symmetry in the bulk spacetime, which makes the great simplification of reducing the extremising equations from partial to ordinary differential equations. The experience from the geodesics in higher dimensions, where the boundary $O(1)=\mathbb{Z}_2$ symmetry of swapping insertion 
 points need not be respected by the shortest geodesics in the bulk with given equal time endpoints, shows that this may be a genuinely restrictive assumption. 
However, regarding the surfaces more naturally as a generalisation of geodesics in 3 dimensions counters this concern.
Moreover, in the case of extremal surfaces, the symmetry is continuous, which perhaps makes it less likely to be broken than for the discrete symmetry for geodesics.

The surface is parametrised by $d-2$ `longitudes' $\phi$ on $S^{d-2}$, along with one other parameter, for now generically denoted by $s$, such that $v$, $r$, and the colatitude $\theta$ depend only on $s$. The area functional of the surface is given by

\begin{equation}
A=V(S^{d-2})\int ds \, (r\sin\theta)^{d-2} \, \sqrt{-f(r,v) \, \dot{v}^2+2 \, \dot{r} \, \dot{v}+r^2 \,  \dot{\theta}^2}
\end{equation}
where dots denote differentiation with respect to $s$, and $V(S^{d-2})$ is the volume of the unit $S^{d-2}$. The surfaces of interest are extrema of this functional, so the integrand acts as the Lagrangian from which the equations of motion my be obtained\footnote{
One may worry that an extremum of this functional may not be a true extremal surface, since there are variations away from spherical symmetry. However, we have verified that stationarity with respect to variations preserving the symmetry implies stationarity with respect to all variations.}.
We have complete freedom in choice of the parameter $s$, setting it to be equal to one of the coordinates, or another convenient choice, for example a parameter analogous to the arc length of geodesics.

The surface must meet the poles of the $S^{d-1}$ at $\theta=0$ or $\pi$ exactly once (excluding self-intersecting surfaces), so we can set boundary conditions at the North pole (WLOG), specifying that the surface must be smooth there. The equations of motion are singular at these points, so for the purposes of numerics, initial conditions are set by solving the first few terms in a series expansion near $\theta=0$, and using this to begin the integration at a small positive value of $\theta$. An exception to this rule is when the surface passes through the origin $r=0$ (inside the shell, when $v<0$), in which case the symmetry is enhanced, and the surface must lie entirely on the equatorial plane $\theta=\pi/2$. Details of parameters used, equations of motion, series solutions and initial conditions are given in  Appendix \ref{surfApp}.

One useful point from the equations of motion is that, like the geodesics, if $\dot{v}=0$, then $\ddot{v}>0$, so $v$ can never have a local maximum. This tells us that the value of $v$ on a particular surface is largest when $r\to\infty$, smallest at the initial point where it crosses the pole, and monotone between. In particular, a surface anchored to the boundary before the collapse begins can only sample the pure AdS part of the geometry, so entanglement entropy cannot evolve before the quench. It also means that $v$ is a suitable parameter along the surface (unlike $\theta$ or $r$), useful because it reduces the order of the equations of motion to 4.

From the field theory point of view, there are three interesting pieces of data associated with each extremal surface ending on the boundary. The first two, $\theta_\infty$ and $t_\infty$, respectively the colatitude and time at which the surface meets the AdS boundary (i.e.~the asymptotic values of $\theta$ and $v$ respectively as $r\to\infty$), characterise the region with which it is associated. The third, the area $A$, is a candidate for the entanglement entropy of the region, according to the conjecture in~\cite{Hubeny:2007xt}. If there are several candidates for a given boundary region, the proposal specifies that we take the minimal area,\footnote{
A summary of alternate definitions and subtleties in specification of the covariant entropy proposal, especially the role of the homology constraint, have been recently discussed in \cite{Hubeny:2013gta}.
} so comparison of areas in particular will be important for our purposes. As in the case of geodesics, we can think of each surface as giving a point in the `boundary parameter space' $(\theta_\infty,t_\infty,A)$.

The area itself is divergent due to the portion of the surface heading to $r=\infty$. We regulate by cutting off the surface at a large but finite $r=r_c$, corresponding to a UV cutoff in the field theory. The leading order divergence can be computed as $V(S^{d-2})(r_c \sin\theta_\infty)^{d-2}/(d-2)$, though for $d>3$ there are additional divergences: logarithmic for $d=4$, and stronger as $d$ increases. Hence, as a simple universal prescription to renormalise the area, we use background subtraction. We compute the area as a function of the cutoff radius $r_c$, and subtract the same quantity computed for an extremal surface in the static AdS spacetime with the same $\theta_\infty$, which can be computed analytically. As the cutoff is taken to infinity, this tends to a constant, which is defined as the renormalised area. The practicalities of this are outlined in the appendix.

The equations of motion along with smoothness condition at $\theta=0$ have a 2-parameter family of solutions, labelled by the point $(r_0,v_0)$ where the surface crosses the pole of the sphere. Similarly to the case of geodesics, we can think of this as a 2-dimensional `initial condition space', parametrizing the set of spherically symmetric extremal surfaces.

Integration of the equations proceeds from this initial point, and either ends in the singularity, or continues to the AdS boundary $r\to\infty$. We are interested only in the latter, so one requirement is to find the region of the initial condition space $(r_0,v_0)$ corresponding to these probe surfaces.

The problem of computing the entanglement entropy of regions bounded by such a sphere of constant latitude, according to the proposal in \cite{Hubeny:2007xt}, is then one of identifying the surfaces with appropriate $(\theta_\infty,v_\infty)$, and finding their areas. For any given region, we expect some discrete set of surfaces; those of most interest will be the ones of minimal area amongst that set.

The main problem thus amounts to finding first the domain of the function $(r_0,v_0)\mapsto(\theta_\infty,t_\infty,A)$, for which the minimal surface ends at the boundary rather than in the singularity, and then understanding the image, a surface in $(\theta_\infty,t_\infty,A)$-space. We must bear in mind that for the purposes of computing entanglement entropy, there is an equivalence $\theta_\infty \sim\pi-\theta_\infty$, corresponding to a choice of which side surfaces may pass round the origin and related by a rotation. In particular, this is not disrupted by a homology constraint as in the globally static case \cite{Hubeny:2013gta,Takayanagi:2010wp}, 
and only connected surfaces need be considered, which is a reflection of the CFT being in a pure state.\footnote{
Geometrically, the homology constraint being satisfied (even in the strong form of there existing a bulk codimension-one smooth achronal surface whose only boundaries are the anchoring region and the extremal surface), follows from the fact that from arbitrarily late section of the boundary, there exists a smooth spacelike surface stretching to pre-singularity-formation era.
This is most readily apparent by considering the Penrose diagram, and follows directly from the spacelike nature of the curvature singularity and the fact that there is only a single asymptotic region.
}

The result must interpolate between the two static geometries, namely pure AdS for $t_\infty<0$ and \SAdS\ for large $t_\infty$. The surface in the $(\theta_\infty,t_\infty,A)$ boundary parameter space must smoothly match up these early and late parts.

For pure AdS, this is all known analytically: the surfaces lie on constant time slices, and in fact are surfaces of revolution of geodesics. Explicitly, the map is given by $\theta_\infty=\cot^{-1}r_0$, $t_\infty=v_0+\cot^{-1}r_0$, and $A=0$ as a consequence of the renormalisation prescription. The domain of relevant initial conditions is the whole spacetime, and the resulting surface in $(\theta_\infty,t_\infty,A)$ space is the plane $A=0$, for $0<\theta<\frac{\pi}{2}$.

\begin{figure}
\begin{center}
\includegraphics[width=.8\textwidth]{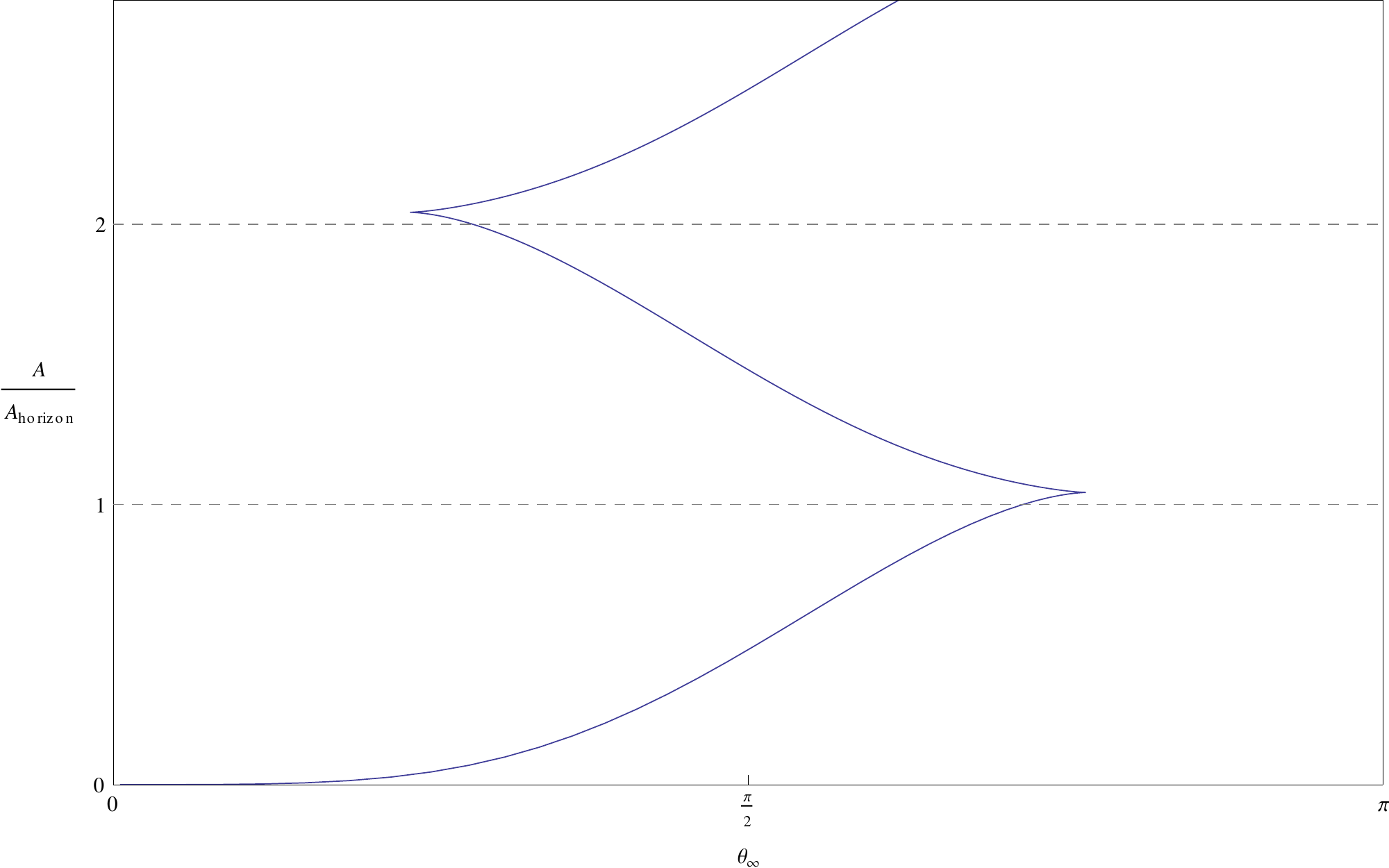}
\caption{
Renormalised areas of connected minimal surfaces relative to the area of the event horizon plotted against their colatitude $\theta_\infty$ at the boundary. The tower of surfaces continues indefinitely, repeating in an essentially periodic pattern.
}
\label{f:schwAreas}
\end{center}
\end{figure}

In the case of \SAdS, again the staticity greatly simplifies, since the surfaces are orthogonal to the timelike Killing field. This causes the dependence to decouple, so that $\theta_\infty$ depends only on $r_0$, and $t_\infty$ reduces to the value of the static coordinate $t_o$ on which the surface lies. The na\"ive expectation here is that any boundary region should be matched by two minimal surfaces, one passing round each side of the event horizon, but the actual story is more complicated, and is described in detail in \cite{Hubeny:2013gta}. There is an infinite tower of surfaces for sufficiently large boundary regions, shown in \fig{f:schwAreas}. Only the lowest branch of this tower, up to $\theta_\infty=\pi/2$, is directly relevant for computing entanglement entropy in the static case, since the others all have larger area. The domain of relevant initial conditions is the exterior of the event horizon, and the surface in $(\theta_\infty,t_\infty,A)$ space is a translation in the $t$-direction of the multi-branched shape of \fig{f:schwAreas}.

It is useful to bear these static cases in mind, since during the collapse, the surface in $(\theta_\infty,t_\infty,A)$ must interpolate between these regions, and join the simple plane pre-collapse to the complicated folded surface post-collapse. To solve the puzzle of how this can occur, we turn to numerical studies. The computations were made almost exclusively in $\text{AdS}_5$ (i.e.~$d=4$), with horizon radius $r_+=1$, though comparisons with other parameters indicate that the results are generic. Details of the methods are given in Appendix \ref{surfApp}.

\paragraph{Domain of relevant initial conditions, and an interesting class of surfaces}

\begin{figure}
\begin{center}
\includegraphics[width=.4\textwidth]{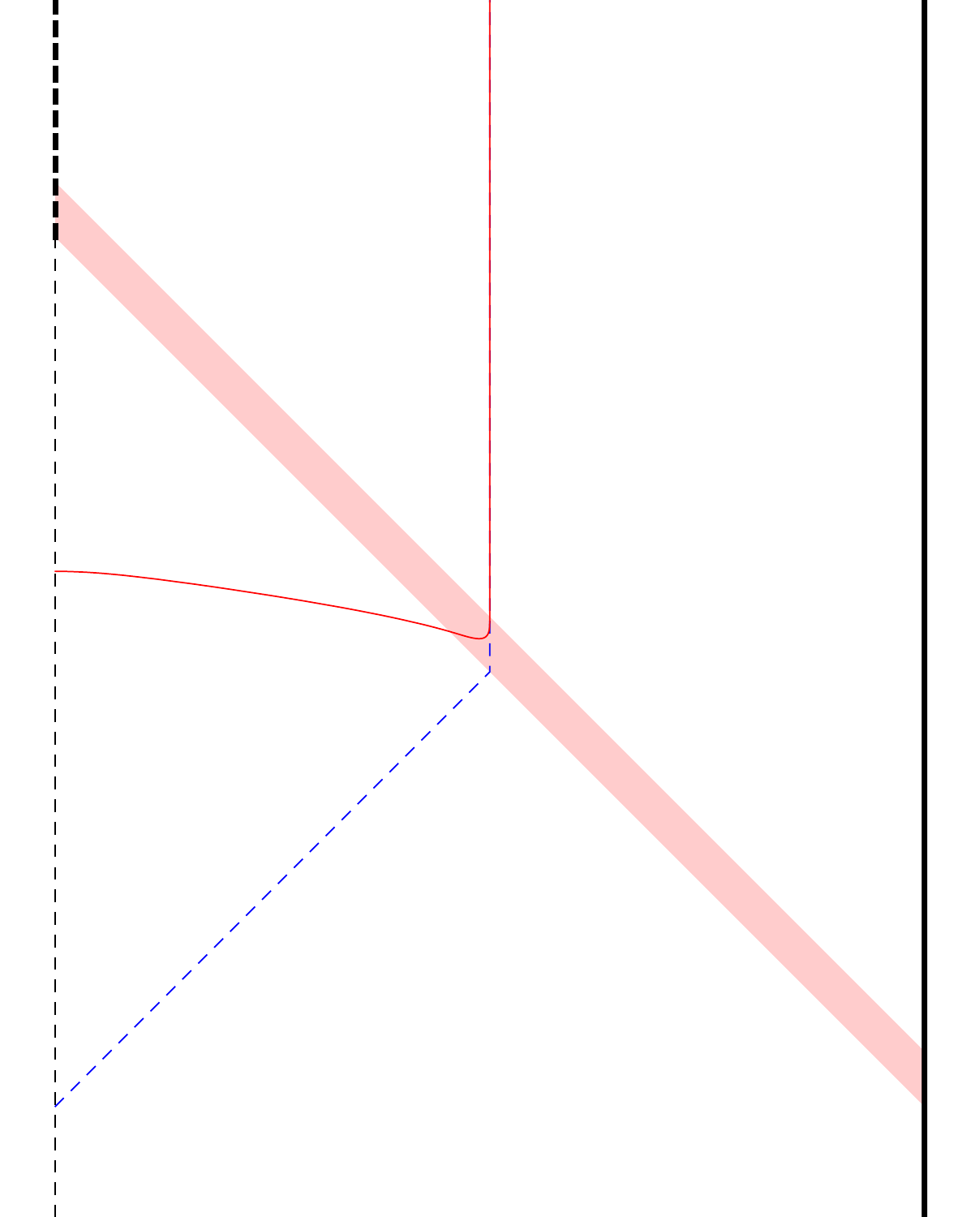}
\hspace{1cm}
\includegraphics[width=.4\textwidth]{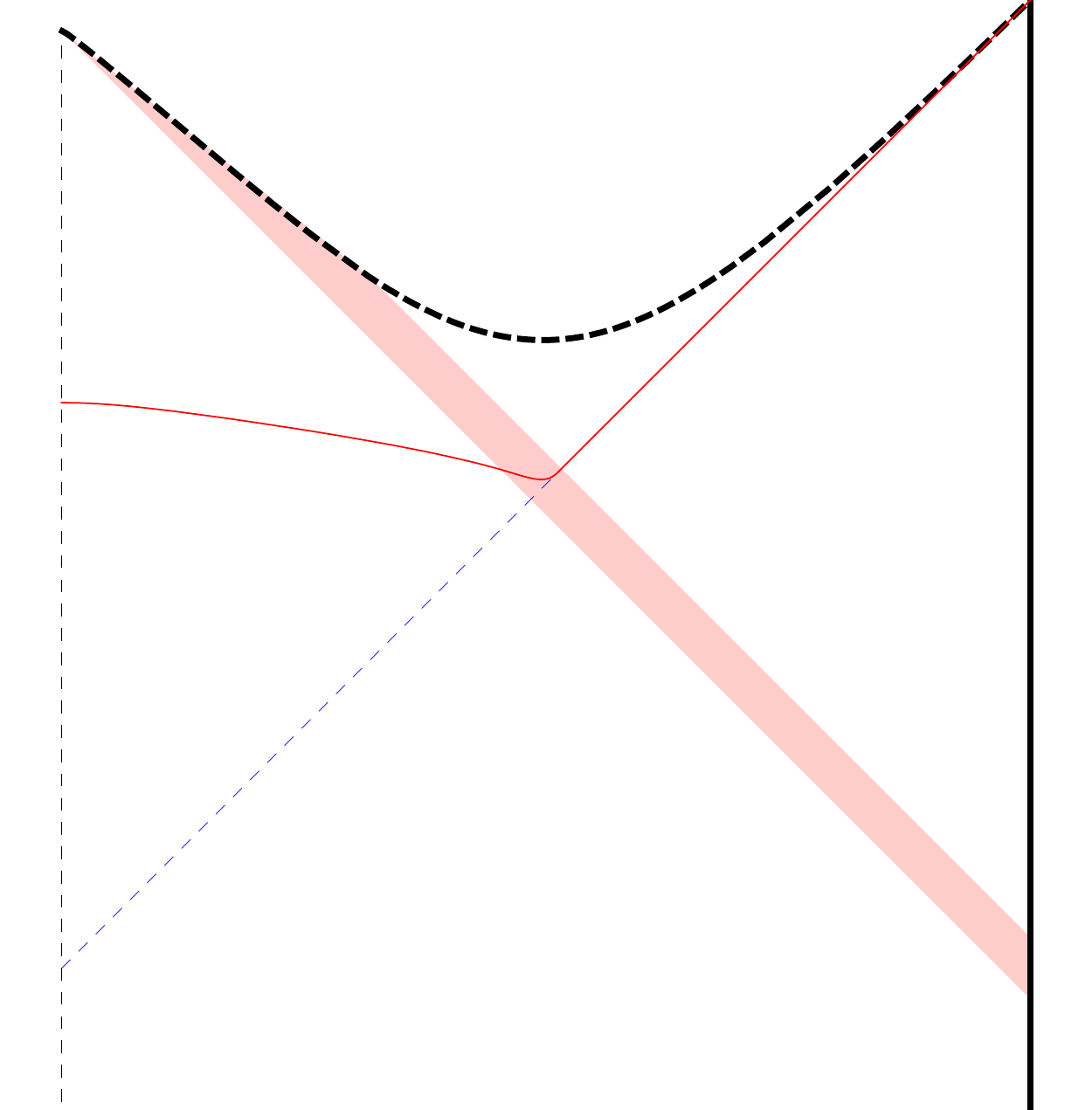}
\caption{
 The region of relevant initial conditions. Surfaces whose earliest point (with respect to the ingong time $v$) lie outside or inside the curve reach the boundary or end in the singularity respectively.
}
\label{f:ICRegion}
\end{center}
\end{figure}

Numerical studies to identify the region of relevant initial conditions show that it is characterised by a very simple critical curve, which can be defined by a function $r_c(v_0)$. This is shown in \fig{f:ICRegion}. Initial conditions $(r_0,v_0)$ give a surface ending on the boundary when $r_0>r_c(v_0)$.
(Note that the actual surfaces themselves can penetrate past $r_c(v_0)$, as we will see momentarily, but their `initial condition point' is restricted by $r_c(v_0)$.) 
After the collapse of the shell, the critical curve coincides with the event horizon, so $r_c$ equals $r_+$. Before the collapse, the curve meets the origin, so $r_c$ vanishes at a particular value of $v_0$, before which time all surfaces will reach the boundary, for any $r_0$. 
This part of the critical curve lies entirely inside the event horizon. In particular, this shows already that codimension-two extremal surfaces anchored on the boundary do reach within the horizon.

\begin{figure}
\begin{center}
\begin{minipage}{.3\textwidth}
  \begin{subfigure}{\linewidth}
    \centering
    \includegraphics[width=.6\linewidth]{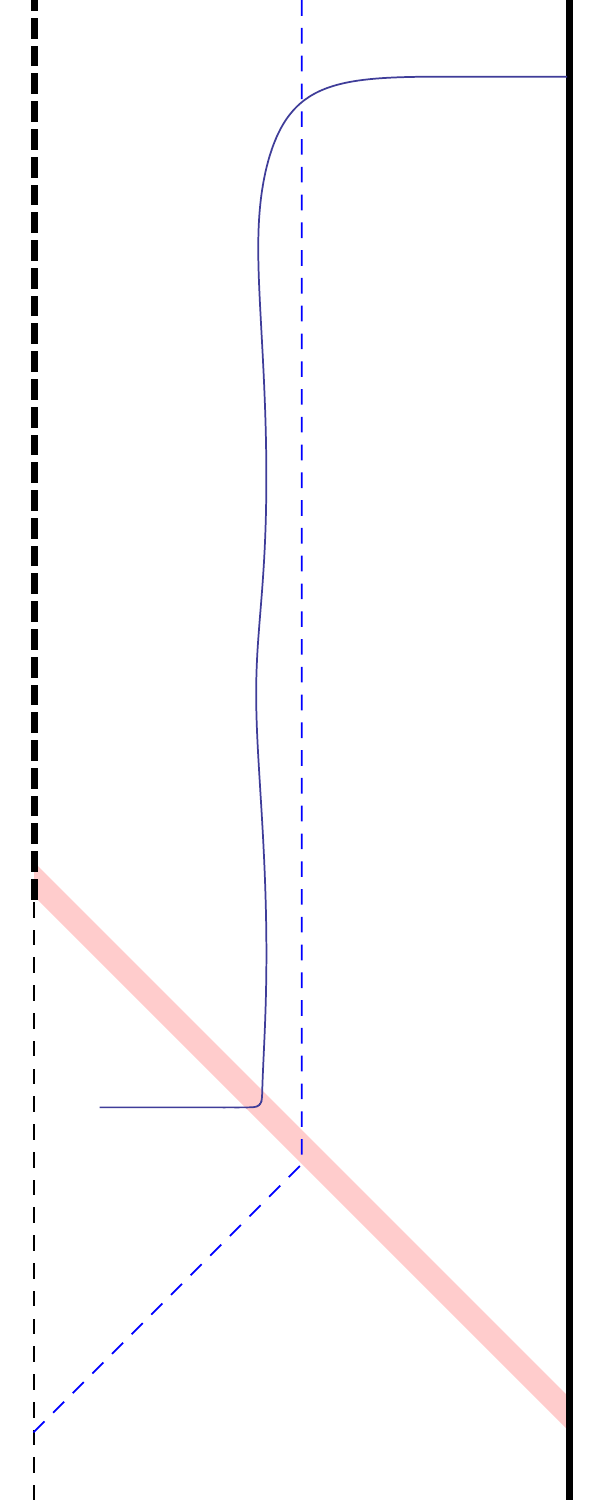}
    \caption{Eddington}
  \end{subfigure}
\end{minipage}%
\begin{minipage}{.4\textwidth}
  \begin{subfigure}{\linewidth}
    \centering
    \includegraphics[width=.7\linewidth]{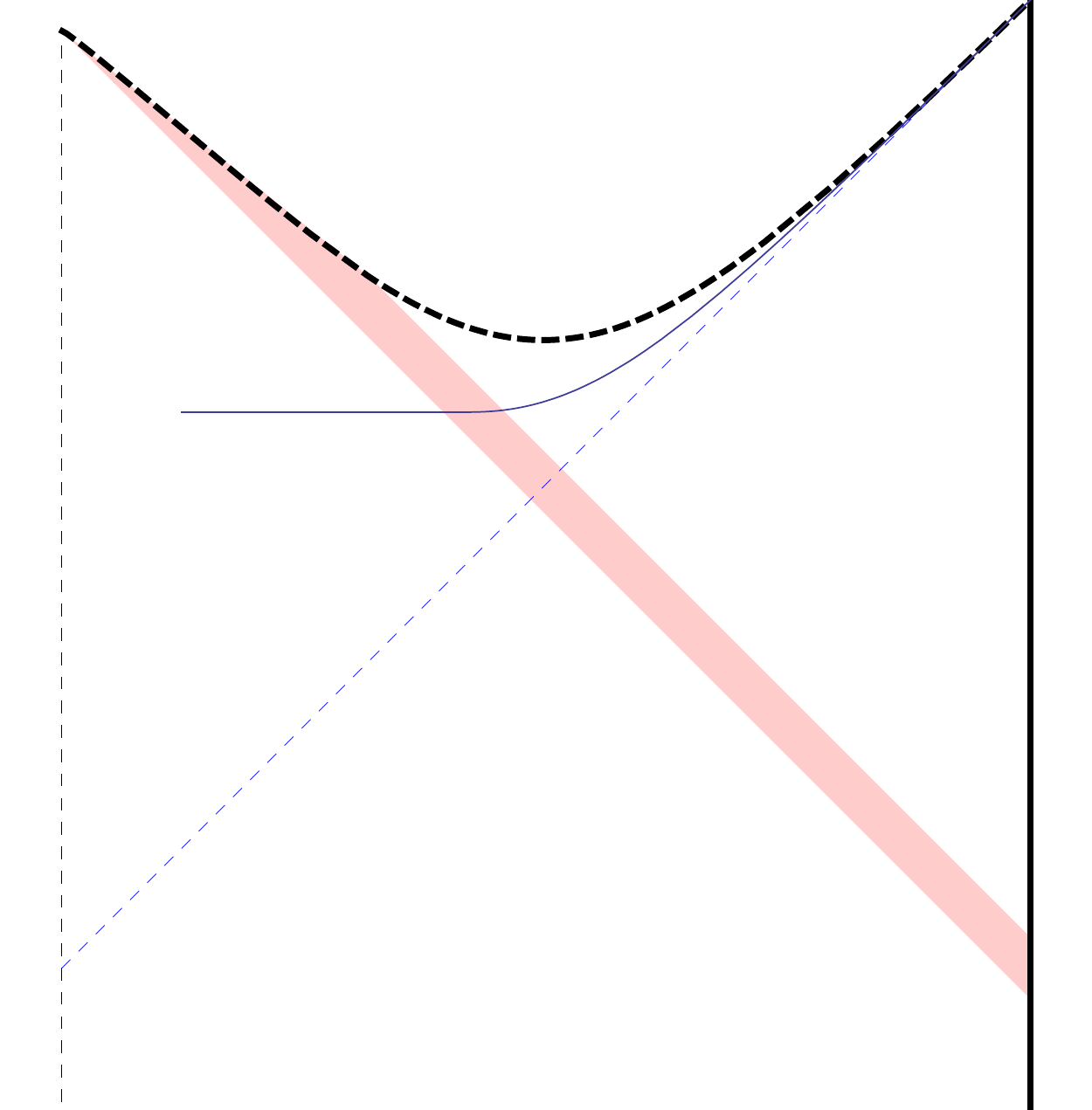}
    \caption{Penrose}
  \end{subfigure}\\[1ex]
  \begin{subfigure}{\linewidth}
    \centering
    \includegraphics[width=.7\linewidth]{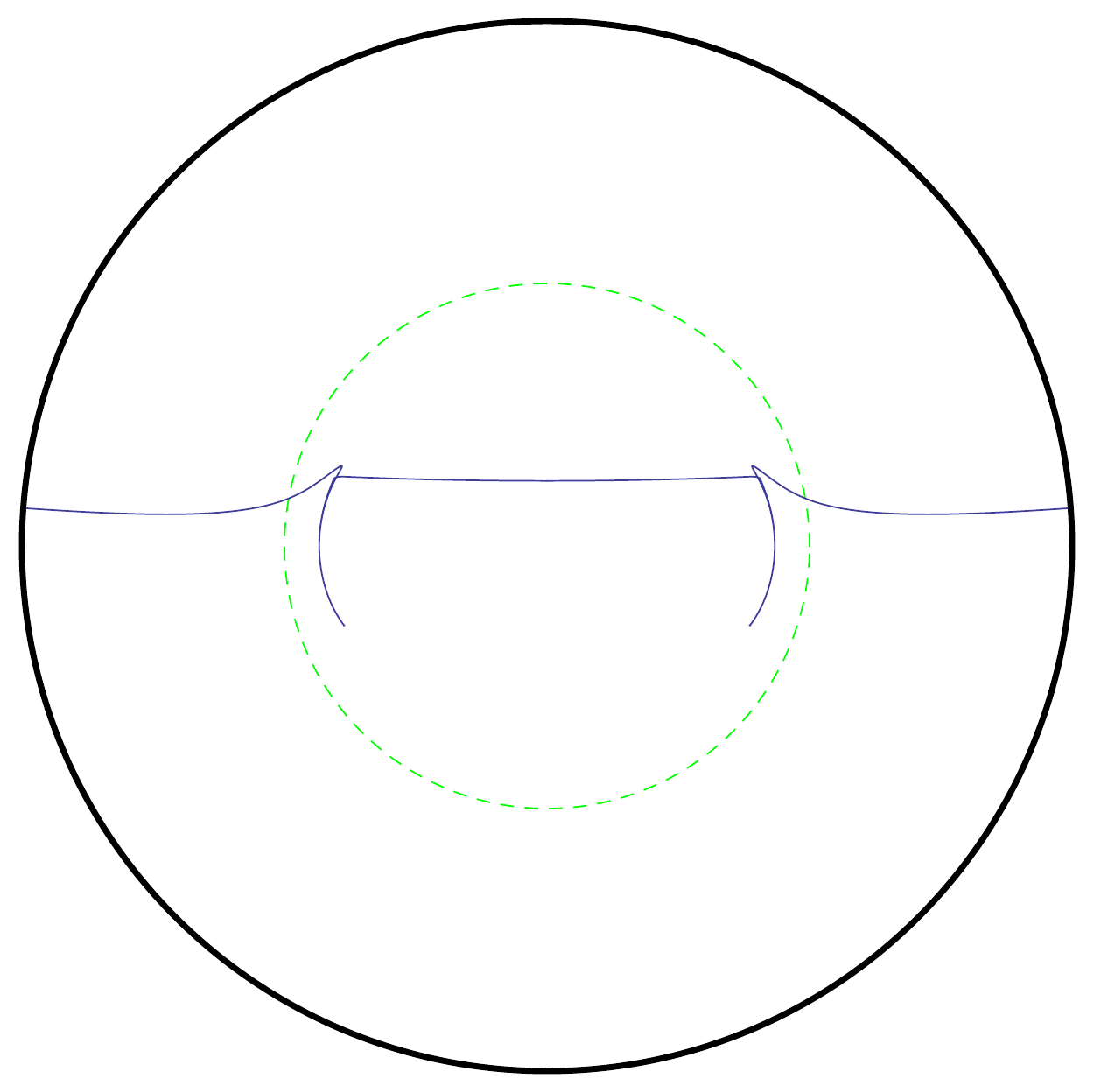}
    \caption{Poincar\'e}
  \end{subfigure}
\end{minipage}%
\begin{minipage}{.3\textwidth}
  \begin{subfigure}{\linewidth}
    \centering
    \includegraphics[width=.9\linewidth]{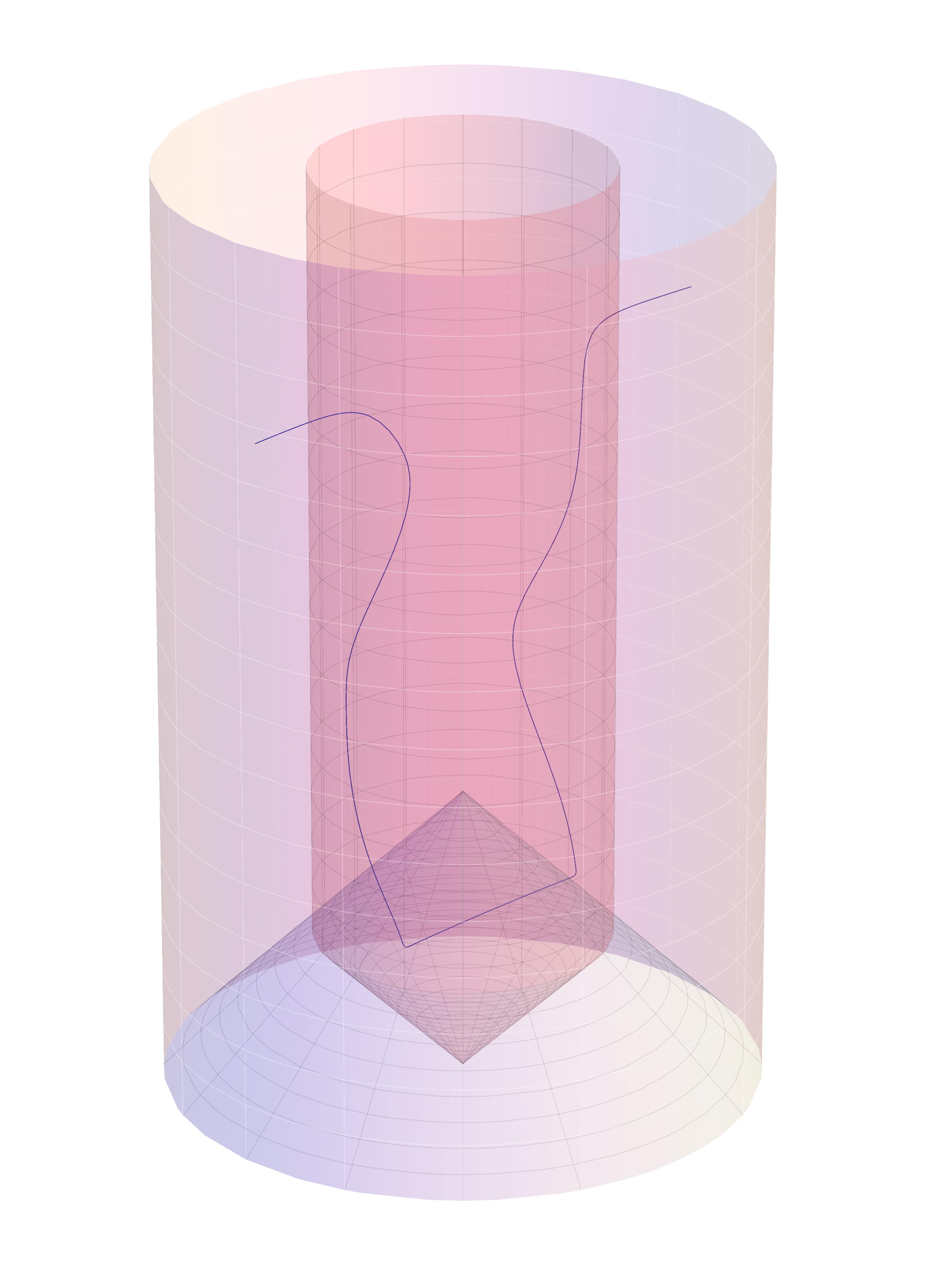}
    \caption{3-d Eddington}
  \end{subfigure}
\end{minipage}%
\caption{
 A surface whose initial conditions are very close to the critical value in the \fig{f:ICRegion}. By tuning this closeness to be exponentially small, the boundary time $t_\infty$ may be made as late as desired.
}
\label{f:esLate}
\end{center}
\end{figure}

In the process of finding this critical surface, we found a particularly interesting class of solutions, a typical example of which is shown in \fig{f:esLate}. If the initial conditions are chosen close to the critical curve, the resulting solution lies inside the event horizon along much of its length, at nearly constant $r$, moving outside the horizon and to the boundary at a late time. By careful tuning of the initial conditions, extremal surfaces meeting the boundary at arbitrarily late times can be constructed. These surfaces link up to the multiply wrapping surfaces in the static black hole geometry; indeed it was these surfaces in the dynamic Vaidya geometry which led to the discovery of the tower of surfaces in the static case. They can be understood from an analytic study, to which we now turn.

Consider first surfaces in the \SAdS~part of the spacetime which lie at constant radius $r$, on the equator $\theta=\pi/2$, extended in the $v$ (or $t$) direction. Inside the event horizon, these are spacelike. Taking $r$ to be small, approaching the singularity, their area (per unit extent in the $v$ direction) reduces to zero by virtue of the shrinking in the spherical directions (for $d>2$). Taking $r$ larger, approaching the event horizon, the area also reduces to zero, this time by virtue of the surface becoming null. Between these two extrema, there must be a maximal area surface\footnote{
The former effect of shrinking the sphere is absent for geodesics, which is essentially why they are different. We would expect surfaces between the two cases discussed here, with dimension larger than one, but smaller than $d-1$, to behave more like the codimension-two surfaces than the geodesics for essentially this reason.}.

Indeed, the equations of motion for extremal surfaces in the static black hole geometry admit an exact solution at constant $r=r_*<\rh$, on the equatorial plane $\theta=\pi/2$. In the globally static spacetime, a perturbation of this surface must either meet the singularity, or form a tube connecting the two asymptotically AdS regions, as in \cite{Hartman:2013qma}.
In constrast, the Vaidya geometry allows for extremal surfaces lying close to this critical radius for much of their extent, but smooth everywhere, and terminating on the boundary, as seen in the numerics.

We look next at the equations of motion, in the \SAdS~geometry, linearised around this constant radius solution. We parameterise by $v$, primes denoting differentiation with respect to $v$. The radius of maximal area $r_*$ satisfies
\begin{equation}
 f'(r_*)+2 \, \frac{d-2}{r_*} \, f(r_*)=0,
\end{equation}
which has a unique solution (when $d>2$). Linearising around this, with $r=r_*+\rho$ and $\theta=\frac{\pi}{2}+\eta$, the equations of motion decouple, and reduce to
\begin{align}
 \rho''-\lambda^2 \, \rho&=0\\
 \eta''+\omega^2 \, \eta&=0
\end{align}
with parameters given by
\begin{align}
 \lambda^2&=\frac{(2d-3)(d-2)f(r_*)^2}{r_*^2}-\frac{1}{2}f(r_*)f''(r_*)\\
 \omega^2&=-\frac{(d-2)f(r_*)}{r_*^2},
\end{align}
which are both positive when $f$ has the form of \SAdS.

A case of particular interest is when the surface meets the boundary at late times. For these, the growing mode for $\rho$ will be tuned close to zero, so $\rho$ will be much smaller than $\eta$, and $\eta^2$ terms become relevant at leading order for $\rho$. We include this forcing for $\rho$ by solving for $\eta$ as $\theta=\pi/2+a\cos(\omega v)$, keeping terms up to order $a^2$, to get
\begin{equation}
 \rho''-\lambda^2\rho=-a^2 (d-1)(d-2)\frac{f(r_*)^2}{r_*} \sin^2(\omega v),
\end{equation}
which has a particular solution
\begin{equation}
 \rho=a^2 (d-1)(d-2)\frac{f(r_*)^2}{2r_*}\left(\frac{1}{\lambda^2}-\frac{\cos(2\omega v)}{\lambda^2+4\omega^2}\right).
\end{equation}
This explains the oscillations in the radial direction visible in \fig{f:esLate}, showing that $r$ is largest when $\theta=\pi/2$, and smaller where the surface is further from the equator. Of particular note is that this particular solution is strictly positive, which constrains the surface to lie outside $r=r_*$, a point to which we return later.

Not every possible solution in \SAdS~will give an extremal surface when continued into the full spacetime, since generically this continuation will not be smooth at the pole of the sphere: indeed, in \SAdS, there are four parameters describing the solutions, but there is only a two parameter family smooth solutions in Vaidya. Consider starting at an initial point inside the shell, with the two initial condition parameters $(r_0,v_0)$, and integrating until reaching the outside of the shell. After this, our surface is well described by our analytic solution. In this way, the initial conditions map to values for the four constants of integration: the amplitude $a$ and phase of the angular oscillations, and the growing and shrinking modes of $\rho$. Further, this map should be smooth. For surfaces for which the linearisation is a good approximation, this observation alone will tell us much.

The fate of an extremal surface is characterised, in this approximation, by the sign of the coefficient of the growing mode $e^{\lambda v}$ of $\rho$, which we denote by $g$. It will escape to the boundary if $g$ is positive, or end in the singularity if $g$ is negative. The limiting case, when $g=0$, will give the critical curve in the $(v_0,r_0)$ initial conditions plane separating surfaces ending on the boundary from those ending in the singularity.

\paragraph{Boundary region from initial conditions}

Having characterised the domain of relevant initial conditions, we would now like to understand how they correspond to a region on the boundary. Initially, the analytic study will take us a long way towards this goal.

When $g>0$, it will roughly tell us the time at which the surface meets the boundary, since we expect $g\approx e^{-\lambda v_\infty}$. Since $g$ changes smoothly with the initial conditions, barring any coincidences, it should go to zero linearly as the critical curve is approached, along a line of constant $v_0$, say. This implies that $t_\infty$ should diverge logarithmically near the edge of initial condition parameter space:
\begin{equation}
t_\infty \sim -\frac{1}{\lambda}\log (r_0-r_c(v_0))
\end{equation}
In particular, the surface can reach the boundary at arbitrarily late time, by tuning initial conditions exponentially accurately.

This also corresponds nicely to what happens for inital conditions near the horizon in the static black hole part of the geometry. Surfaces fully in this region lie on a constant-time slices with respect to the static coordinate $t_o$, and, approaching the future horizon, $t_o$ blows up logarithmically.

Close to the critical curve, changes in the other parameters will be unimportant relative to the effect of $g$ going to zero. Fixing $v_0$, and considering $r_0$ very close to $r_c$, the shape of the surface should be almost unchanged, except for the time at which the growing mode of $\rho$ kicks in and drives the solution out to the boundary. Hence, we expect the value of $\theta_\infty$ to be determined largely by the phase of $\eta$ when the growth of $\rho$ begins.

Since $\eta$ undergoes oscillations in $v$, this means that as the critical curve is approached, the blow-up of $v_\infty$ is accompanied by increasingly rapid oscillations in the value of $\theta_\infty$. The angular frequency in $t_\infty$ will be $\omega$, so we expect
\begin{equation}
\theta_\infty \sim a\cos\left(\frac{\omega}{\lambda}\log (r_0-r_c(v_0))+\text{phase}\right)
\end{equation}
with the phase depending on the choice of $v_0$ along which the critical curve is approached. The range of $\theta_\infty$ covered is determined by the amplitude $a$ of the $\eta$ oscillations, which depends most strongly on $r_0$. It must vanish, by the enhanced symmetry, when $r_0=0$, and should increase with increasing $r_0$, as the surfaces depart further from the equatorial plane $\theta=\pi/2$.

Again, this matches the behaviour seen in the static \SAdS~spacetime, where there are similar oscillations as the initial conditions approach the event horizon.

The numerical results show all these features. In particular, the details close to the edge of the initial condition space match the expectation from the analytic calculations, including the rate of blow-up of $t_\infty$, and the period of oscillations of $\theta_\infty$.

\begin{figure}
\begin{center}
\includegraphics[width=.4\textwidth]{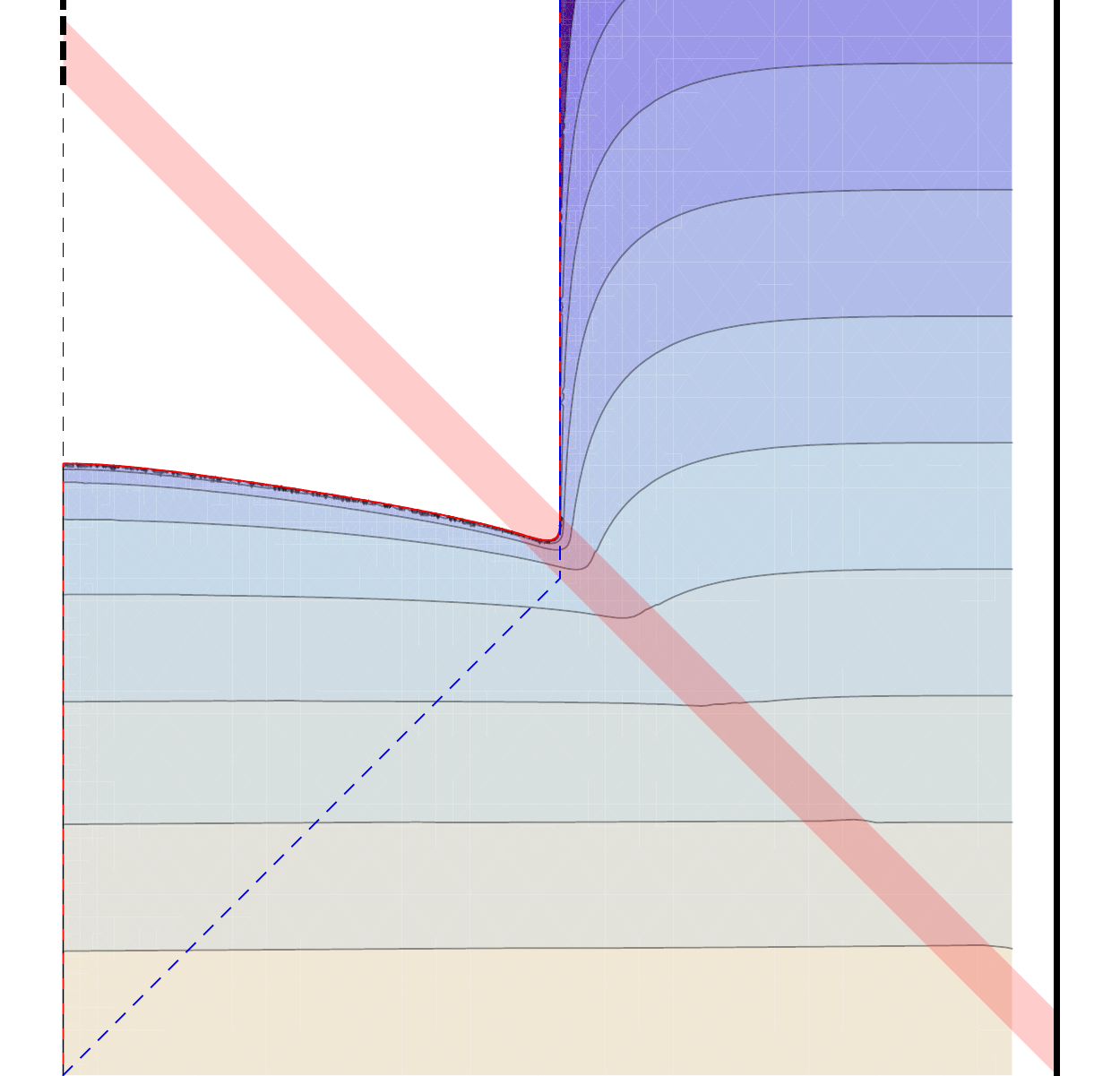}
\includegraphics[width=.4\textwidth]{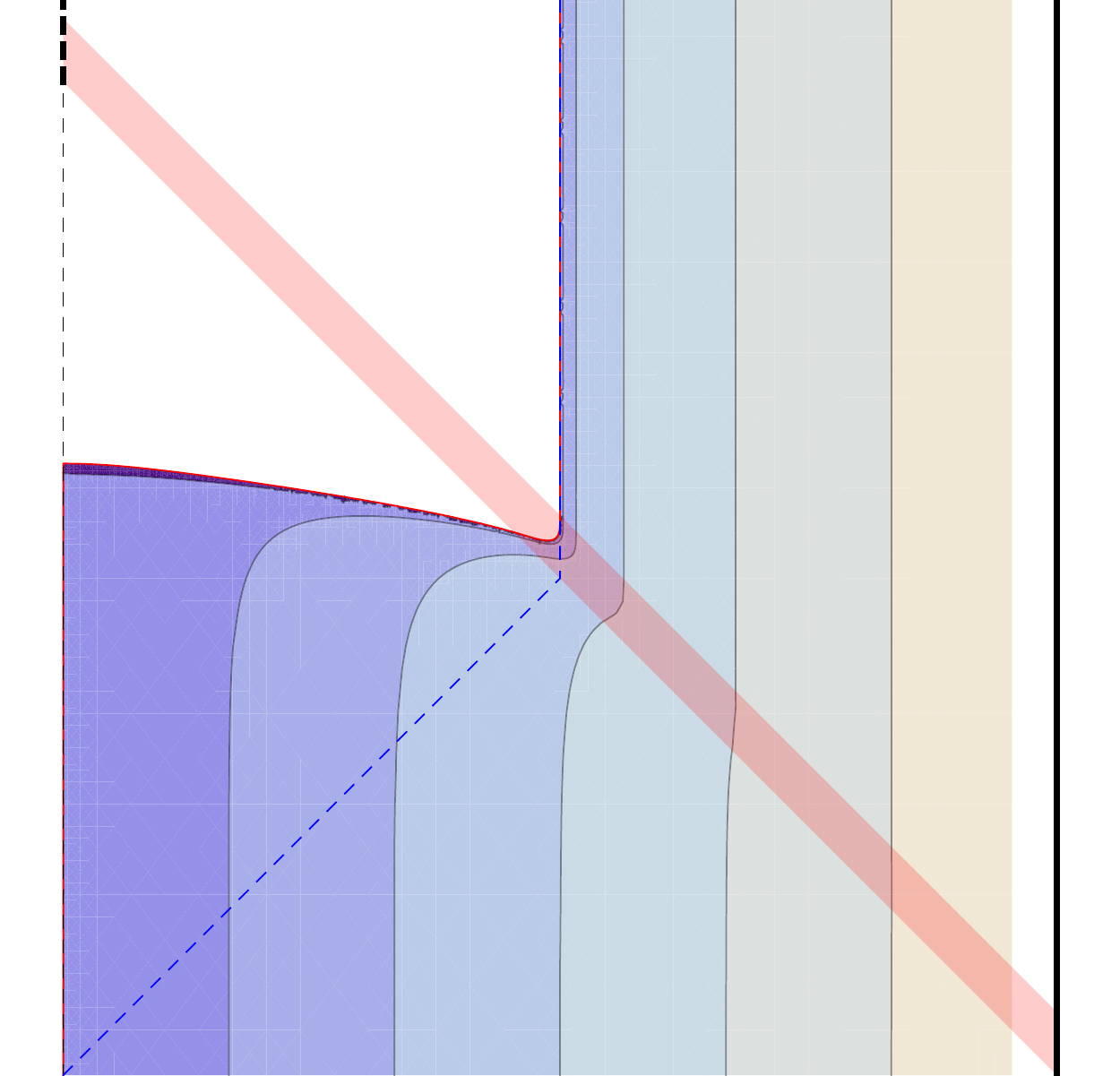}
\\
\includegraphics[width=.4\textwidth]{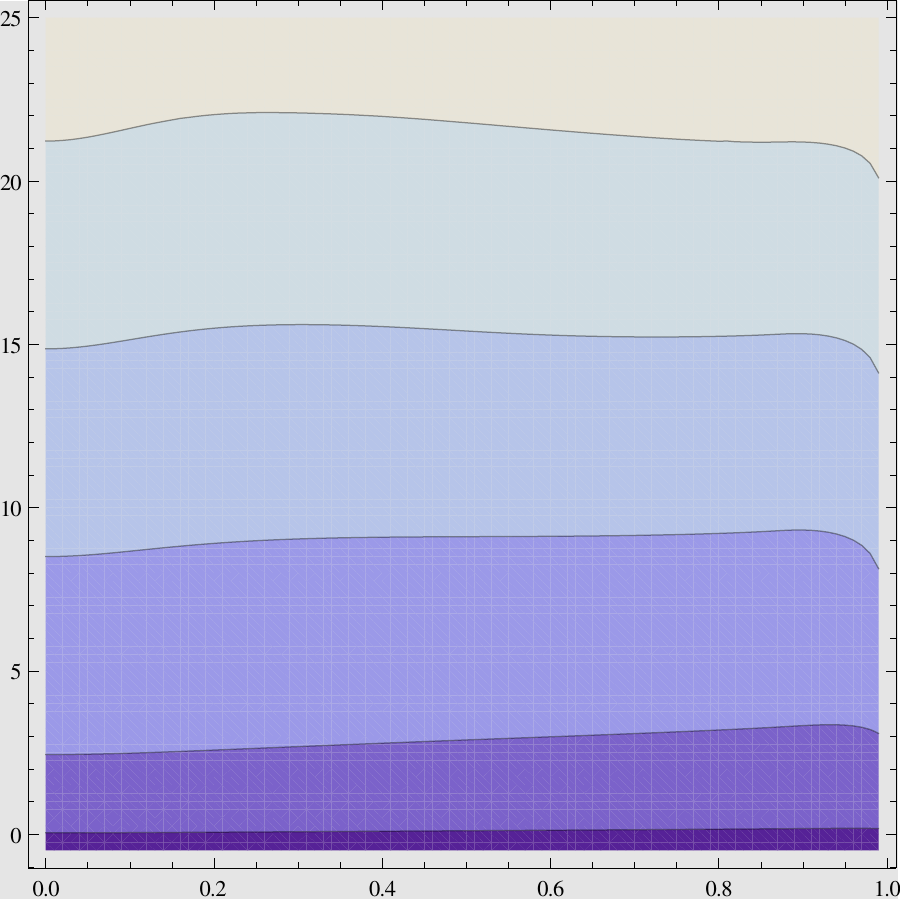}
\includegraphics[width=.4\textwidth]{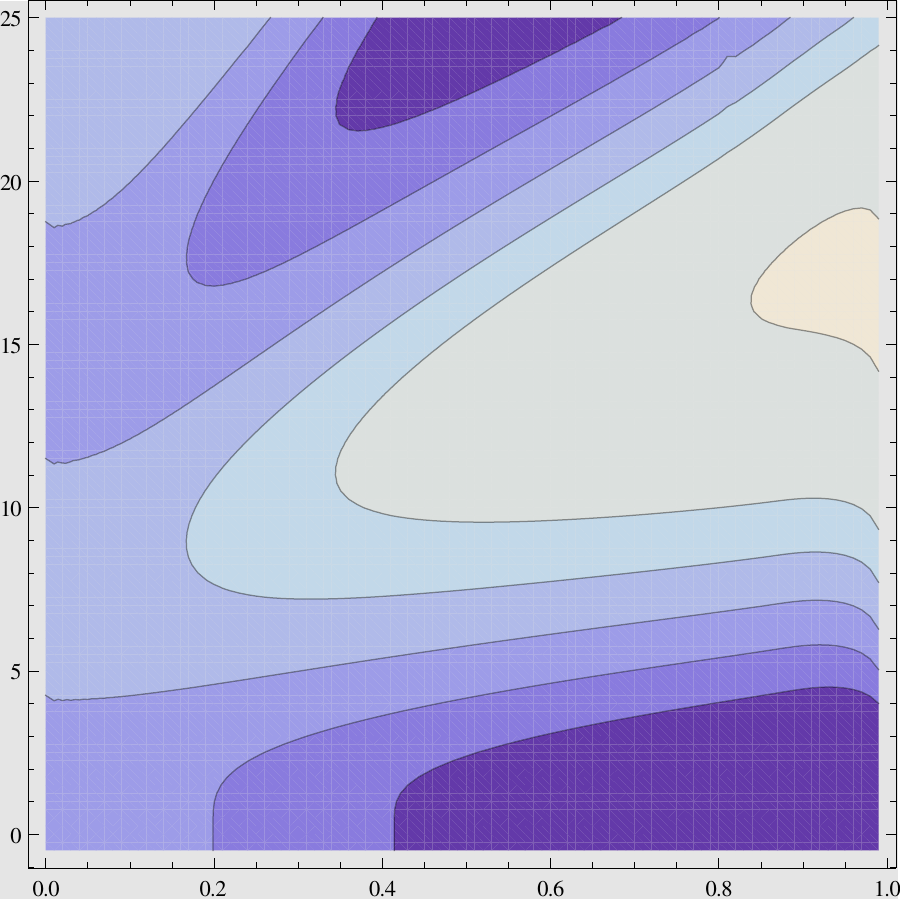}
\caption{
  Contour plots for $t_\infty$ (left) and $\theta_\infty$ (right) as a function of initial conditions $(v_0, r_0)$. The red curve is the edge of the relevant initial conditions (as in \fig{f:ICRegion}). The values of the contours in the top right figure are at multiples of $\pi/12$; part of the outermost contour for $\theta_\infty=\pi/2$ is just visible. The bottom figures show the detail close to the edge of the region of relevant initial conditions. The horizontal coordinate is $r_0$, and the vertical coordinate $-\log(v_c(r_0)-v_0)$, where $v_c(r_0)$ gives the latest relevant initial $v_0$ for given $r_0$.
}
\label{f:Contours}
\end{center}
\end{figure}

Contour plots, representing how the boundary region associated to a given surface corresponds to its initial point where it crosses the pole of the $S^3$, are shown in \fig{f:Contours}.

The contours of constant $\theta_\infty$ are particularly interesting, as they correspond to a family of curves associated with entanglement entropy for a specific region of space in the field theory. We see the first few curves of what we expect to be an infinite collection for a given region (as long as that region is not too small). These continue into the black hole part of the geometry, where they correspond to the tower of surfaces of \cite{Hubeny:2013gta}.

\begin{figure}
\begin{center}
\includegraphics[width=5in]{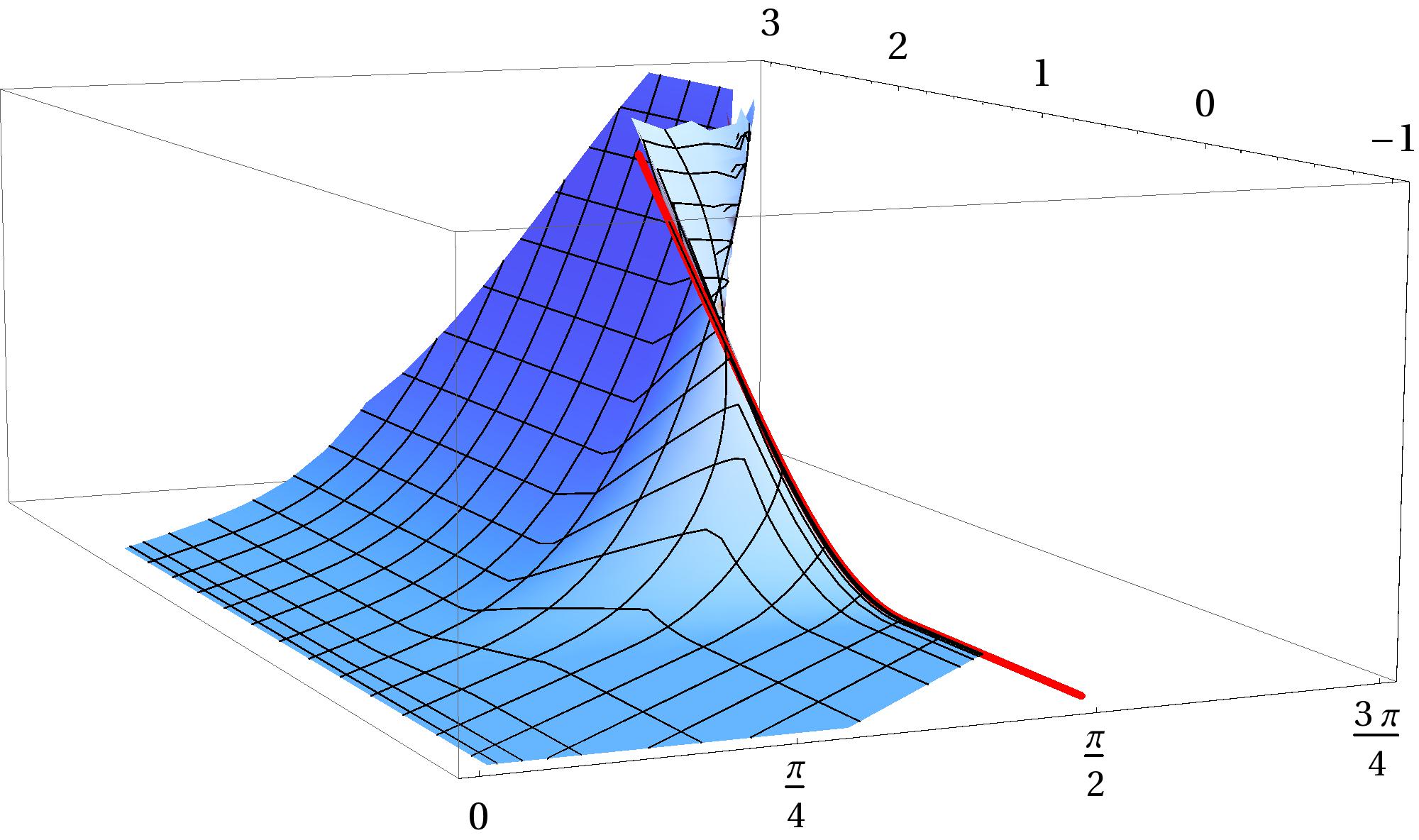}
\begin{picture}(0,0)
\setlength{\unitlength}{1cm}
\put (-4.6,0.2) {$\theta_\infty$}
\put (-3,7) {$t_\infty$}
\put (0,3.5) {$A$}
\end{picture}
\caption{
Part of the surface in boundary parameter space $(\theta_\infty,t_\infty,A)$ induced by the extremal surfaces. The red curve shows the edge of the surface, which corresponds to surfaces lying on the equator $\theta=\pi/2$, and passing through the origin with $r_0=0,v_0<0$.
}
\label{f:surfaceBoundaryData}
\end{center}
\end{figure}

In terms of the surface in boundary parameter space $(\theta_\infty,t_\infty,A)$ we have enough to build a qualitative picture of what goes on. The surface will have an edge corresponding to the equatorial surfaces lying entirely on $\theta=\pi/2$, which is the image of the initial conditions $r_0=0, v_0<0$. The other boundary of initial condition space (the curve of \fig{f:ICRegion}) maps to $t_\infty\to\infty$, so the only other edge of the surface is at $\theta_\infty=0, A=0$, when the initial conditions approach the AdS boundary. The surface thus looks like a strip, which for $v<0$ is just the flat plane $0<\theta<\pi/2, A=0$, and thereafter progressively folds over itself to link up with the tower of \fig{f:schwAreas} one branch at a time. The beginning of the first such folding is shown in \fig{f:surfaceBoundaryData}.

\paragraph{Spacetime region covered}

\begin{figure}
\begin{center}
\includegraphics[width=.4\textwidth]{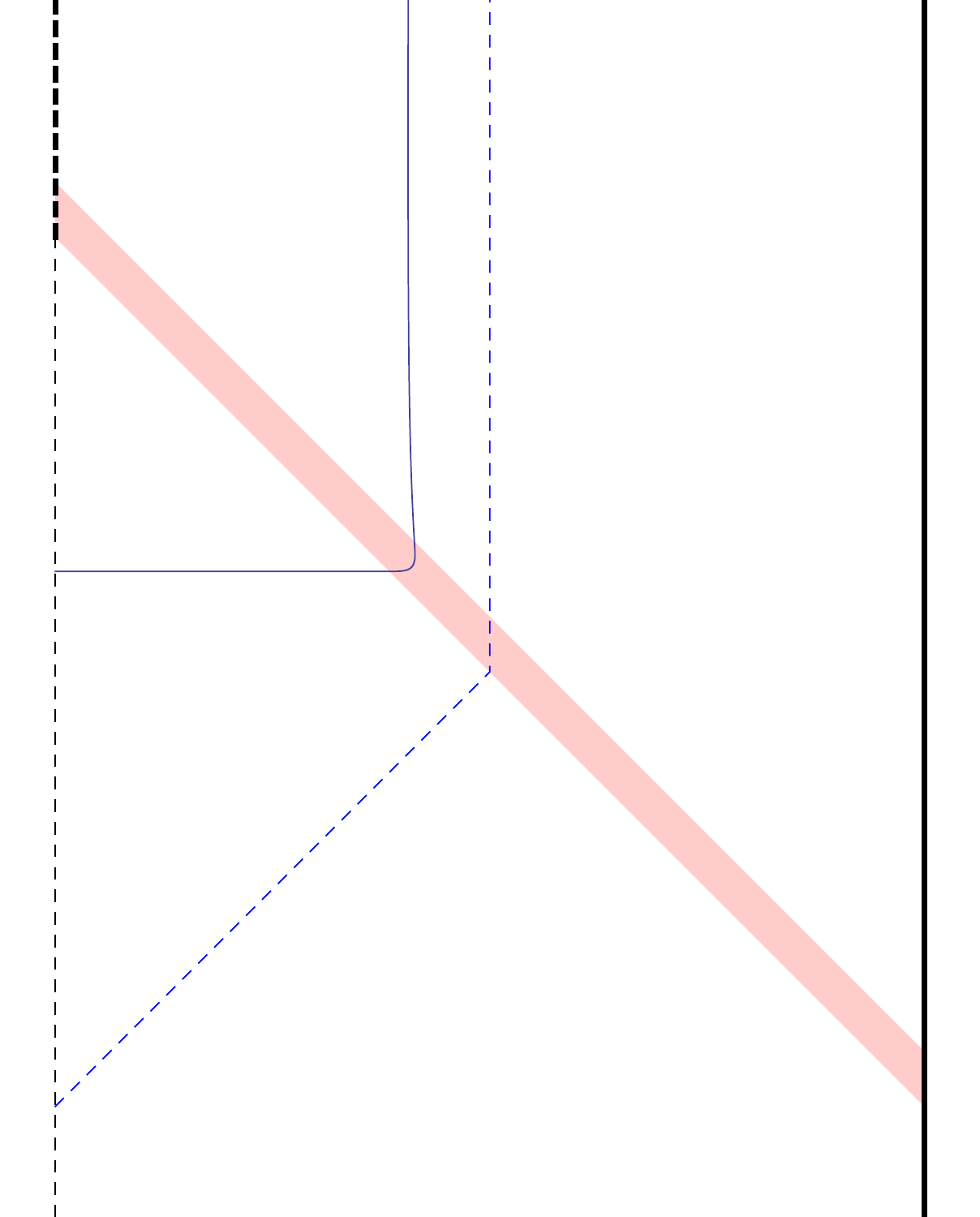}
\hspace{1cm}
\includegraphics[width=.4\textwidth]{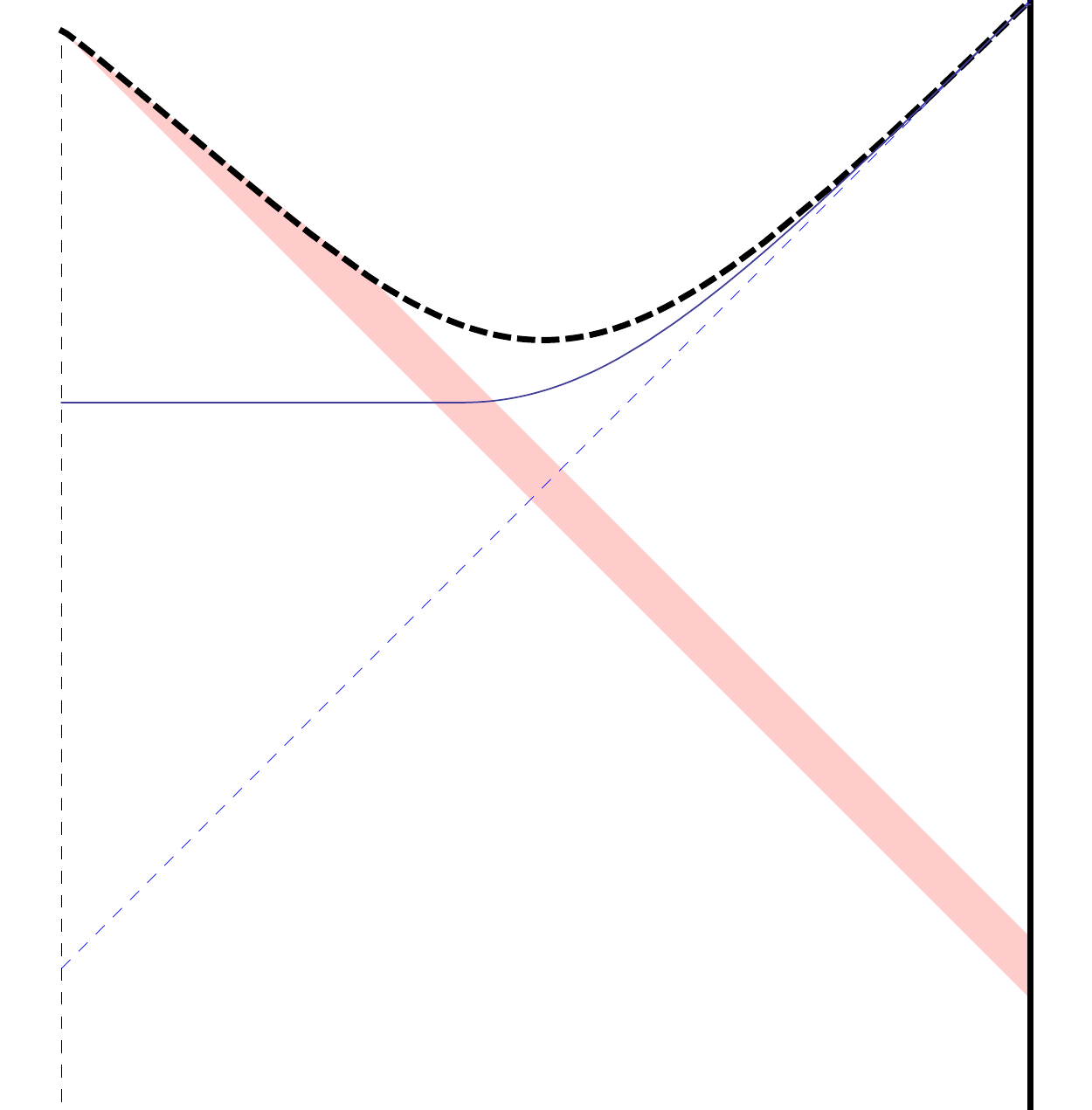}
\caption{
  The region covered by all extremal surfaces reaching the boundary. The edge is the same as the limiting surface between surfaces ending on the boundary or in the singularity, lying on the equatorial plane, and at constant $r=r_*$ after the collapse.
}
\label{f:ESRegion}
\end{center}
\end{figure}

Of particular interest is the region in spacetime covered by the probe extremal surfaces. As long as there are no unexpected departures from the approximation scheme, the analytic study gives us a complete characterisation of this. After the collapse of the shell, extremal surfaces reach precisely the region of the bulk outside the radius $r=r_*$ (perhaps excepting `cutting the corner' very near the shell, allowing for a negative decaying mode of $\rho$). Given this, it is not unreasonable to expect that the covered region of spacetime is bounded by the surface on the equator $\theta=\pi/2$ (so $r_0=0$), with $v_0$ chosen such that $r\to r_*$ as $v\to\infty$. This is the critical value between surfaces ending on the boundary or in the singularity. This expectation is borne out by the numerics, and the region covered is shown in \fig{f:ESRegion}.

\paragraph{Surfaces of minimal area, and entanglement entropy}

We now turn to the measurement of the field theory quantity of interest, namely the area of the surfaces. This is interesting for at least two reasons. Firstly, we can learn about the thermalization of entanglement entropy for a field theory on a sphere. Secondly, we would like to compare areas of surfaces anchored to the same boundary region, because the extremal surface of least area is most directly associated with the field theory observable. This will allow us to refine our description of the spacetime region covered by surfaces, to include only those of minimal area for a given boundary region.

The simplest possibility is that the relevant surfaces of minimal area are those arising from a continuous deformation between the static parts of the geometry. In terms of the initial conditions, this will correspond to the outside of the outermost $\theta_\infty=\pi/2$ contour (see \fig{f:Contours}). In principle, this could be spoilt by the higher branches of surface: for example, one could imagine a case where the folding portion of the surface in boundary parameter space (\fig{f:surfaceBoundaryData}), where $\theta_\infty>\pi/2$, dips below the corresponding piece under the equivalence $\theta_\infty\sim\pi-\theta_\infty$. If this were to occur, it would allow for some novel behaviour of entanglement entropy, such as kinks, discontinuities, and non-monotonicity as a function of time, where two branches of surfaces exchange dominance or new branches appear. This turns out not to be realised in the case at hand, though we know of nothing which would prevent it; it would not be entirely dissimilar to what we have found in the case of geodesics. It would be interesting to see if these possibilities can be excluded, or instead found to be present in an altered geometry.

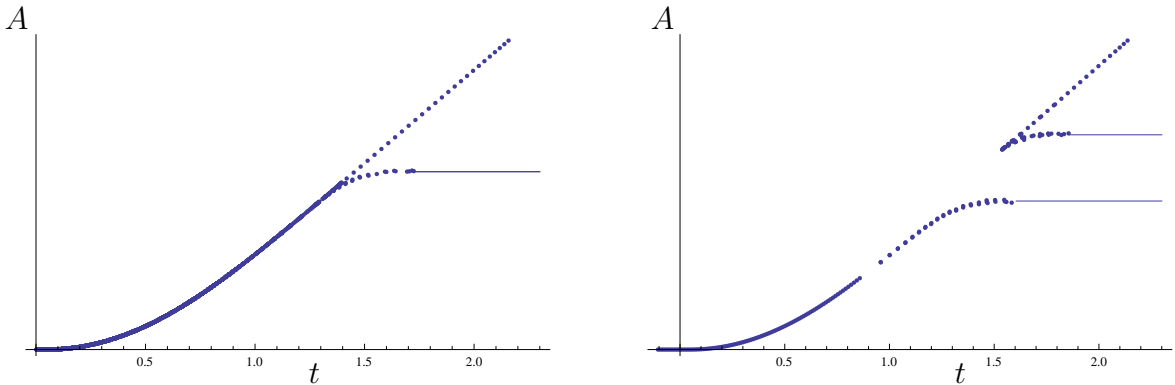
\begin{figure}
\begin{center}
\includegraphics[width=.4\textwidth]{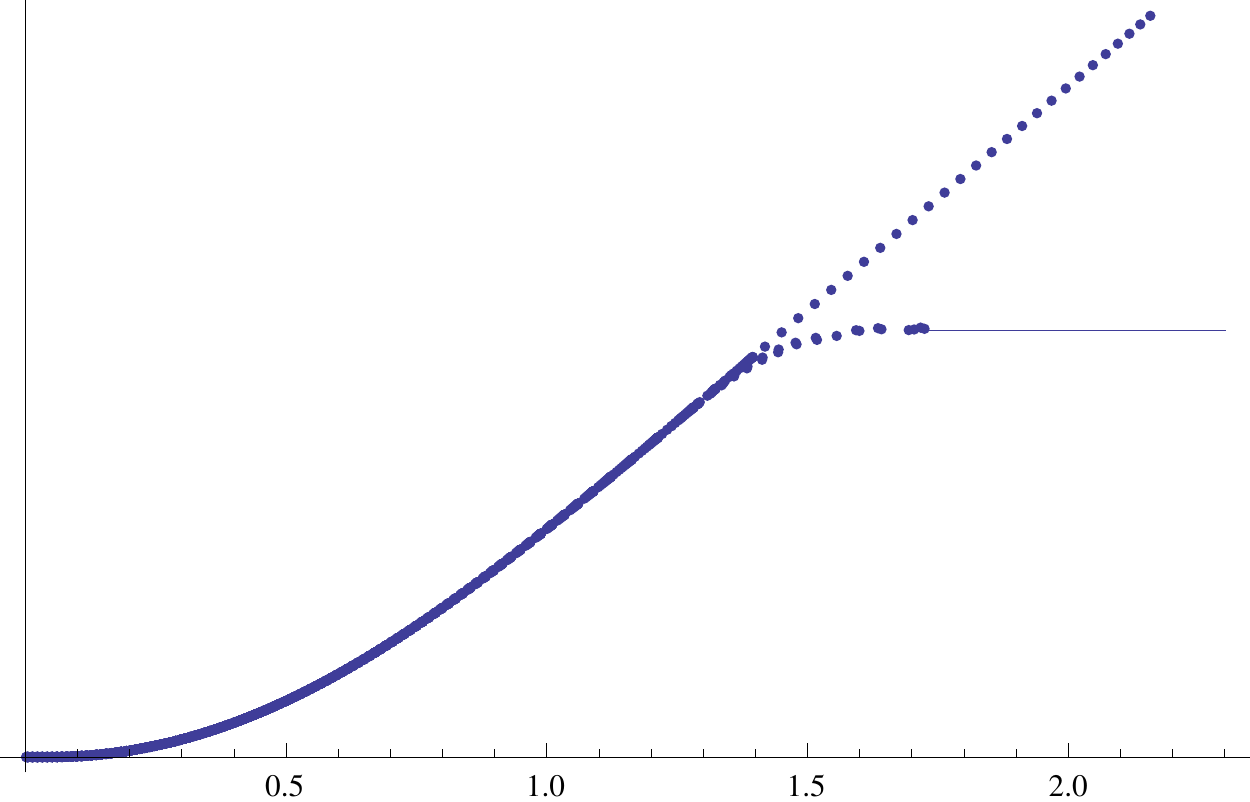}
\hspace{1cm}
\includegraphics[width=.4\textwidth]{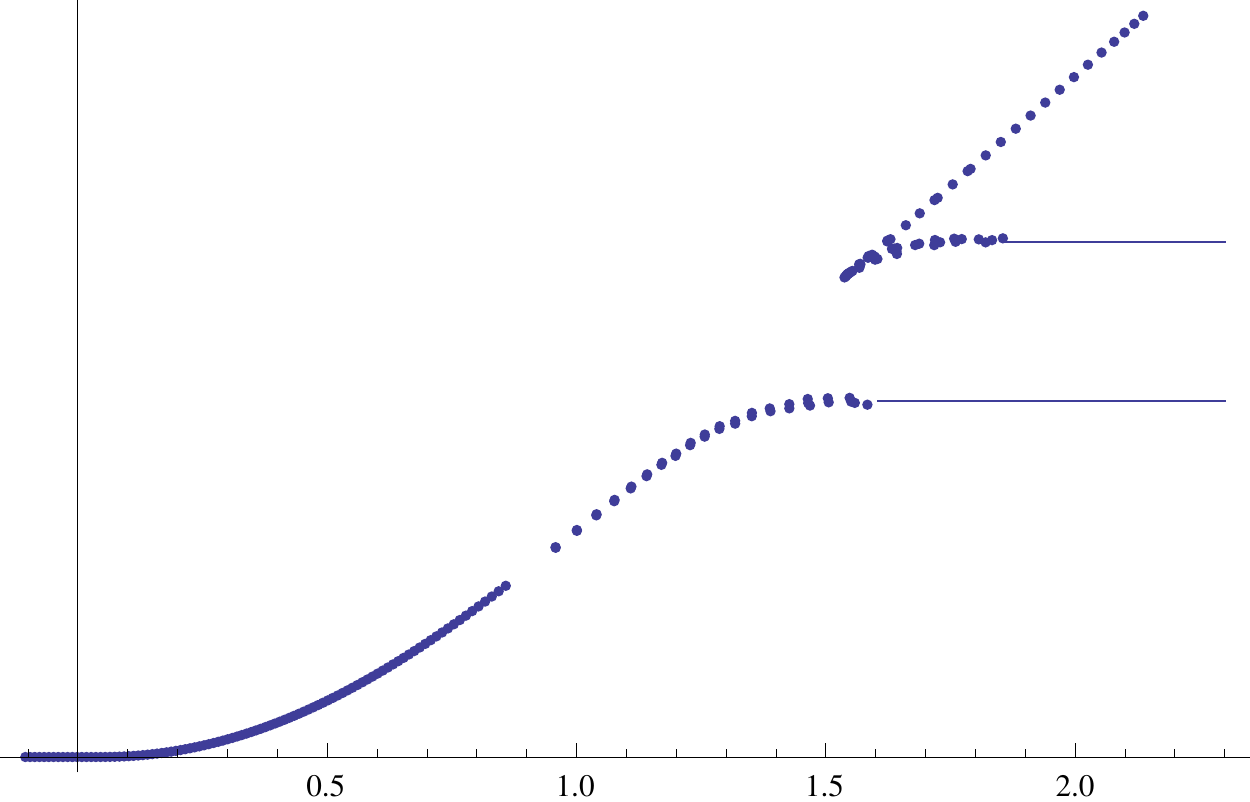}
\begin{picture}(0,0)
\setlength{\unitlength}{1cm}
\put (-3,-0.2) {$t$}
\put (-11.5,-0.2) {$t$}
\put (-7,4.5) {$A$}
\put (-15.5,4.5) {$A$}
\end{picture}
\caption{
  Thermalization of entanglement entropy for a hemisphere (left), and a slightly smaller region (right). Data points for extra branches of surfaces are shown, and can be seen to have larger areas.
}
\label{f:EEthermalization}
\end{center}
\end{figure}

With the absence of these complications, the thermalization of entanglement entropy offers nothing new, as shown in \fig{f:EEthermalization}, smoothly and monotonically increasing from the vacuum to the thermal value.
This matches well with the findings of \cite{Liu:2013qca,Liu:2013iza}, who undertake a similar study in the planar case.  Here we might a-priori have expected the physics to be much richer, but that expectation has not been realised.
In particular, there is an intermediate regime where the area grows linearly, controlled by the surface extending along the critical radius $r=r_*$.  Additionally, if the is shell sufficiently thin, at early times there is a quadratic growth with known coefficient, proportional to the area of the bounding region (in our case an $S^{d-2}$) and the energy density, since the calculation of \cite{Liu:2013iza} goes through unchanged. Unsurprisingly, this is modified to a slower growth if the collapse is more gradual.
We found no clear robust law found governing the final approach to equilibrium, excepting that it appears to be smooth, though a more thorough investigation of this would be worthwhile.

\begin{figure}
\begin{center}
\includegraphics[width=.4\textwidth]{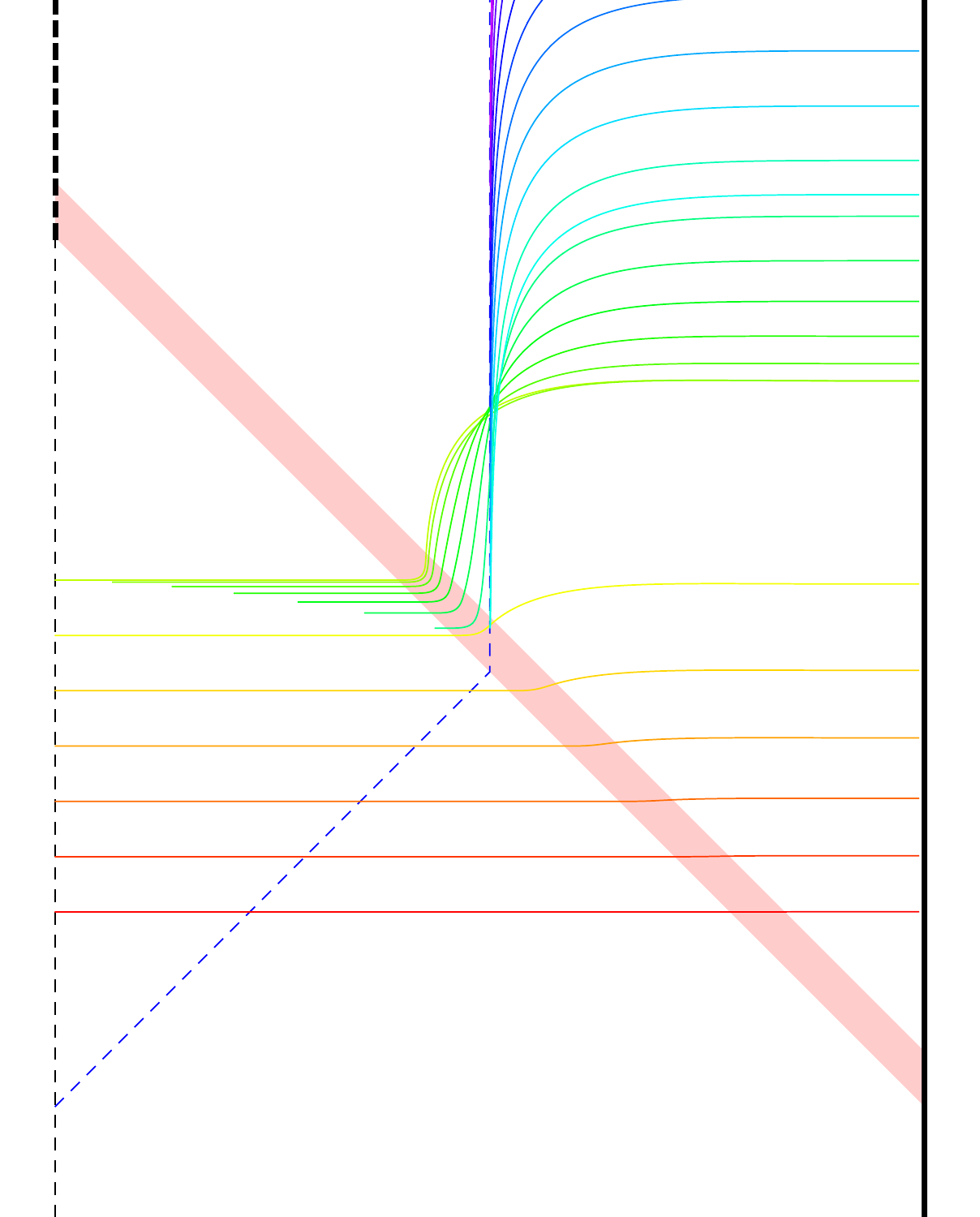}
\hspace{1cm}
\includegraphics[width=.4\textwidth]{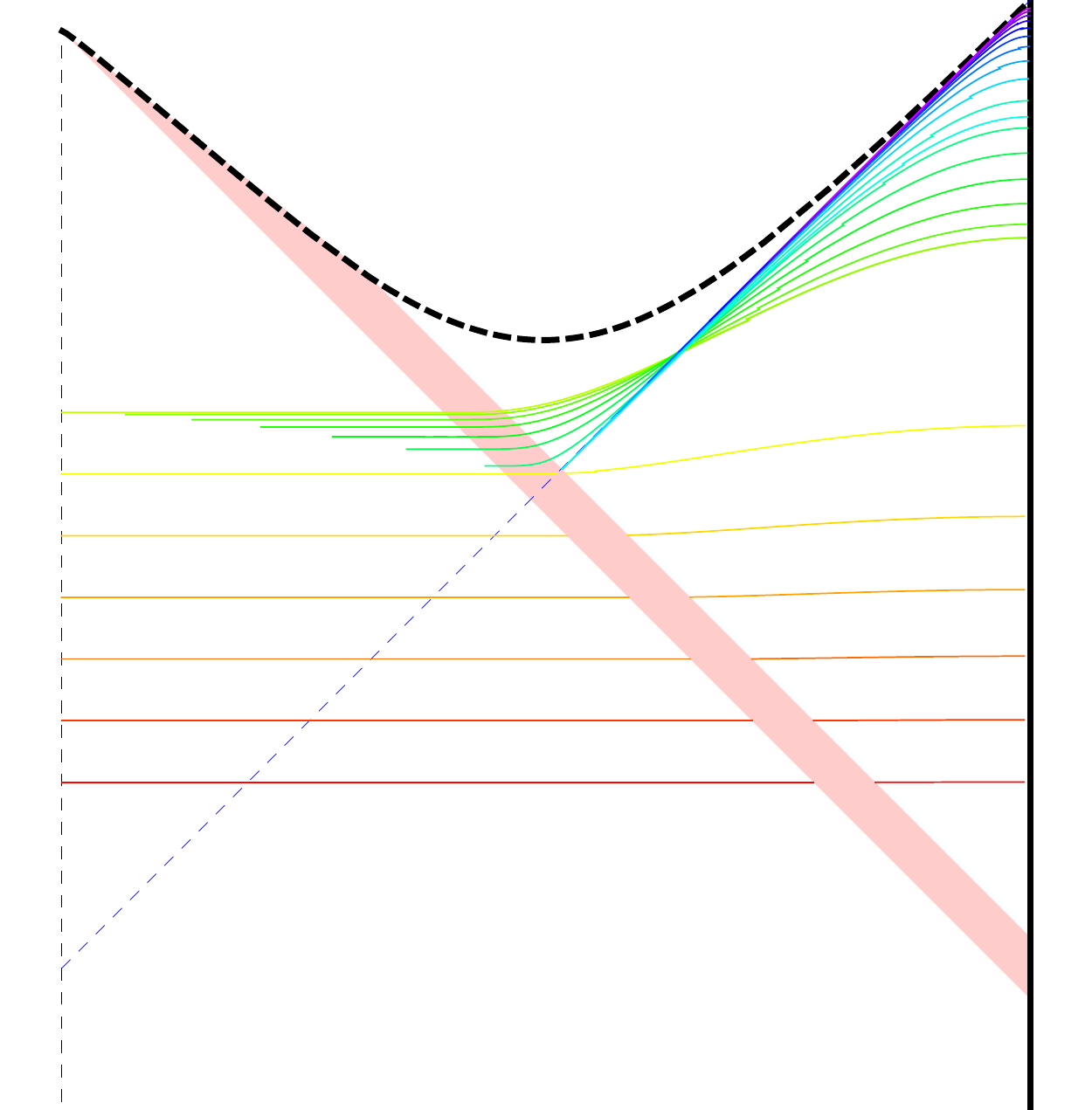}
\caption{
  A selection of the surfaces corresponding to entanglement between hemispheres. The innermost of these characterises the region of the bulk probed by these observables.
}
\label{f:minSurfaceRegion}
\end{center}
\end{figure}

The part of spacetime covered by the extremal surfaces of minimal area is accurately characterised as the outside of either of two regions. The first is given by the deepest point of the minimal surface giving the entanglement between hemispheres in \SAdS, which is a value of $r$ strictly larger than $\rh$, important after the collapse. The second is the latest surface giving entanglement between hemispheres which passes through the origin. The initial conditions of this are given by the meeting of the outermost $\theta=\pi/2$ contour of \fig{f:Contours} with the origin. This surface samples the inside of the horizon, including for a significant time outside the shell, but is bounded well away from the singularity and hence is protected from regions of strong curvature. This is illustrated in \fig{f:minSurfaceRegion}.

%
\section{Discussion}
\label{s:disc}

We have explored the behaviour of spacelike boundary-anchored geodesics and codimension-two
extremal surfaces in a spherically symmetric Vaidya-AdS bulk spacetime. 
The main goal was to assess how deep past the horizon can such surfaces reach in this simple model
of a collapsing black hole.  This was motivated in part by the observation of \cite{Hubeny:2012ry}, that unlike for a static spacetime wherein such boundary-anchored extremal surfaces cannot penetrate a black hole,\footnote{
In fact, there is an interesting subtlety in this argument, partly analogous to the geodesic behaviour we exploited in the present context: \cite{Hubeny:2012ry} used conservation of `energy'  along a geodesic to argue that in static spacetimes,   once a geodesic enters the future event horizon, it cannot exit back through the future horizon.  This argument didn't prevent the geodesic from reaching the same boundary via the past horizon, but \cite{Hubeny:2012ry} argued that in order to do that, it would have to turn around in the `right' exterior region, which, {\it assuming it has the same radial profile of the metric as the left region}, would contradict the assumption that the geodesic reached from the left boundary to the horizon.  One may, however, consider a more contrived spacetime with a static shell on the right side of \SAdS\ beyond which the right boundary is replaced by e.g.\ de Sitter region with smooth origin  as in \cite{Freivogel:2005qh}.  In such a situation, as discussed in that work, there certainly do exist boundary-anchored spacelike geodesics passing through the black hole.
} in dynamically evolving spacetime, the event horizon, being globally defined, does not pose a fundamental obstruction.
We realized this expectation both with geodesics as well as with codimension-two extremal surfaces, and in all cases there are even surfaces passing inside the horizon at arbitrarily late times. Having said this, the actual bulk regions probed are qualitatively rather different. 

Geodesics with endpoints on the boundary at equal times are able to reach arbitrarily close to the singularity when it forms, accessing regions of arbitrarily strong curvature. The same is not true of codimension-two extremal surfaces, which remain bounded away from the singularity. However, these access a much larger region later on, being constrained only to lie outside the surface of maximal area at constant radius, while the geodesics probe a region shrinking to the horizon at later times. 

The exception to this is in the special case of 3-dimensional bulk, where the two classes coincide, and the behaviour depends crucially on the final size of the black hole. For black holes no larger than the critical radius $\rh=1$ in AdS units, even the geodesics with endpoints at equal times cover the whole spacetime. For black holes larger than this, there is a region inaccessible to all geodesics with both endpoints on the boundary, even without the restriction to equal-time endpoints, around the singularity when it first forms. At late times, the geodesics still can approach arbitrarily close to the singularity.

In all cases, restricting considerations to the surfaces expected to dominate the CFT variable, of smallest area or shortest length, puts constraints on the probed region. In particular, none of these surfaces reach inside the horizon for an extended time after the shell passes.

Unsurprisingly, the surfaces reaching the deepest are consistently those corresponding to the largest possible length scales, namely geodesics connecting antipodal points, and surfaces corresponding to entanglement between hemispheres. Access to the largest part of the bulk requires knowledge of the correlations on the biggest scales.

Let us pause briefly to consider what attributes of the geometry enabled the novel features described above such as \ETEBA\ geodesics penetrating to regions of arbitrarily high curvature.  The most obvious feature is the rapid time-dependence.  Indeed, the shell has a dramatic effect on geodesics which traverse it, especially near the point of implosion.
But compared to previous studies of extremal surface probes in Vaidya-AdS, further novel features arise due to the compactness of the horizon, i.e.\ by considering collapsing black hole with spherical, rather than planar, symmetry.  From the field theory point of view, this enables us to access finite-volume effects.  Indeed, we have seen that some of the surprising features arise only when the boundary endpoints are sufficiently nearly antipodal.
From the bulk standpoint, there are two different effects of the spherical geometry.  The one which is most crucial for the asymmetric radial geodesics is the fact that prior to the shell, the geometry has a smooth origin through which the geodesic may pass, heading back out to the boundary rather than through the Poincar\'e horizon which replaces it in the planar geometries.  The other effect, which is most crucial for the extremal surfaces, is that surfaces can pass around the black hole.

In \cite{Hubeny:2013gta} we considered the question of whether in a fixed bulk geometry, the area of smallest-area extremal surface can jump discontinuously as a function of the parameters specifying the surface (namely the size and time of the boundary region on which this bulk surface is anchored).  We argued that in the case of static spacetimes, where the extremal surface is in fact a {\it minimal} surface on a constant-time slice as required by the Ryu-Takayanagi prescription  \cite{Ryu:2006bv,Ryu:2006ef}, the area must vary continuously, which implies that correspondingly the entanglement entropy must vary continuously.\footnote{
See also the discussion in \cite{Headrick:2013zda} which appeared concurrently with our work.}  The argument however relied on minimality, so that the corresponding issue was not clear for the broader class of extremal surfaces which is relevant for time-dependent bulk geometries.

Indeed, the most general statement to the effect that areas of smallest-area extremal surfaces vary continuously with the boundary conditions is false. The results of \sect{s:geods} provide a manifest counter-example in the case of geodesics, being one-dimensional extremal surfaces, since we saw in \fig{f:antipodal} that the minimal length $\length(t)$ is discontinuous.  This is in sharp contrast to the naive thermalization picture where we would expect this quantity to grow monotonically and thermalize.  

On the other hand we found that found that, in all situations we considered, codimension-two extremal surfaces do vary continuously, thus rescuing entanglement entropy from the bizarre contingency of discontinuous jumps. The question of whether this is true in general remains open, though its failure for geodesics provides some guidance for any attempt at a proof, by ruling out arguments that would encompass all extremal surfaces. On the other hand, if there are situations in which the area may vary discontinuously, looking at asymptotically locally AdS spacetimes with non-planar boundary topology seems a good place to search for counterexamples. As seen here, this allows for a richer structure, with multiple branches of surfaces, which is likely to be a minimum requirement for a discontinuous exchange of dominance.

We now turn to the question of interpreting the results obtained for the lengths of the geodesics.
Using spacelike geodesics to probe the black hole has a long history; see e.g.\ 
\cite{Balasubramanian:1999zv},
\cite{Louko:2000tp},
\cite{Kraus:2002iv},
\cite{Fidkowski:2003nf},
\cite{Kaplan:2004qe},
\cite{Festuccia:2005pi},
and more recently revisited in 
\cite{Shenker:2013pqa},
\cite{Shenker:2013yza},
\cite{Andrade:2013rra}.
In the present work, perhaps the most fascinating result is the striking contrast between the conventional thermalization picture and the non-monotonic, discontinuous behaviour of the length $\length(t)$ along shortest \ETEBA\ geodesics with endpoints at time $t$, antipodally-separated, as illustrated in \fig{f:antipodal}.  Translated directly to the corresponding expectation for the equal-time CFT correlator of high-dimension operators, $\langle \phi \phi \rangle (t) \sim e^{-m \, \length(t)}$, this would be very bizarre.  However, we do not expect this to hold due to the subtlety that the geodesic may not lie on the path of steepest descent.

As argued in \cite{Fidkowski:2003nf} in the eternal \SAdS\ context (see also \cite{Festuccia:2005pi} for a complementary approach and \cite{Shenker:2013pqa} for more recent discussion in a broader context, more germane to the present case of interest), if the CFT correlator were dominated by the shortest spacelike geodesic (which bounces off the black hole singularity), the correlator would become singular when the insertion points are such that the joining geodesic approaches being null, which is ruled out by direct considerations in the CFT.  The resolution of this apparent puzzle comes from the fact that there are multiple (complexified) geodesics connecting the boundary points.  At the time-reflection-symmetric point they coincide, signalled by a branch point in the correlator.  By considering the resolution of this branch point, \cite{Fidkowski:2003nf} was able to show that the correlator is given by a sum of two complex branches.  But since the correlator is an analytic function in the position and time of the insertion points, one can recover the expected `light cone singularity' by analytic continuation.  

Here the situation is more complicated, since these methods are explicitly inapplicable if the spacetime itself is not analytic -- as in the present case of the shell having compact support.  This by itself might be circumvented by considering an analytic spacetime (which can be arbitrarily close to the present geometry and therefore the behaviour of the real geodesics will likewise be arbitrarily close to the present case), by making the shell profile analytic.  However, in this dynamical situation we will have lost the standard crutch of being able to use the Euclidean continuation.  
In \cite{Shenker:2013pqa} the authors consider such a situation, involving a shock wave in BTZ which is nonanalytic, and analytic approximations do not have real Euclidean continuations.  
For small non-analytic perturbations of the metric, the authors argue that indeed the saddle point represented by the perturbed geodesic continues to give the dominant contribution to the two-point function when the unperturbed one does.  
On the other hand, in the higher dimensional case where the shortest nearly-null geodesic of \cite{Fidkowski:2003nf} did not dominate the correlator of the eternal  \SAdS\ geometry, a `corresponding' geodesic continued from \SAdS\  to our Vaidya-AdS geometry will probably likewise not dominate.  
It would be useful to develop a robust and universal criterion for directly determining when a given shortest geodesic dominates the corresponding CFT correlator, without recourse to solving the wave equation.

In the second part we focused on spacelike surfaces which are anchored on a round $d-2$ sphere at constant time in $d$-dimensional boundary.  The motivation for this restriction was two-fold: from the pragmatic standpoint, this is the case which is simplest to solve when the bulk spacetime is spherically symmetric.  Although the specification of the entangling surface necessarily breaks the full symmetry,  for spherical regions we retain $SO(d-1)$, which  is inherited by the full extremal surface.  This is an enormous simplification, since the extremal surface is determined by coupled ODEs rather than PDEs.  By itself, this is a looking under the lamppost type motivation; however, choosing spherical regions has a separate reason, based on the expectation that for a fixed extent of the entangling surface on the boundary, the corresponding extremal surface reaches the deepest into the bulk.\footnote{
This was argued in \cite{Hubeny:2012ry} in case of planar AdS: deforming the entangling surface on the boundary, while keeping its extent (or volume enclosed) fixed, makes the bulk extremal surface recede towards the boundary.}
Hence for the question of how deep into the black hole can extremal surfaces reach, spherical entangling regions seem like the `best bet'.  However it would be useful to verify this expectation explicitly, by considering other entangling surfaces.
It would also be interesting to consider disconnected boundary regions, with multiple entangling surfaces.  Is there a constellation allowing the corresponding extremal surfaces to probe still deeper?

We have seen that the two types of probes we focused on, spacelike geodesics and codimension-two extremal surfaces, both probe inside the genuine black hole, but that each class accesses a different region inside the horizon.   One might then ask if there is a natural geometrical characterization of the region probed, without making direct reference to the probes.
 In other words, is there any special meaning to this region, especially from the CFT standpoint?  Such a characterization cannot be global like the event horizon, nor can it be quasi-local (spacetime foliation-dependent) like the apparent horizon.\footnote{
 A similar issue for planar charged collapsing shell was recently considered in \cite{Caceres:2013dma} which discussed connection between surfaces reaching past apparent horizon and strong subadditivity violation.
 }  It should also be something which only requires knowledge of the local part of the geometry, but at the same time it should allow for the richness of changing with dimension.
 
A weaker version of this question is whether there may be simple considerations giving bounds on the accessible region. In the present work, one likely candidate for such a bound was the critical surface corresponding to the maximal-area constant-$r$ surface inside the black hole, which we saw characterized the region covered very simply. 
Some steps in this direction have been made in \cite{Engelhardt:2013tra}, which provided criteria for surfaces bounding the region accessible by extremal surfaces anchored on a boundary. While no bound exists for geodesics, which access the whole spacetime, in may be possible to use their results to say something about codimension-two surfaces. It would be interesting to see how much can be said on such general grounds, and in particular how closely any bounds thus constructed come to characterizing the actual region probed.

Apart from the understanding they might provide in their own right, such bounds have a possible practical purpose in a study of thermalization requiring numerical evolution of the spacetime, such as formation of a black hole after sourcing some specific CFT operator for a time. The numerics do not allow for evolving the spacetime all the way to the singularity, but to study entanglement entropy via extremal surfaces, stopping at the horizon will be inadequate. The radius of maximal area, or any other more general bound that could be found, give a natural intermediate place to stop integration.\footnote{We thank M.\ Headrick for this suggestion.}

Since geodesics (1-dimensional extremal surfaces) and codimension-two (in $d+1$ dimensional bulk) extremal surfaces have such different behaviour in terms of their reach when $d>2$, one might naturally ask what happens to $n$-dimensional extremal surfaces with $1<n<d-1$ when $d>3$.  For example in $d=4$, string worldsheets corresponding to a Wilson loop would constitute such an intermediate case. While we expect the qualitative behaviour to be close to the case of the codimension-two surfaces, a more full comparison would be needed to check whether this is borne out.
Another natural generalization to consider would be more general spacetimes, for example adding charge, as in \cite{Caceres:2013dma}, and even causally trivial spacetimes may give interesting results (for example, see work on scalar solitons as in \cite{Nogueira:2013if,Gentle:2013fma}).

\acknowledgments 
It is a pleasure to thank Matt Headrick, Gary Horowitz, Hong Liu, Juan Maldacena, Don Marolf, Mukund Rangamani, Steve Shenker, and Tadashi Takayanagi for various illuminating discussions.  VH would like to thank CERN, ITF, Amsterdam, ICTP, Centro de Ciencias de Benasque Pedro Pascual and the Isaac Newton Institute for hospitality during this project.
HM is supported by a STFC studentship. 
 VH is supported in part by the STFC Consolidated Grant ST/J000426/1 and by FQXi Grant RFP3-1334.

\appendix

\section{Geodesics in Vaidya-BTZ}
\label{VaidyaBTZapp}

In this appendix we collect the details of the calculations of geodesics in the $d=2$ case of BTZ, in the limit of a thin shell, used in \sect{VaidyaBTZgeods}.

Firstly, we give the change of coordinates to the Penrose diagram as described in \sect{s:Vaidya}. In the BTZ part of the spacetime, after the collapse, for $v>0$, the coordinate transformation is given by
\begin{align}
v &= \frac{2}{\rh} \coth^{-1}\left(\frac{1}{\rh} \cot\frac{V}{2}\right) \\
r &= -\frac{\rh^2\tan\frac{V}{2}+\tan\frac{U}{2}}{1+\tan\frac{U}{2}\tan\frac{V}{2}}
\end{align}
and in the pure AdS by
\begin{align}
v &= V \\
r &= \tan\left(\frac{V-U}{2}\right)
\end{align}
Writing $T=\frac{V+U}{2}$ and $R=\frac{V-U}{2}$, the metric is
\begin{equation}
ds^2=\frac{-dT^2+dR^2}{\cos^2R}+r(T,R)^2 d\phi^2.
\end{equation}
One point that this choice of coordinates makes clear is that the metric is in fact continuous, which is not evident from the original coordinates. This implies that the tangent vectors of geodesics will change continuously across the shell, with no kink.

A striking feature of these coordinates is that the $T-R$ part of the metric is identical to pure AdS. It should be emphasized that this does not happen in higher dimensions, but is special to the BTZ case.

The most useful equations of motion will be:
\begin{align}
\dot{r}^2 &= E^2+\left(1-\frac{L^2}{r^2}\right)f(r) \\
\dot{v} &= \frac{\dot{r}+E}{f(r)}
\end{align}

To match cross the shell, we use the fact that  $\dot{v}$ is continuous. To get energy after shell crossing, we eliminate $\dot{r}$ from the above and use
\begin{equation}
E=\frac{f(r)\dot{v}}{2}-\frac{1}{2\dot{v}}\left(1-\frac{L^2}{r^2}\right).
\end{equation}

We first record the solutions for symmetric geodesics in the pure AdS part of the geometry, corresponding to zero energy there. The initial condition will be parameterized by the angular momentum $L$, which is also the minimum of $r$, and the static time slice it starts on, labelled by $\tau=\tan^{-1}r_0-v_0$, lying in the range $(\tan^{-1}L,\pi/2)$. For radial geodesics, starting at the origin, this gives the time before the formation of the shell; it is larger for starting points further in the past. The solution is

\begin{align*}
r &= \sqrt{L^2 \, \cosh^2 s + \sinh^2 s }
\\
\dot{v} &=\frac{\tanh{s}}{ \sqrt{L^2 \, \cosh^2 s + \sinh^2 s }}
\\
v &=  \tan^{-1} \sqrt{L^2 \, \cosh^2 s + \sinh^2 s } - \tau
\\
\ph &=  - \tan^{-1} ( L \,  \coth s )
\end{align*}

The shell is hit at 
\begin{equation}
s= \cosh^{-1} \frac{\sec \tau}{\sqrt{1+L^2}}
	= \frac{1}{2} \, \cosh^{-1} \frac{1-L^2 + 2 \, \tan^2 \tau}{1+L^2}
\end{equation}
at which point $v=0$, $r = r_s = \tan \tau$, and 
\begin{align*}
\dot{v} &= \cos \tau \, \sqrt{1-L^2 \, \cot^2 \tau}
\end{align*}

The next step is to extend into the BTZ part of the spacetime. Note that the $s$ in what follows is not the same parameter, but differs by some shift. This only matters for measuring length, where the two parts need to be added separately.

It turns out that it's very convenient to parametrize the radius of the BTZ horizon as $r_+ =\sec \mu + \tan\mu$, with $-\frac{\pi}{2}<\mu<\frac{\pi}{2}$. With this, we get the energy outside the shell as
\begin{equation}
E = - \frac{\cos\tau \, \sqrt{1-L^2 \, \cot^2 \tau}}{1-\sin \mu}.
\end{equation}
A useful piece of information is also the value of $\dot{r}$ after shell crossing, which is
\begin{equation}
\dot{r}(\rs) =  \frac{\sqrt{1-L^2 \, \cot^2 \tau}}{\cos\tau} \, \frac{(\sin^2 \tau - \sin \mu)}{(1-\sin \mu)}
\end{equation}  
 
\subsection{Symmetric radial geodesics}
We now specialize further to consider just the radial geodesics, with $L=0$. Outside the shell, the radial equation of motion can be obtained from the energy, and is
\begin{align*}
\dot{r}^2 &= r^2 +\frac{\sin^2\mu-\sin^2\tau}{(1-\sin\mu)^2}
\end{align*}
Combining this with the value of $\dot{r}$ after the shell, we find that the boundary is reached if and only if $\tau>\mu$, as argued in the text. In particular, if $\mu<0$ (corresponding to $r_+<1$), the geodesics will reach the boundary for all positive $\tau$.

We must now split into two cases, depending on whether $\tau$ is greater than or less than $|\mu|$.

\paragraph{$\tau>|\mu|$:} We begin with the case of earlier geodesics, relevant for any size of black hole. The simpler parts of the solution to obtain are

\begin{align*}
r &= \frac{\sqrt{\sin^2\tau-\sin^2\mu}}{1-\sin\mu}\cosh s \\
\dot{v} &= \frac{1-\sin\mu}{\sqrt{\sin^2\tau-\sin^2\mu}\sinh s+\cos\tau}
\end{align*}
 
and at the shell, we have
\begin{equation}
s = \log\left[\frac{1+\sin\tau}{\cos\tau} \sqrt{\frac{\sin{\tau}-\sin{\mu}}{\sin{\tau}+\sin{\mu}}}\right].
\end{equation}

Note that this can be positive or negative, depending on whether the geodesic is going inwards or outwards after crossing the shell.

By integrating $\dot{v}$ from this value to $\infty$, using the substitution
\begin{equation}
x=\frac{\sqrt{\sin^2\tau-\sin^2\mu}\,e^s+\cos\tau}{\cos\mu},
\end{equation}
we eventually get the time at which the boundary is reached:
\begin{equation}
t_\infty=\frac{1-\sin\mu}{\cos\mu}\log\left[\frac{\cos\left(\frac{\tau+\mu}{2}\right)}{\sin\left(\frac{\tau-\mu}{2}\right)}\right],
\end{equation}
or back in terms of $r_+$,
\begin{equation}
t_\infty=\frac{1}{\rh} \log\left[\frac{\sec\tau+\tan\tau+\rh}{\sec\tau +\tan\tau-\rh}\right].
\end{equation}

Finally, we extract the length from this, by the value of $s$ at the (large) cutoff $r=R$, minus the value of $s$ at the shell, plus the length from the pure AdS region:
\begin{align*}
\length &=2\log\left[\frac{2(1-\sin\mu)R}{\sqrt{\sin^2\tau-\sin^2\mu}}\right]
-2\log\left[\frac{1+\sin\tau}{\cos\tau}\sqrt{\frac{\sin{\tau}-\sin{\mu}}{\sin\tau+\sin\mu}}\right]\\ &\quad+2\log(\sec\tau+\tan\tau)\\
&= 2\log\left(\frac{1-\sin\mu}{\sin\tau-\sin\mu}\right)+2\log(2R),
\end{align*}
where the second term in the last line is the result in vacuum.

We can eliminate $\tau$ between these, to get
\begin{equation}\label{BTZradiallengths}
\length=2\log\left(\cosh^2\frac{r_+t}{2}+\frac{1}{r_+^2}\sinh^2\frac{r_+t}{2}\right)+2\log(2R)
\end{equation}
for $t>0$.

\paragraph{$\tau<|\mu|$:} The second case is only relevant for small black holes ($\mu<0$). The solution is

\begin{align*}
r &= \frac{\sqrt{\sin^2\mu-\sin^2\tau}}{1-\sin\mu}\sinh s \\
\dot{v} &= \frac{1-\sin\mu}{\sqrt{\sin^2\mu-\sin^2\tau}\cosh s+\cos\tau}
\end{align*}

and at the shell, the parameter is
\begin{equation}
s = \log\left[\frac{1+\sin\tau}{\cos\tau} \sqrt{\frac{\sin{\mu}-\sin{\tau}}{\sin{\mu}+\sin{\tau}}}\right].
\end{equation}

The time at which the boundary is reached, as well as the length, are computed in a similar way to the first case, and the resulting expressions are identical.

An alternative way to reach the same results is to calculate directly in the Penrose coordinates, which reduces to the simpler computation in pure AdS. The only extra work required is to check that the geodesics remain away from the singularity, and to work out how the coordinates transform at the boundary, to obtain $t_\infty$ and to make the correct regularization of the lengths.

\subsection{Region covered by geodesics}

We here flesh out the arguments in the text describing the regions of spacetime accessible to geodesics of various classes.

The radial motion in the BTZ part of the geometry is described by
\begin{equation}
\dot{r}^2 = 
r^2  - 
L^2 - \frac{\cos^2 \mu \, \left( 1-\frac{L^2}{r^2}  \right) -\cos^2 \tau \, (1-L^2 \, \cot^2 \tau)}{(1-\sin \mu)^2},
\end{equation}	
and $\dot{r}^2$ has a minimum at $r=r_0=\sqrt{L \, \rh}$. This minimum value $\dot{r}^2_\mathrm{min} \equiv \dot{r}^2(r_0)$ is given by 
\begin{equation}
\dot{r}^2_\mathrm{min} = \frac{\cos^2 \tau \, (1-L^2 \, \cot^2 \tau) - ( \cos \mu - L \, (1 - \sin \mu))^2}{(1-\sin \mu)^2}
\end{equation}	
which can take either sign, depending on the specific values of the parameters.

The fate of a given geodesic depends on the interplay of $\rh$, $L$ and $\rs$.  In particular, it can make it out towards the boundary only if either $\dot{r}^2_\mathrm{min} > 0 $ and $\dot{r}(\rs)>0$, or if $\dot{r}^2_\mathrm{min} \leq 0 $ but $\rs > r_0$.  The geodesic can remain for an arbitrarily long time $\Delta v$ in the vicinity of $r_0$, which happens in the former case as $\dot{r}^2_\mathrm{min} \to 0^+$, or in the latter case when $\dot{r}(\rs)<0$ and  $\dot{r}^2_\mathrm{min} \to 0^-$.

In either case, the desired fine-tuning is one which makes the magnitude of $\dot{r}^2_\mathrm{min}$ very small.  For a fixed black hole size $\mu$, the condition $\dot{r}^2_\mathrm{min}=0$ specifies a curve in the $\tau - L$ plane.  Solving for $L=L_0(\tau)$, we obtain
\begin{equation}
L_0(\tau; \mu) = 
\frac{\pm\cot\tau(\sin^2\tau-\sin\mu)-\cos \mu(1-\sin\mu) }
{(1-\sin\mu)^2 + \cos^2\tau \, \cot^2\tau}.
\label{}
\end{equation}	
In particular, asymptotically as $\tau \to 0$, we find that 
$L_0 \sim \pm (\sin \mu) \, \tau .
$

This motivates us to look at the family of geodesics with $L=-(\sin\mu)\tau$ for small black holes. We expand the relevant quantities for small $\tau$, and conclude that the whole spacetime is accessed by this family of geodesics as described in the text.

\section{Details of extremal surface computations}
\label{surfApp}

Extremal surfaces with the appropriate symmetries, parametrized by generic parameter $s$, are found from the Lagrangian
\begin{equation}
\mathcal{L}=(r\sin\theta)^{d-2}\sqrt{-f(r,v)\dot{v}^2+2\dot{r}\dot{v}+r^2}.
\end{equation}
The equations of motion, at a point where $\dot{v}=0$, give $\ddot{v}=(d-1)r\dot{\theta}^2>0$, which implies that any critical point for $v$ must be a minimum. This, along with the fact that $v$ must be increasing as the boundary is approached, tells us that $v$ has exactly one local minimum. Furthermore, the symmetries imply that this must occur when the surface crosses the pole of the sphere at $\theta=0$. This makes $v$ an appropriate candidate for a parameter along the surface.

Denoting differentiation with respect to $v$ by primes, the Lagrangian with the parameter $v$ is
\begin{equation}
\mathcal{L}=(r\sin\theta)^{d-2}\sqrt{-f(r,v)+2r'+r^2\theta'^2},
\end{equation}
which gives the equations of motion
\begin{align} 
 r'' &= \frac{3r'-f}{2}\frac{df}{dr}+(f-r')\left((d-1)r\theta'^2+\frac{d-2}{r}(-f+2r')\right) \\
 \theta''&=\left(\frac{1}{2}\frac{df}{dr}-\frac{f}{r}\right)\theta'-\left(\frac{d-1}{r}\theta'-\frac{d-2}{r^2}\cot\theta\right)(-f+2r'+r^2\theta'^2).
\end{align}
Initial conditions are chosen as the values of $v$ and $r$ as the pole $\theta=0$ is crossed. Since this is a singular point of the equations, for numerics we must start integration slightly away from this point, picking initial conditions by a series solution:

\begin{equation}
\theta(v_0+h)=\sqrt{\frac{2h}{r_0}}\,\left(1+O(h)\right), \quad r(v_0+h)=r_0+f(r_0,v_0)\, h+O(h^2).
\end{equation}

This is altered in the special case where the surface is equatorial; this means that it passes through $r=0$ and has $\theta=\frac{\pi}{2}$. The solution will always be the same close to the origin, namely lying on a static slice of pure AdS until it meets the shell, which, and this is how initial conditions are specified.

The areas are found by numerically integrating the difference between the Lagrangian and
\begin{equation}
\frac{(r \sin\theta)^{d-2}}{\sqrt{1+r^2\sin^2\theta}}(r'\sin\theta +r\theta' \cos\theta),
\end{equation}
which is the derivative of the function which gives the area of a minimal surface in AdS passing through the point $r,\theta$. This automatically regularizes the area by subtracting off the background vacuum value.

\bibliographystyle{JHEP}
\bibliography{vaidya}

\end{document}